\documentclass[12pt,a4paper]{article}
\usepackage{amsmath,geometry}
\usepackage{amsmath}
\usepackage{amsfonts}
\usepackage{amssymb}
\usepackage{verbatim}
\usepackage{graphicx}
\usepackage{multirow}
\usepackage{tablefootnote}
\usepackage{float}
\usepackage{empheq}
\usepackage[all]{xy} 
\usepackage{color}
\usepackage{hyperref}
\usepackage{indentfirst}
\usepackage{cite}
\usepackage{array}

\usepackage[ruled,vlined]{algorithm2e}
\SetKw{Continue}{continue}
\SetKw{Break}{break}
\SetKw{And}{and}

\newcommand{\nc}{\newcommand}
\nc{\beq}{\begin{equation}}
	\nc{\eeq}{\end{equation}}
\nc{\bea}{\begin{eqnarray}}
	\newcommand{\bal}{\begin{aligned}}   \newcommand{\eal}{\end{aligned}}
	\nc{\eea}{\end{eqnarray}}

\def\cO{{\cal O}}
\def\IP{\mathbb{P}}
\def\cO{\mathcal{O}}

\def\cI{\mathcal{I}}
\def\cG{\mathcal{G}}
\def\bW{\mathbf{W}}

\def\ra{\rightarrow}
\def\fto{\longrightarrow}

\def\clap#1{\hbox to 0pt{\hss#1\hss}}

\newcolumntype{P}[1]{>{\centering\arraybackslash}p{#1}}

\newdimen\csize\csize=1.5ex
\def\young#1{\tiny\vcenter{\hbox{\vrule\vtop{\hrule
				\offinterlineskip\halign{&\vbox
					{\hbox to\csize {\strut\hss##\hss\vrule}\hrule}\cr#1 \crcr}}}}}

\def\fnote#1#2{\begingroup\def\thefootnote{#1}\footnote{#2}
	\addtocounter{footnote}{-1}\endgroup}

\def\IP{\mathbb{P}}

\def\IZ{\mathbb{Z}}

\def\IC{\mathbb{C}}

\def\cA{\mathcal{A}}

\def\cF{\mathcal{F}}
\def\cI{\mathcal{I}}

\def\cN{\mathcal{N}}
\def\cO{\mathcal{O}}

\def\Calabi-Yauan{\color{Calabi-Yauan}}

\def\dburl{\url{https://github.com/GroupofXG/anewcydatabase/}}

\begin{document}
	
	\vspace*{-1.5cm}
	\begin{flushright}
		{\small
		}
	\end{flushright}
	
	\vspace{1.5cm}
	\begin{center}
		{\LARGE  Orientifold Calabi-Yau Threefolds: \\ \vskip0.2cm Divisor Exchanges and Multi-Reflections}
	\end{center}
	
	\vspace{0.75cm}
	\begin{center}
		{Xu Cao$^{\dagger}$, Hongfei Gao$^{\dagger}$, and Xin Gao$^{\dagger}$}
	\end{center}
	
	\vspace{0.1cm}
	\begin{center} 
		{\small 
			{\small 
				$^{\dagger}${\it College of Physics, Sichuan University, Chengdu, 610065, China}\\
			}

			\fnote{}{
				\hskip-0.65cm 
			        2023222020001@stu.scu.edu.cn\\
				2021222020001@stu.scu.edu.cn\\
				xingao@scu.edu.cn
				}
		}
	\end{center}
	
	\vspace{0.2cm}

	\vspace{1cm}
	
	\begin{abstract}
		\noindent

Using the Kreuzer-Skarke database of 4-dimensional reflexive polytopes, we systematically constructed a new database of orientifold Calabi-Yau threefolds with  $h^{1,1}(X) \leq 12$. Our approach involved  non-trivial $\IZ_2$ involutions, incorporating both divisor exchanges and multi-divisor reflections acting on the Calabi-Yau threefolds. Each proper involution results in an orientifold Calabi-Yau threefolds and we constructed $320,386,067$ such examples. We developed a novel algorithm that significantly reduces the complexity of determining all the fixed loci under the involutions, and clarifies the types of O-planes. Our results show that under  proper involutions, the majority of cases end up with $O3/O7$-plane systems, and most of these further admit a naive Type IIB string vacua. Additionally, a new type of free action was determined. We also computed the smoothness and the splitting of Hodge numbers in the $\IZ_2$-orbifold limit for these orientifold Calabi-Yau threefolds.

	\end{abstract}
	
	\clearpage

	\tableofcontents
	
	\pagebreak
	
	\section{Introduction}
	\label{sec:intro}

String theory, aiming to describe the fundamental forces of the universe within a single theoretical framework, often requires compactification from higher dimension to lower dimension to properly describe observable phenomena. Among various compactification methods, four-dimensional $\cN=1$ supersymmetric compactifications are particularly well-studied due to their theoretical tractability. Within this framework, the compactification manifold usually is  a Calabi-Yau threefold  $X$. Compactifying type IIA or type IIB string theories on Calabi-Yau threefolds $X$ yields an $\cN=2$ supersymmetry theory in four dimensions. Therefore, to break half of the supersymmetry to $\cN=1$, and to address the necessity of O-planes for tadpole cancellation when including open string modes such as D-branes and fluxes, an additional orientifold projection with a proper $\IZ_2$ involution $\sigma$ acting on the Calabi-Yau threefold $X$ is required. In this paper, we employ a new strategy and a novel algorithm, significantly extending our previous work \cite{Gao:2013pra, Altman:2021pyc},  to construct orientifold Calabi-Yau threefolds with $h^{1,1}(X) \leq 12$, considering both non-trivial divisor exchange involutions and multi-divisor reflection involutions.

Our investigation focuses specifically on Type IIB orientifold geometries, where an orientifold projection $\mathcal{O}$ involves two key components: the worldsheet parity $\Omega_p$ and a diffeomorphism map $\sigma$ acting on the Calabi-Yau threefolds $X$, referred to as the involution. Notably, the involution $\sigma$ such that $\sigma^2=1$ must meet certain criteria to preserve supersymmetry, thus it has to be isometric and  holomorphic \cite{Acharya:2002ag, Brunner:2003zm}. Under this action, the compactification manifold may feature fixed loci, corresponding to orientifold planes (O-planes), whose structure depends on the specific orientifold system, such as the $O5/O9$ and $O3/O7$ systems:
\begin{eqnarray}
\label{eq:orientifold}
 {\cal O}= \begin{cases}
                       \Omega_p\, \sigma \qquad &{\rm with} \quad
                       \sigma^*(J)=J\,,\quad  \sigma^*(\Omega_3)=+\Omega_3,  \quad  $O5/O9$\,\, {\rm system} ,\\[0.1cm]
                       (-1)^{F_L}\,\Omega_p\, \sigma\qquad & {\rm with} \quad
        \sigma^*(J)=J\,, \quad \sigma^*(\Omega_3)=-\Omega_3, \quad    $O3/O7$\,\, {\rm system},
\end{cases}
\end{eqnarray}
Each of the involution $\sigma (\sigma^2=1)$ defines a new Calabi-Yau in the orbifold limit unless it is a free action.  In general, the involution $\sigma$ splits the cohomology groups $H^{p,q}(X/\sigma^{*})$ into eigenspaces of even and odd parity:
 \bea
 H^{p,q}(X/\sigma^{*})=H^{p,q}_{+}(X/\sigma^{*})\oplus H^{p,q}_{-}(X/\sigma^{*}).
 \eea
For reflection involution, the equivariant cohomology $h^{1,1}_-(X)$ is always vanishing while for  divisor exchange involution it is usually   $h^{1,1}_-(X) > 0$:
  \begin{eqnarray}
 \sigma =    \begin{cases}
                 { {\rm Reflection:}}\,\, \{\,x_i \leftrightarrow -x_i, \cdots\,\} &  h^{1,1}_- (X)= 0  \\[0.1cm]
                        { {\rm Exchange\,\, involution:}}\,\, \{\,x_i \leftrightarrow x_j, \cdots \,\} & {h^{1,1}_- (X) > 0} 
\end{cases}
\end{eqnarray}

Orientifold Calabi-Yau threefolds  play an important role in string phenomenology for both particle physics and string cosmology.
Recently, in the context of the swampland conjecture(first proposed in \cite{Vafa:2005ui, ArkaniHamed:2006dz} and see \cite{Brennan:2017rbf, Obied:2018sgi, Palti:2019pca, vanBeest:2021lhn} for a detail review), various corrections, such as warping correction, loop correction, and $\alpha'$ correction, put a constraint on the orientifold Calabi-Yau that can be used for compactification to construct de-Sitter vacua in the Large Volume Scenario (LVS) \cite{Balasubramanian:2005zx, Bena:2020xrh, Junghans:2022exo, Gao:2022fdi, Junghans:2022kxg, Gao:2022uop, McAllister:2024lnt}.  
In order to solve the chirality issue when combining the local particle physics and moduli stabilization \cite{Blumenhagen:2007sm}, tuning on fluxed-instanton on the divisors was considered \cite{Grimm:2011dj,Hebecker:2011hk,Kerstan:2012cy}.
There is also a selection rule determining which kinds of  orientifold Calabi-Yau threefolds can  support the global embedding of Standard Model at the toric singularity in the geometry, where both reflections with $h^{1,1}_-(X) =0$ and divisor exchange involutions with $h^{1,1}_- (X)> 0$ were considered \cite{Wijnholt:2002qz, Cicoli:2011qg, Balasubramanian:2012wd, Cicoli:2012vw, Cicoli:2016xae, Cicoli:2017shd, Cicoli:2017axo, Cicoli:2021dhg, Bera:2024sbx, Bera:2024zsk}.

Recently, several databases of orientifold Calabi-Yau threefolds  have been established.  Initially, the orientifold Calabi-Yau threefolds with divisor exchange involutions were constructed for $h^{1,1}(X)\leq 4$  \cite{Gao:2013pra} using maximal triangulations from the Kreuzer-Skarke dataset of reflexive four-dimensional polyhedra \cite{Kreuzer:2000xy}.  This toric construction was extended to  $h^{1,1}(X)\leq 6$, with explicitly fixed loci and types of O-planes  in \cite{Altman:2021pyc} based on the Calabi-Yau database \cite{Altman:2014bfa}.  Given the fact that Calabi-Yau manifold compatible with the proper exchange involution are rare in the entire database, it is a very good signal to apply machine learning tecnique to identify those polytopes which can result in an orientifold Calabi-Yau threefold \cite{Gao:2021xbs}. Such orientifold structure from the polytope perspective was  later studied in \cite{Moritz:2023jdb}. The orientifold  Calabi-Yau threefolds with single divisor reflection involutions were considered in \cite{Crino:2022zjk}. More general free quotient in the toric Calabi-Yau with $h^{1,1}(X)\leq 3$  were systematically explored  in~\cite{Braun:2017juz}.
 In the context of Complete Intersection Calabi-Yau 3-folds (CICYs) embedded in products of projective spaces \cite{Candelas:1987kf}, people start to construct a landscape of orientifold vacua \cite{Carta:2020ohw} from the most favorable description of CICY 3-folds database \cite{Anderson:2017aux}.  General free quotients have been classified and studied in the case of CICY 3-folds ~\cite{Braun:2010vc, Candelas:2015amz, Constantin:2016xlj, Gray:2021kax}.

Despite significant progress in constructing orientifold Calabi-Yau threefolds, technical challenges persist when extending to higher 
 $h^{1,1}(X)$.  The first obstacle is the huge number of Calabi-Yau threefolds itself. Using graph theory and neural network techniques, it has been shown that the number of fine, regular, star triangulations (FRSTs) of four-dimensional reflexive polytopes is bounded by $\cO(10^{928})$, with topologically inequivalent ones  bounded by $\cO(10^{428})$ \cite{Demirtas:2020dbm}.
Although divisor exchange involution compatible with the geometry are rare in the database, one can always perform the multi-divisor reflection involutions, which is missing in literatures. Each  involution  results in an orientifold Calabi-Yau manifold for which we must determine the fixed loci and types of O-planes. However, the computational complexity of determining these fixed loci often exceeds the capabilities of current computer.  Therefore, developing a more efficient algorithm is crucial  for  extending our previous work to a much larger database with higher $h^{1,1}(X)$, including both divisor exchange involutions and reflections,  and  clarifying  ambiguities in determining the fixed loci. 

In this paper, we extend and improve upon previous work in several key areas, including large-scale construction, increased efficiency, and clarification of ambiguities:
\begin{enumerate}
\item
We extend our  classification  bound up to $h^{1,1}(X)=12$ in the Calabi-Yau database constructed from the Kreuzer-Skarke list~\cite{Kreuzer:2000xy}.  For  $h^{1,1}(X) \leq 7$,  we expand our analysis to hypersurfaces in all possible maximal projective crepant partial (MPCP) desingularizations. For $8 \leq h^{1,1}(X) \leq 12$, we randomly choose some favorable polytopes for each hodge number as in \cite{Crino:2022zjk}. The number of toric triangulations we will analyze increases by one order of magnitude from $653,062$ \cite{Altman:2021pyc} to $6,696,714$.
 
\item We expand our classification to cover both divisor exchange involutions and multi-divisor reflection involutions.
First,  we determine the individual topology of divisors in each  toric Calabi-Yau threefold and  refine our algorithm to identify all proper divisor exchange involutions, resulting in a total of {$156,709$} proper divisor exchange involutions.
For multiple reflection involutions, we explore all possible single, double, and triple divisor reflections for Calabi-Yau threefolds with $h^{1,1}(X) \leq 6$. For $h^{1,1}(X) = 7$, we analyze all single reflections, and randomly chosen 15 double and 15 triple reflections. For $8 \leq h^{1,1}(X) \leq 12$, we randomly select 15 triple reflections in addition to single reflections. In total, we examine $320,229,358$ different types of reflections. Consequently, we determine their equivariant cohomology (Hodge number splitting) in the $\mathbb{Z}_2$-orbifold limit.

Since each of the such involutions results in an orientifold Calabi-Yau manifold, combining both  types of involutions,  we constructed a total $320,386,067$ of orientifold Calabi-Yau threefolds in our new database (\dburl), which is three orders of magnitude ($\cO(10^3)$) larger than our previous work \cite{Altman:2021pyc}.

\item
We employed new strategy to identify all possible fixed loci under divisor exchanges and reflection involutions.  This enables us to identify the positions of various types of O-planes, which are crucial for D-brane constructions. The new algorithm significantly reduces the calculation complexity for determining fixed loci by more than five order  of magnitude ($\cO(10^5)$) for a standard Calabi-Yau threefolds with $h^{1,1}(X) = 7 $ and will be more efficient for higher $h^{1,1}(X)$. 

\item
We classify freely acting involutions on Calabi-Yau threefolds into two categories:  those with a fixed locus in the ambient space that does not intersect with the Calabi-Yau threefold and those even without a fixed locus in the ambient space.
\item 
In orientifold Calabi-Yau threefolds featuring the $O3/O7$-system, we proceed to classify the so-called  \lq\lq naive orientifold Type IIB string vacua" by considering the D3 tadpole cancellation condition when putting eight $D7$-branes on top of $O7$-plane. 
\end{enumerate}

This paper is organized as follows: In Section \ref{subsec:review PTT}, we briefly review the construction of Calabi-Yau threefolds as hypersurfaces in toric varieties and how to compute the Hodge numbers of individual toric divisors. In Section \ref{subsec:involution}, we identify all pairs of ``Non-trivial Identical Divisors'' (NIDs) and then present the proper divisor exchange involutions  and multi-divisor reflections. The fixed-point loci  on the ambient space are then identified in Section \ref{subsec:newalgorithm} and then restricted on the Calabi-Yau threefolds in Section \ref{subsec:fixedX}. Computation complexity of the previous algorithm was described  in Section \ref{subsubsec: complexity} and great efforts were performed to introduce a new strategy to reduce the computation complexity in Section \ref{subsubsec:newalgorithm}. 
This information is used to classify the involutions as either non-trivially or freely acting in Section \ref{subsec:smoothfree}. %
We illustrate the procedures via detailed examples for both divisor exchange involutions and multi-divisor reflections in Section \ref{sec:example} and clarify some ambiguities there. Then we summarize our results in Section \ref{sec:result}.


	\section{Construct Orientifold Calabi-Yau Threefolds}
	\label{sec:construct}
	\subsection{Polytope, Triangulation and Divisors}
	\label{subsec:review PTT}

The toric Calabi-Yau threefolds $X$  can generically be obtained by taking the anticanonical hypersurface in an ambient four-dimensional Gorenstein toric Fano variety, denoted by  $\mathcal{A}$~\cite{Batyrev:1993}. 
First, we require a desingularized  representation of our original four-dimensional ambient space $\mathcal{A}$ from the Kreuzer-Skarke database  \cite{Kreuzer:2000xy}.  Achieving this entails smoothing out some irregularities in a four-dimensional ambient space $\mathcal{A}$,  by blowing up enough of its singular points through a process called maximal projective crepant partial (MPCP) desingularization. 
This method involves triangulating the polar dual reflexive polytope, denoted as $\Delta^{*}$, to ensure it contains at least one fine, regular, star triangulation (FRST)\footnote{A triangulation is \lq\lq fine" if all points not interior to facets appear as vertices of a simplex. \lq\lq regularity" is needed so that variety is projective and K\"ahler.  It is \lq\lq star" if the origin is vertex of all full-dim simplices. }. Such a desingularized four-dimensional ambient  toric variety could be expressed as:
\bea\label{eq:defamb}
\cA=\frac{\mathbb{C}^{k}- Z}{\left(\mathbb{C}^{*}\right)^{k-4}\times G}\, ,
\eea
where $Z$ is the locus of points in $\mathbb{C}^{k}$ ruled out by the Stanley-Reisner ideal $\mathcal{I}_{SR}(\cA)$, and $G$ is the stringy fundamental group. 
There are only 16 manifolds in the Kreuzer-Skarke list contain a non-trivial first fundamental group $G$.
 In many applications in physics the group $G$ can be taken to be trivial and we are left with the simple split torus action $\left(\mathbb{C}^{*}\right)^{k-4}$ in the denominator of the quotient. This torus $\IC^*$ action  can also take the role of an abelian gauge group acting on some two-dimensional field theory as was explained in \cite{Witten:1993yc} and may therefore also be called the $U(1)$ action.
  
 We will consider the ambient space $\mathcal{A}$ as a resolved four-dimensional Gorenstein toric Fano variety whose anticanonical divisor $X=-K_{\mathcal{A}}$ represents a Calabi-Yau threefold hypersurface. 
Here we restrict  ourselves to the so-called \lq\lq {\it favorable}" description, in which  the toric divisor classes on the Calabi-Yau hypersurface $X$ are all descended from ambient space $\cA$ \footnote{There exits a stronger notion of \lq\lq K\"ahler favorability" where K\"ahler cones on $X$ descend from an ambient space in which they are embedded \cite{Anderson:2017aux}. This involves a detailed argument regarding the descent of the effective, nef, and ample cones of divisors,  which we refer the reader to see \cite{Oguiso}. In some cases, the \lq\lq favorable” geometry is not \lq\lq K\"ahler favorable” because the K\"ahler cone of $X$ is actually larger than the positive orthantg. For example in gCICY cases~\cite{Anderson:2015iia}  with negative entries in the defining configuration matrix, the K\"ahler cone of $X$ is usually enlarged.}.
The following short exact sequence and its dual sequence
 \bea
  0 \ra TX \ra T\cA|_X \ra \cN_{X/\cA} \ra 0, \\\nonumber
    0 \ra \cN^*_{X/\cA} \ra T^*\cA|_X \ra T^*X \ra 0,
 \eea
can induce the long exact sequence in sheaf cohomology
 {\small
\bea\label{eq_generalizedkoszulwithdiviso}
      \parbox{0.1cm}{\xymatrix{
          &
           \hspace*{-1cm} \cdots \fto H^{1} (X, \cN^*_{X/\cA})   \xrightarrow{\quad\quad\alpha\quad\quad}    &
         {H^{1} (X, T^*\cA|_{X})}  \xrightarrow{\quad\quad\quad}  &
      H^1(X, T^*X) \ar`[rd]`[l]`[dlll]`[d][dll] &
         \\ &
           \quad  H^{2} (X, \cN^*_{X/\cA})   \xrightarrow{\quad\quad\beta\quad\quad}     &
               {H^{2} (X, T^*\cA|_{X})}  \xrightarrow{\quad\quad\quad}  &
             H^2(X, T^*X)  \fto \cdots\,. &
        }}
    \eea
}
From the above sequence and  Dolbeault's theorem, $H^1(X, T^{*}X) \cong H^{1,1}(X) \cong {\rm coker }(\alpha) \oplus {\rm ker} (\beta)$. This equivalence has two parts. One comes from the restriction of K\"ahler moduli from $\mathcal{A}$ to $X$, and the other comes from the kernel part. If the kernel part is empty, we call the geometry  favorable and $h^{1,1}(X) ={\rm  dim}(H^{1,1} (X)) \cong {\rm dim}({\rm Pic}(\cA))$.  

 Denote $x_i$ as the weighted homogeneous coordinates used to define $X$ inside the ambient space $\cA$. Then the divisor $D_{i}\equiv\{x_i=0\}$ defines a 4-cycle on $X$ and it is dual to a 2-cycle $\omega_i$, i.e., $D_i \in H^{1,1}(X,\IZ)$.
 Due to the favorability description of polytopes and geometries, all such  toric divisors $D_i$ are irreducible on the Calabi-Yau threefolds $X$\footnote{When the geometry is unfavorable,  it contains divisor with disconnected pieces like $\IP^n \cup \dots \cup \IP^n$, $dP_n \cup \dots \cup dP_n$ or others.}. Hence, $H_4(X,\IZ)$ is generated by any basis constructed from $D_i$, $i = 1,..., k = h^{1,1}(X) + 4.$
 Now, the Calabi-Yau threefolds on this toric variety can be described by a polynomial in terms of  the projective coordinates $\{x_{1},...,x_{k}\}$. Their torus $\left(\mathbb{C}^{*}\right)^{k-4}$ equivalence classes reads
\bea
(x_{1},...,x_{k})\sim (\lambda^{\mathbf{W}_{i1}}x_{1},...,\lambda^{\mathbf{W}_{ik}}x_{k})\, ,
\eea
where $\mathbf{W}$  is the GLSM weighted matrix of rank $r= k-4$, charged under $r$ $U(1)s$. 

 The internal topology of these divisors play an important role in string compactification and moduli stabilization.
The Hodge numbers of divisors are collectively denoted as
\bea
h^\bullet(D, \cO_{D}) = \{h^{0,0}(D), h^{1,0}(D), h^{2,0}(D), h^{1,1}(D)\}
\eea
For an irreducible divisor $D$, the complex conjugation $h^{p,q}(D)=h^{q,p}(D)$ and Hodge star $h^{p,q}(D)=h^{2-p,2-q}(D)$ dualities constrain the independent Hodge numbers of $D$ down to only $h^{1,0}(D), h^{2,0}(D)$, and $h^{1,1}(D)$. 
This calculation will be performed by using the Koszul extension to the \texttt{cohomCalg} package \cite{Blumenhagen:2010pv, cohomCalg:Implementation} with the \texttt{HodgeDiamond} module. When we encounter difficulties with \texttt{cohomCalg} in calculating Hodge number of a divisor, we first calculate the Euler number $\chi(D)=\int_{D}{c_{2}(D)}$ of the divisor on the hypersurface, and then determine $h^{1,0}(D)$ and $h^{2,0}(D)$ by calculating the trivial line bundle cohomology of the divisor $h^\bullet(D, \cO_{D})$ \cite{Braun:2017nhi}. Then, using the expression
\bea
\chi(D) = \sum\limits_{i=0}^{2} (-1)^i \,{\rm dim}\left(H^i_{\rm DR}(D)\right) = \sum\limits_{p+q=0}^{2}(-1)^{p+q} \,{\rm dim}\left(H^{q}(D,\Omega^{p})\right)\, ,
\eea
we can fix $h^{1,1}(D)$ and get the full Hodge diamond for any divisor.
 In our procedure for scanning divisor involutions, several  types of divisors are of particular phenomenological interest:\\

\noindent {\it Completely rigid divisors:}  The Hodge numbers of these divisors are characterized by 
$h^\bullet(D) = \{1,0,0,h^{1,1}(D)\}$  such that  $h^{1,1}(D) \neq 0$. 
This group of divisors falls into two categories: del Pezzo surfaces, denoted by $\{\IP^2 \equiv dP_0$, $dP_{n}$, with $n=1, \dots , 8\}$ and $n=h^{1,1}(D)-1$.  These del Pezzo divisors are usually shrinkable depending on the diagonalizability of their intersection tensor. The shrinkable del Pezzo surface plays a crucial role to generate a non-perturbative superpotentail \cite{Witten:1996bn}, which is important for KKLT \cite{Kachru:2003aw} and LVS \cite{Balasubramanian:2005zx} construction. For those divisors with $h^{1,1}(D)>9$, they are always referred as  \lq\lq non-shrinkable rigid divisors".\\

\noindent {\it``Wilson'' divisors:} The Hodge numbers of these divisors are characterized by
$h^\bullet(D) = \{h^{0,0}(D), h^{0,1}(D), $ $h^{0,2}(D), h^{1,1}(D)\} =\{1, h^{1,0},0, h^{1,1}\}$ with $h^{1,0}(D),\, h^{1,1}(D) \neq 0$.
We will also further specify the ``Exact-Wilson'' divisor as $h^{\bullet}(D)=\{1,1,0,h^{1,1}\}$ with $h^{1,1} (D)\neq 0$ which are crucial for supporting poly-instanton inflation\cite{Cicoli:2011ct,Blumenhagen:2012kz,Blumenhagen:2012ue}. \\

\noindent {\it Deformation divisors:}  These divisors are characterized simply by $h^{2,0}(D) \neq 0$.
\begin{itemize}
\item
A K3 divisor is a deformation divisor with Hodge numbers $h^{\bullet}(D)=\{1,0,1,20\}$, which is used to generate fiber inflation \cite{Cicoli:2008gp}.
\item
A  deformation divisor resembles a K3 divisor but includes an additional $h^{1,1}(X)$ deformation degree of freedom, i.e., $h^{\bullet}(D)=\{1, 0, 1, 21\}$.  We refer to this type of divisor as a type-1 special deformation divisor,  denoted by $SD1$.
\item
In our scan, a type-2 special deformation divisor, denoted as $SD2$, frequently appears with Hodge numbers $h^{\bullet}(D)=\{1, 0, 2, 30\}$.
\end{itemize}

\subsection{Proper Involutions from Divisor Exchanges and Reflections}
\label{subsec:involution}

We expand our classification of involutions to cover both divisor exchange involutions and multi-divisor reflection involutions compared with previous work \cite{Gao:2013pra, Altman:2021pyc}.
For divisor exchange involutions, the map $\sigma:  x_i \leftrightarrow x_j$, which swaps two homogeneous coordinates in the ambient toric variety $\cA$,  induces a holomorphic involution  $\sigma^{*}: D_i \leftrightarrow D_j$ on the corresponding toric divisor cohomology classes. 
On favorable manifolds, this involution restricts in a straightforward way to the Calabi-Yau hypersurface $X$. We then define the even and odd parity eigendivisor classes $D_\pm \equiv D_i \pm D_j \in H^{1,1}_{\pm}(X/\sigma^{*})$. In general, a given geometry may allow multiple disjoint involutions $\sigma_1, \sigma_2,\dots,\sigma_{n}$. In this case, the full involution is given by $\sigma\equiv\sigma_{1}\circ\sigma_ {2}\circ\cdots\circ\sigma_{n}$. 

Consequently, it is necessary to identify the proper involution $\sigma$ that exchanges one or more pairs of divisors. These divisors should share the same topology but have different charge weights. We call such pairs of divisors as non-trivial identical divisors (NIDs) and it can be summarized as:
\bea
\sigma: x_i \leftrightarrow x_j \quad \Longleftrightarrow \quad \sigma^*: D_i \leftrightarrow D_j. \nonumber\\
H^{\bullet} (D_i) \cong H^{\bullet} (D_j),  \quad  \cO(D_i) \neq \cO(D_j)  
\eea
Furthermore, such involution should satisfy the symmetry of Stanley-Reisner ideal  $\cI_{SR} (\cA)$ and the symmetry of  the linear ideal $\cI_{lin} (\cA)$. 
The first symmetry ensures that the involution is an automorphism of $\cA$, preserving the exceptional divisors from resolved singularities. The second symmetry ensures that the defining polynomial of the Calabi-Yau manifold remains homogeneous under the involution.
Putting these two together, the involution should be a symmetry of the Chow-group:
\bea
\label{eq:chow}
A^{\bullet}(\cA)  \cong \frac{\IZ ( D_1, \cdots, D_k)}{\mathcal{I}_{lin}(\cA)+ \mathcal{I}_{SR}(\cA)}\, , 
\eea
Due to the favorability condition on the Calabi-Yau threefold hypersurface we have
\bea
\label{eq:favor}
A^1(\cA) \cong H^{1,1}(\cA) \cong  {\rm Pic} (\cA) \cong {\rm Pic} (X) \cong H^{1,1}(X)  \cong A^1(X)\, ,
\eea
and thus the toric triple intersection number defined in the Chow ring $A^4(X)$  should also be required to be  invariant under the involution $\sigma$. Only when an involution, exchanging pairs of NIDs, satisfying all these requirements described above, can be called a \lq\lq proper" involution.

For reflection involutions on divisors, these are pulled back from the coordinate reflections on the ambient space $\cA$:
\bea
\sigma: x_i \leftrightarrow - x_i \quad \Longleftrightarrow \quad \sigma^*: D_i \leftrightarrow -D_i.
\eea 
 The situation is much simpler since it always 
satisfies the conditions for \lq\lq proper" involutions.  So we will explore  possible single, double and triple divisor reflections for Calabi-Yau threefolds.

One crucial difference between divisor exchange  and reflection involutions is the Hodge number splitting structure. The holomorphic condition requires that the pullback $\sigma^{*}$ maps $(p,q)$-forms on $X$ to $(p,q)$-forms on $\cA$. This is also true at the level of cohomology as the Dolbeault operator $\bar\partial$ commutes with the pullback $\sigma^*$. This implies that in the orientifold limit,  the dimensions of equivariant cohomology split as:
\bea
\label{eq:hodgesplit}
h^{p,q}(X/\sigma^{*})=h^{p,q}_{+}(X/\sigma^{*}) \, + h^{p,q}_{-}(X/\sigma^{*})\, .
\eea
The reflection involution $\sigma^{*}: D_i \leftrightarrow - D_i$ acts trivially on the divisor classes and thus  manifestly does not contribute to $h^{1,1}_-(X/\sigma^{*})$. 
However, the divisor exchange involution  $\sigma^{*}: D_i \leftrightarrow  D_j$   acts non-trivially on the divisor classes and thus contributes to the non-trivial odd cohomology $h^{1,1}_-(X/\sigma^{*})$.  In order to determine whether the $h^{1,1}_\pm (X/\sigma^*)$ split,  we should expand the K\"ahler form $J$, which has even parity under $\sigma$ and thus $J \in H^{1,1}_+(X)$,  in terms of  those  divisor classes.  By defining even and odd parity eigendivisors $D_{\pm}=D_{i}\pm D_{j}$, one can expand $J$  in the new divisor basis including $D_+ \in H^{1,1}_+$. This leads to a specific form for $J$ and we can read off the Hodge number splitting of $h^{1,1}_\pm$ by the number of independent expansion coefficients.

Furthermore, we can determine the Hodge number splitting of $h^{2,1}_\pm(X/\sigma^*) $ in the $\IZ_2$-orbifold limit by 
 Lefschetz fixed point theorem \cite{Shanahan, Blumenhagen:2010ja}.  In general the $\IZ_2$ involution $\sigma$ induced  a fixed-point set ${\cal F}$.  Due to the hodge number splitting eq.(\ref{eq:hodgesplit}), we can 
define the Leftschetz number of $\sigma^*$ as $ L(\sigma^*, X)$:
\bea
 && L(\sigma^*, X) \equiv \sum_i (-)^i (b^i_+-b^i_-) = \chi({\cal F})\,, \nonumber\\
 {\rm  where\quad\quad}
 &&  \chi(\cal F)  \supset \begin{cases}
                    \, \, \chi(O7) = \int_{O7} \,c_2(O7)\, ,  \\
                    \, \, \chi(O5) = \int_{O5} \,c_1(O5)\, , \\
                   \,\,   \chi(O3) = \int_{O3} \,c_0(O3) \,=  N_{O3}\,  .
\end{cases}
\eea
 $b_\pm^i$ is the splitted Betti numbers and $ \chi({\cal F})$ is the Euler number of the fixed locus ${\cal F}$. 
There is a very useful theorem to calculate the Euler number of the $\IZ_2$-orbifold space:
\bea
\label{eq:chi}
    \chi(X/{\sigma^*}) =  \frac{1}{2}\,\big(L(\sigma^*, X) + \chi(X)\big) \, = \sum_i  (-)^i (b^i_+).
\eea
This number is the average of the Lefschetz number and the Euler number of $X$.
Then we can determine the $h^{2,1}_-(X/\sigma^*)$ as:
\bea
\label{eq:h21split}
h^{2,1}_-  (X/\sigma^*) = h^{1,1}_-  (X/\sigma^*) + \frac{L(\sigma^*, X)  - \chi (X) }{4} -1,
\eea 
where for reflections $h^{1,1}_-  (X/\sigma^*) = 0$.

For a consistent orientifold, we must ensure both
$\sigma^{*}J=J$ and  $\sigma^{*}\Omega_{3}= \pm \Omega_{3}$ as shown in eq.(\ref{eq:orientifold}), 
 where $\Omega_{3}$ is the unique holomorphic (3,0)-form on $X$.  The holomorphic (3,0)-form $\Omega_{3}$ can be constructed using the homogenous coordinates \cite{Denef:2008wq},
\bea
\label{eq:3form}
\Omega = \frac{1}{2\pi i}\oint_{P=0}\frac{\omega \cdot \Pi_{a} V^{a}}{P},
\eea
where $P$ is the  hypersurface polynomial and $\omega =  dx^{1} \wedge \cdots \wedge dx^{k}$. The $V^{a}$ are the holomorphic vector fields that generate the gauge symmetries, determined in terms of the weights $Q^{a}_{i}$ as follows:
\bea
\label{eq:V}
V^{a}=\Sigma_{i} Q^{a}_{i}x_{i}\frac{\partial}{\partial x_{i}}.
\eea
Since  the hypersurface polynomial $P$ must be invariant under $\sigma$, the numerator of the integrand, denoted as $\mathcal{Q}$, determines the parity.   The parity of $\mathcal{Q}$ serves as a useful cross-check to verify if the correct O-plane system is obtained under   the involution.

Next, we need to determine the fixed locus under the involutions and identify the corresponding types of O-planes. This process is more technical and will be introduced in the next subsection. For now, we assume the data of O-planes is known, allowing us to verify if the orientifold Calabi-Yau manifold supports a string vacuum. In this context, we consider a simple case where the $D7$-brane tadpole cancellation condition is satisfied by placing eight $D7$-branes on top of the $O7$-plane. Consequently, we only need to check the $D3$-brane tadpole condition, which is simplified to:
\bea
\label{eq:tadpole}
N_{D3} + \frac{N_{\text{flux}}}{2}+ N_{\rm gauge}= \frac{N_{O3}}{4}+\frac{\chi(D_{O7})}{4}\, \equiv - \, Q_{D3}^{loc}.
\eea
with $N_{\text{flux}}=\frac{1}{(2\pi)^4 \alpha^{'2}}\int H_3\wedge F_3 $, $N_{\text{gauge}}=-\sum_{a} \frac{1}{8\pi^2} \int_{D_a} \text{tr}{\cal F}_a^2$, and $N_{D3}$, $N_{O3}$ the number of D3-branes, $O3$-planes respectively.  
The D3-tadpole  cancellation condition requires the total D3-brane charge $Q_{D3}^{loc}$ of the seven-brane stacks and $O3$-planes to be an integer. If the involution passes this naive tadpole cancellation check, we will denote this geometry as a \lq\lq  {\it naive orientifold  Type IIB string vacuum}".

\subsection{Putative Fixed Locus on Ambient Space $\cA$}
\label{subsec:newalgorithm}

A smooth Calabi-Yau  hypersurface $X=-K_{\cA}$ is defined by the vanishing locus of a homogeneous polynomial $P$. The polynomial $P$ can be expressed in terms of the known vertices $m\in \Delta, n\in \Delta^*$ of the Newton and dual polytopes, respectively
\bea
P = \sum_{m\in\Delta}{a_{m}M_{m}}=0,\hspace{5mm}\text{where}\hspace{5mm}M_{m}=\prod_{i=1}^{k}{x_{i}^{\langle m,n_{i}\rangle +1}} \, .
\label{eq:hypersurface}
\eea
where $k= h^{1,1}(X)+4$ due to the favorability. In order for the Calabi-Yau hypersurface to be invariant under the involution $\sigma$, we must restrict to the subset of moduli space in which the defining polynomial is invariant. 
The first step is to fix the invariant polynomial such that $P_{sym} = \sigma (P_{sym})$ in addition to $\sigma^{*}J=J$.   
 Mathematically, this could be done by a regulation of the coefficients of the original polynomial $P$ by:
  \bea
\left\{
\begin{aligned}
&a_m = 0\qquad && \sigma(M_m) \notin \{M_m|m\ \in\ \Delta\} \\
&a_m = a_n\qquad && \sigma(M_m) = M_n, n \neq m\\
&a_m\ {\rm \,\, generic}\qquad && \sigma(M_m) = M_m
\end{aligned}
\right.
\eea
 Clearly,  imposing these restrictions requires some tuning in the complex structure moduli space and this tuning may introduce singularities into the invariant polynomial $P_{sym} $.  For reflection involutions, the invariant polynomial $P_{sym} $ is simply the sum of invariant monomials, where the coordinates involved in the reflections $\sigma: \{x_i \leftrightarrow -x_i, x_j \leftrightarrow - x_j \dots\}$ have even powers in total, i.e., $\{x_i^2..., x_j^2...,  x_i^2 x_j^2..., x_ix_j ...,\dots\}$.
 
Now we start to search for the set of points fixed under $\sigma$. We first locate the fixed-point set in the ambient space $\cA$, and then restrict this set to the Calabi-Yau hypersurface $X$. 
In the following we will first describe how to find the fixed point loci for divisor exchange involution and then treat the reflection cases as  special ones.

For divisor exchange involution $\sigma$,  following \cite{Altman:2021pyc}, 
 we first construct the minimal generators $\mathcal{G}$  generated  by homogeneous polynomials  that are  (anti-)invariant under $\sigma$:
\bea
\mathcal{G}=\mathcal{G}_{0}\cup\mathcal{G}_{+}\cup\mathcal{G}_{-} \, .
\eea
where $\mathcal{G}_{0}$ is the collection of unchanged coordinates under $\sigma$.
The unexchanged coordinates in $\mathcal{G}_{0}$ are known  from our choice of involution. 
If the involution $\sigma$ exchange $n$ pairs of coordinates, to find the non-trivial even and odd parity generators in $\mathcal{G}_{+}$ and $\mathcal{G}_{-}$,  we must consider not only $\sigma$, but all possible non-trivial sub-involutions given by the nonempty subsets of $\{\sigma_{1},...,\sigma_{n}\}$ of size $1\leq m\leq n$, with $\sigma_m : x_{i_m} \leftrightarrow x_{j_m}$. Then we denote the new coordinate in $\mathcal{G}\equiv\{y_{1},...,y_{k'}\} $ as:
\bea
\label{eq:segre}
y_{\pm}(\mathbf{a})=  x_{i_{1}}^{a_{1}}x_{i_{2}}^{a_{2}}\dotsm x_{i_{m}}^{a_{m}} \pm x_{j_{1}}^{a_{1}}x_{j_{2}}^{a_{2}}\dotsm x_{j_{m}}^{a_{m}}\, ,
\eea
where  $\mathbf{a} = (a_{1}, a_{2}, \dotsc, a_{m}) \in \mathbb{Z}^{m}$,  $k' = |\mathcal{G}|$ is the number of definite parity polynomial generators, related to $k$ and $\sigma$, and could be smaller or bigger than $k$.  The condition for { homogeneity}, in terms of the   columns $\mathbf{w}_{i_{s}}$ and $\mathbf{w}_{j_{s}}$ of the weight matrix $\mathbf{W}$ is given by 
\bea
\label{eq:homo}
a_{1}(\mathbf{w}_{i_{1}}-\mathbf{w}_{j_{1}}) + a_{2}(\mathbf{w}_{i_{2}}-\mathbf{w}_{j_{2}}) + \dotsb + a_{m}(\mathbf{w}_{i_{m}}-\mathbf{w}_{j_{m}})=0\, .
\eea

The second step is to   perform a  Segre embedding transforming the origianl coordinates $\{x_i\}$ into the new (anti-)invariant generators defined in eq.(\ref{eq:segre}): 
\bea 
\{x_{1},...,x_{k}\}\mapsto\{y_{1},...,y_{k'}\}\equiv \mathcal{G},
\eea
which constructs a new weight matrix $\tilde{\mathbf{W}}$ for $\{ y_i \}$. Now we have transform the original $(k-4) \times k$ GLSM matrix $\mathbf{W}$ to a $r \times k'$ matrix  $\tilde{ \mathbf{W}}$ with  $r=\text{rank}(\tilde{\mathbf{W}})$, $k = h^{1,1}(X) + 4$ and  $k' = |\mathcal{G}|$. Then we can find out the naive fixed point loci in the new coordinates.  

After Segre embedding, we have transformed the divisor exchange involution to the reflection $\sigma : y_i \mapsto -y_i $. 
We can show that  the corresponding coordinate exchange must force the codimension-1 subvariety defining polynomial to vanish  so that $D_i=\{y_i=0\}$ is fixed.   It also implies that the polynomial of every point-wise fixed codimension-1 subvariety can be generated by odd-parity generators in $\mathcal{G}_{-}$.

In general, we need  to check whether the involution allows a subset of generators $\mathcal{F}\subseteq\mathcal{G}$ to vanish simultaneously. 
Meanwhile, the  torus $\mathbb{C}^{*}$ actions provide additional $r=\text{rank}(\tilde{\mathbf{W}})$  degrees of freedom for the generators to avoid being forced to zero.  
More precisely, the requirement for a locus to be  fixed  could be represented as forcing the related generators vanish simultaneously while leaving the others transform under $\IZ_2$ symmetry as usual. 
This is achieved by checking each subset of generators $\mathcal{F}\subseteq\mathcal{G}$ whether the following $\lambda$ system have a solution:
\bea
\label{eq:system}
\lambda_{1}^{\tilde{W}_{1i}} \lambda_{2}^{\tilde{W}_{2i}} \dotsb \lambda_{r}^{\tilde{W}_{ri}} = \sigma(y_{i})/y_{i},\quad i = 1, \dots, k' \,{\rm and }\, \,y_i \notin \cF 
\eea
where $\lambda_i \in \IC^*$ is the torus action and the right-hand side is equal to $\pm 1$ depending on whether the reflection involving the coordinates or not. 
 
 For reflection involution, there is no need for Segre embedding and we only need to consider a simpler system with  $r=\text{rank}({\mathbf{W}})=h^{1,1}(X) $:
 \bea
\label{eq:system2}
\lambda_{1}^{{W}_{1i}} \lambda_{2}^{{W}_{2i}} \dotsb \lambda_{r}^{{W}_{ri}} = \sigma(x_{i})/x_{i},\quad  i = 1, \dots, k \,{\rm and }\, \,x_i \notin \cF \, 
\eea
where $k= h^{1,1}(X)+4$ and the right-hand side is equal to $\pm 1$ depending on whether $\sigma$ include $x_i \leftrightarrow -x_i$.   

For both divisor exchange involutions and reflections,  the  set  in $\cF$ is point-wise fixed on the ambient space $\cA$ if and only if the complex $\lambda$ system equations eq.(\ref{eq:system}) and eq.(\ref{eq:system2}) are solvable  after  imposing the related set in $\cF$ to vanish simultaneously.  We call these subsets $\cF$  passed these check as {\it putative fixed locus} in the ambient space $\cA$.

 \subsubsection{Calculation Complexity}
\label{subsubsec: complexity}

There are two aspects of calculation complexities for determining the putative fixed locus on $\cA$.  One is the total number of  subsets  $\mathcal{F}\subseteq\mathcal{G}$ may be large. For divisor exchange involutions and reflections respectively, in principle there are as much as $\displaystyle \sum_{i=1}^{k'} C_{k'}^{i}=2^{k'}$  and $\displaystyle \sum_{i=1}^{k} C_{k}^{i}=2^{k}$ subsets $\cF \subseteq \cG$ we need to solve the $\lambda \in \IC^*$ system eq.(\ref{eq:system}) and eq.(\ref{eq:system2}).  Another complexity comes from solving the $\lambda$ system in complex field itself which we will describe below.

For divisor exchange involutions as in \cite{Altman:2021pyc} and each of the putative fixed locus $\cF$, in order to check whether the system eq.(\ref{eq:system}) is solvable, we define
$ \lambda_{i} \equiv e^{i\pi u_i}  \in \IC^*$ with $0 \le u_{i} < 2$
so that the equation (\ref{eq:system}) is converted to:
\bea
\label{transformation}
\tilde{W}_{1i}u_{1} + \dotsb + \tilde{W}_{ri}u_{r}\equiv\left\{\begin{array}{rl}
0\pmod{2},&y_{i}\in\mathcal{G}_{0}\cup\mathcal{G}_{+}\\
1\pmod{2},&y_{i}\in\mathcal{G}_{-}
\end{array}\right. \, .
\eea
and further becomes:
\bea
\tilde{W}_{1i}u_{1} + \dotsb + \tilde{W}_{ri}u_{r}-2q_{i}\equiv\left\{\begin{array}{rl}
0,&y_{i}\in\mathcal{G}_{0}\cup\mathcal{G}_{+}\\
1,&y_{i}\in\mathcal{G}_{-}
\end{array}\right.
\label{eq:oldalgorithm}
\eea
with  $0 \le u_{i} < 2$ and  $q_i \in \IZ$. If the new GLSM weight matrix $\tilde{W}_{ji}$ contains negative entries for a given line-bundle $\cO(D_i)$ corresponding to the coordinate $y_i$,  we need to check the solvability of eq.(\ref{eq:oldalgorithm}) by scanning $q_i \in \IZ$ in a range of
\bea
\left\{\begin{array}{rl}
&{ \displaystyle \sum_j^{r}}\eta(-{\tilde W}_{ji})*\tilde{W}_{ji} < q_{i} < { \displaystyle \sum_j^{r}} \eta({\tilde W}_{ji})*\tilde{W}_{ji},\,\,y_{i}\in\mathcal{G}_{0}\cup\mathcal{G}_{+}\\
&{ \displaystyle \sum_j^{r}}\eta(-{\tilde W}_{ji})*\tilde{W}_{ji} \leq q_{i} < { \displaystyle \sum_j^{r}} \eta({\tilde W}_{ji})*\tilde{W}_{ji},\,\,y_{i}\in\mathcal{G}_{-}
\end{array}\right.
\label{eq:qspace}
\eea
where $\eta(x)$ is a STEP function  such that $\eta(x)=0$ if $x < 0$ and $\eta(x)=1$ for $x \geq 0$. Thus the summation in the left-hand side of eq.(\ref{eq:qspace})  represents the sum of the negative entries, while  the right-hand side represents the sum of the positive entries. If  the entries of $\tilde{W}_{ji}$ are all positive for a given line-bundle $\cO(D_i)$,  the scanning range is simplified to: 
\bea
0 \leq q_i < \sum_j^{r} \tilde{W}_{ji}
\label{eq:qspace2}
\eea  
Thus, the problem of determining the solvability of the $\lambda$ system transforms into determining the solvability of $u_i$ in  eq.(\ref{eq:oldalgorithm}) for a given lattice point $q_i$ within the range defined by  eq.(\ref{eq:qspace}) and eq.(\ref{eq:qspace2}).  If any solution is found for any lattice point $q_i$, then the set of generators $\mathcal{F}$ has a point-wise fixed  locus.

However, this method will exhaust computational resources as $h^{1,1}(X)$ increases.  
For example, if  all the entries of $\tilde{W}_{ji}$ are  positive then  the upper limit of points in the $q_i$ lattice that need to be tested can be estimated as:
	\begin{equation}
    		\prod \limits_{i=1}^{\hat k'} (\sum_j^r \tilde W_{j,i}) \xlongrightarrow{h^{1,1}\,\,\,  {\rm large}} \cO(1-100)^{\hat k'},\label{eq:old}
	\end{equation}
	where $\hat{k'}$ indicates  the number of elements in the  index set $\{1, \dots {\hat i} \dots, k' \}$ excluding the index $i$ for $y_i \notin \cF_i$.
 The total computational complexity we encounter is given by eq.(\ref{eq:old}), multiplied by the number of all possible fixed sets that need to be tested, which is  $\prod \limits_{i=1}^{\hat k'} (\sum_j^r \tilde W_{j,i}) \times 2^{k'}$.
 
The problem of computational complexity is the same for reflection involutions except there is no need for Segre embedding.
Therefore,  we change the GLSM weighted matrix  in eq.(\ref{transformation} - \ref{eq:old}) to $W_{ji}$ and maintain the coordinate system as $\{x_i\}$ with $k = h^{1,1}(X)+4$ and $\hat k$ the number of index set excluding the coordinates involved in reflections.  Consequently, the total complexity for reflection involutions becomes $\prod \limits_{i=1}^{\hat k} (\sum_j^r  W_{j,i}) \times 2^{k}$.

Combining these two complexities in determining putative fixed locus on $\cA$, one may encounter  limitations in computational power. For instance,  consider a favorable Calabi-Yau with  $h^{1,1}(X)=7$ and single reflection, i.e., $k=11$ and $r= h^{1,1}(X)=7$. Even with a small average summation  $\displaystyle \sum_j^r  W_{j,i} = 3$ for all $q_i$,  we need  to test  up to $3^{10}=59,049$ $q_i$ lattice points  for each of the  $2^{11}=2,048$  possible  fixed locus sets $\cF$. This results in a total of $\cO(10^{10})$ lattice points to be tested.  For an exchange involution switch two pairs of NIDs in the same geometry,  consider again a small average summation $\displaystyle \sum_j^r \tilde W_{j,i} = 3$ for all $q_i$ with $k'=15$ and $r = 3$.  We need to test  up to $4,782,969$  $q_i$ lattice points for each of the  $2^{15}=32,768$ possible fixed locus sets $\cF$,  totaling $\cO(10^{12})$ points. For a geometry with  $h^{1,1}(X) = 20$ and moderate average summation $\displaystyle \sum_j^r  \tilde W_{j,i} = 10$, exchanging five pairs of NIDs  (which often results in $k' \sim 34$ and $r \sim 10$)  forces us to  check up to $\cO(10^{33})\times \cO(10^{10})$ lattice points in  extreme cases,  which is  beyond  current computational capabilities. Even though in many cases we don't need to exhaust all the possible lattice points, this remains a significant challenge for general scans and we need to solve these problems..

\subsubsection{New Algorithm}
\label{subsubsec:newalgorithm}
  There are two directions to reduce the complexity of calculation. One is to reduce the scanning space of $q_i$ lattice points, the other is to reduce  the number of the possible fixed loci needed to be test. For the first purpose, we initially solve the $\lambda$ system in the real number field to identify some putative fixed loci and then search in the complex number field to check whether we miss some solutions. In this procedure,  based on the fixed loci obtained in the real number field, we can significantly reduce the number of possible fixed loci we need to check. 
  
  \subsubsection*{Real number $\lambda$ system}

Solving the $\lambda$ system  in  real number field  first can extremely reduce the number of $q_i$ lattice points that need to bw checked. This new algorithm focuses on $\lambda$, i.e., eq.(\ref{eq:system}-\ref{eq:system2}), rather than eq.(\ref{eq:oldalgorithm}) with focus on $u_i$ and $q_i$ with a larger parameter space.  By concentrating on real $\lambda$ values, we only need to consider two relevant value: $\lambda = \pm 1$, or equivalently,  setting $u_i$ in $\lambda= e^{i \pi u_i} \in \IC^*$ to  $0$ and $1$.   Consequently, the  largest dimension of  parameter space for solving the $\lambda$ system is reduced to 
 \bea
\label{eq:new}
2^{r} \quad {\rm  with\,\,\,} \left\{\begin{array}{rlr}
& r=\text{rank}(\tilde{\mathbf{W}}) \leq h^{1,1}(X) &  {\rm exchange \,\, involution} \\
& r=\text{rank}({\mathbf{W}}) = h^{1,1}(X) &  {\rm reflection \,\, involution}
\end{array}\right. \, 
\eea
Now,  let us consider the example with $h^{1,1}(X)=7$ described above to see the extent of the parameter space reduction.  Compared to $59,049$ or $4,782,969$ parameter spaces that need to be scanned for each of possible fixed  loci set in the complex space, we only need to consider $2^7=128$ different values of $\lambda$ at most. For the case with $h^{1,1}(X)=20$,  we reduce the dimension of parameter space from $\cO(10^{33})$ to $2^{20} = 1,048,576$.

The advantage of the new approach is  that it reduce the parameter space of solving system significantly.  This reduction  depends on the rank of  $\tilde{\mathbf{W}}$  through eq.(\ref{eq:new}), which is smaller than $h^{1,1}(X)$,  rather than eq.(\ref{eq:old}) which is sensitive to the GLSM  matrix and $k' $ (which in most cases is larger than $h^{1,1}(X)$). 
For reflection involutions, the only difference is that we  treat ${\mathbf{W}}$ with $r= h^{1,1}(X)$  and $k= h^{1,1}(X)+4$ instead.

However, this is not the end of the story. At this stage,   some putative fixed loci may be missed if, for a given test set $\cF$, there is no choice of  $\lambda = \pm 1$ that satisfy eq.(\ref{eq:system}-\ref{eq:system2}), but solutions may exist for complex $\lambda = e^{i \pi u} \in \IC^*$ with $0 \leq u_i <2$.  Therefore the next step is to search in complex number field to identify any missed fixed loci. Fortunately,  
the fixed loci computed in the real number system can be used to reduce the number of possible loci that need to be tested.

Before we go to the detail of how to reduce the possible fixed locus we need to check, we give a remark on the GLSM weighted matrix in our new algorithm. In \cite{Altman:2021pyc}, the entries of GLSM matrix are chosen to be non-negative by restricting the matrix to the positive orthant $\IZ_{\geq0}^k$. This is achieved by intersecting  two polyhedrons:  the first is generated by lines specified by the elements of kernel of vectors defining the ambient space $\cA$, and the second is generated by rays specified by the unit basis vectors.  There are two shortages of this approach for determining the fix loci. First,   with non-negative entries of ${ \mathbf{W}}$,  the summation $\displaystyle \sum_j^r  W_{j,i} $ shown in eq.(\ref{eq:old}) will  also be large, contributing to the tardiness of fixed loci computation, making it a nearly impossible. 
Second, when intersecting two polyhedron, the parity of the the first polyhedron may change, affecting the results of finding fixed loci in the real $\lambda$ system.  Although such parity change in GLSM weighted matrix do not affect determination of the fixed loci, since we eventually check all solutions of eq.(\ref{eq:system}-\ref{eq:system2}) in the complex number field, it is much  more convenient  to keep the parity throughout  the entire calculation. 
Therefore in this paper we will start from the standard output of PALP \cite{Kreuzer:2002uu} and SAGE  \cite{sage} to get the GLSM  weighted matrix with negative entries from the vertex in the dual-polytope $\Delta^*$ as shown in the Kreuzer-Skarke list~\cite{Kreuzer:2000xy}.

After utilizing our new algorithm to get the fixed loci in real number system,  yielding the same result for single divisor reflection as  in \cite{Crino:2022zjk}, we use these fixed loci to  simplify the computation of the entire fixed locus on the Calabi-Yau threefold in complex space.

\subsubsection*{Reduce possible sets of $\cF$}
\label{subsubsec:reduce}

There are four classes of subsets of  $\cF$ that can help  reduce the number of  possible fixed loci we need to test in the complex $\lambda$ system.
\begin{enumerate}
\item If a subset $\cF_i$ we get in real $\lambda$ system is a  fixed locus, then any set containing it would also be a fixed locus. 
 \item Apply the Stanley-Reisner ideal $\cI_{SR}$ to  rule out subsets $\cF_i$ which should not vanish simultaneously. 
 \item Apply the linear ideal  $\cI_{lin}$ to rule out some possible combinations of subsets $\cF_i$ to test the solvability.
\item If a subset $\cF_i$ is  not a fixed locus,  then all subsets contained within $\cF_i$ will not be fixed loci.  
\end{enumerate} 
The first three classes of sets $\cF_i$  are  loci that allow us to  rule out  sets containing them when searching for new solutions in complex $\lambda \in \IC^*$ system,  thus belonging to the same type.  The last class of sets $\cF_i$ helps us  rule out those sets contained within it.  Let us first explain how these two types of special loci work and  then  estimate  how effectively they can help  reduce the complexity.  

Consider the type one sets first. For known fixed loci $\cF_i$ in real $\lambda$ system, they remian fixed loci in complex $\lambda \in \IC^*$  system, and  any sets  $\cF_i'$ containing  $\mathcal{F}_i$ will also be fixed loci. 
This is because the real number solution ${\boldsymbol \lambda}$ of eq.(\ref{eq:system}-\ref{eq:system2}) for loci $\mathcal{F}_i$ is also the solution for loci within $\mathcal{F}_i'$, as the later has fewer constraints in complex $\lambda$ system.
One can further reduce the test sets $\cF_i$ by using the SR ideal $\cI_{SR}$.   The SR ideal describes  sets  that cannot vanish simultaneously on the ambient space, so there is no need to test  sets $\cF_i$ contain elements of the  SR ideal.  
The third class of sets applies to  divisor exchange involutions, where new bases  are introduced after Segre embedding.  Here the minimal hypersurface generators  $\{y_i\}$ are not independent, and their linear relations, encoded in the linear ideal $\cI_{lin}$, can be used to rule out some possible loci. For example, suppose there are new basis $y_1,y_2,y_3$ consist of old bases $\{x_i\}$ in such way: $\{y_1=x_1*x_2+x_3*x_4,  y_2=x_1*x_4,  y_3=x_2*x_3\}$, then such linear relations can be used to rule out  sets $\cF_i$ in which any combination of $y_2$ and $y_3$ are fixed, but $y_1$ is not fixed.

Finally, the scanning procedure can be  further simplified by recognizing that if a set of points is not fixed, then neither is any set containing it. Thus, if the simultaneous vanishing of a set of generators is not fixed,  neither is the vanishing of any subset. 
 For example, if $\mathcal{F}_i=\{n_i,n_j,n_k\}$ is an non-fixed locus, then all subsets $\{\{n_i\},\{n_j\},\{n_k\},\{n_i,n_j\},\{n_i,n_k\},\{n_k,n_j\}\}$ contained in $\mathcal{F}_i$ are also not fixed and don't need to be tested. 
 Therefore,  we begin our scan with the largest set of generators and work our way down. Usually, the largest set we can choose has four generators, as their simultaneous vanishing defines a set of isolated points on the ambient space $\cA$. However, if  $\cF_i$ vanish on the invariant polynomial $P_{sym}$, we will relax the number of generators to five or more.

\subsubsection*{Efficiency of new algorithm}

Now let us estimate how many sets can be ruled out when we solve the complex $\lambda$ system. 
By considering the known fixed points from real space and  the substantial number of SR ideal in a triangulation, a significant loci could be ruled out. Given an ambient space $\cA$ with $k$ coordinates,  suppose we identify some fixed loci sets $\{\mathcal{F}_{i}\}$ generated by three coordinates (polynomials in $\{y\}$ system)  $\{x_1, x_2, x_3\}$, or the SR-ideal of the triangulations contains $\{x_1 x_2 x_3\}$, then the minimal number of loci ruled out by these three loci is:
\bea
\sum_{i=0}^{i \leq k-3} C_{k-3}^i = 2^{(k-3)}
\eea
which is $1/8$ of the amount of all possible locus $2^{k}$. If there is another set with other three coordinates  which should be ruled out, then the total number of loci ruled out by  such two sets is:
\bea
\label{eq:ruleout}
\sum_{i=0}^{i \leq k-3} C_{k-3}^i+\sum_{i=0}^{i \leq k-3} C_{k-3}^i-\sum_{i=0}^{i \leq k-6} C_{k-6}^i=2*2^{(k-3)}-2^{(k-6)}.
\eea
We can generalise eq.(\ref{eq:ruleout}) to  $m$ fixed locus cases with three generators each,  then the number of possible fixed sets we can rule out is nearly:
\bea
m*2^{(k-3)}-C_m^2*2^{(k-6)}  \,+ C_m^3*2^{(k-9)}-\dots
\eea
For example, when $m =8$,  the percentage of fixed locus left we need to test  approximately ${C_8^2}/{2^6} - {C_8^3}/{2^9} + \dots \sim \frac{1}{3}$.
In practice, the number of such sets $\cF_i$ is so large that we can rule out nearly all loci  described by more than four divisors.

For a non-fixed sets described by a set of $n$ divisors,  we can rule out all $2^n$ possible loci described by the subsets of those $n$ divisors.
The possible fixed loci ruled out by this method has no overlap with those ruled out by the previous three class of subsets $\cF_i$.
To estimate  how effectively the non-fixed locus can help us to remove unwanted loci, suppose  there are $m$ non-fixed loci described by $n_1$ coordinates (polynormials) in the ambient space, the number of loci ruled out by these non-fixed locus is given by:
\bea
m*2^{n_1}-C_m^2*2^{n_2}+C_m^3*2^{n_3}-\dots
\eea
where $n_{2,3}$ is the number of generators in the intersections of two and three non-fixed loci respectively. It is important to emphasize that loci ruled out by non-fixed sets are  distinct from those ruled out by the first three classes of $\cF_i$ described above. In practice many loci are eliminated, significantly reducing the final number of possible fixed sets that need to be checked. 
 
It should be pointed out that the order in which we compute the fixed point can influence the efficiency of our computation. 
In this work, we first rule out those possible fixed loci that contains the first type of sets, i.e., fixed loci identified in the real $\lambda$ system, SR ideals and linear relation sets. Then we search for possible fixed points, starting from large sets and moving to small  sets,  leveraging
the non-fixed loci to  reduce the  candidate sets. This approach greatly   increased our computational speed.

We will show in Section \ref{sec:example} that combining these two methods,  solving the real $\lambda$ system to get  fixed loci and then checking for new solutions in the complex  field  space after excluding  the four classes of subsets $\cF_i$,  significantly enhances the efficiency of  determining putative fixed loci. This combined approach improves the process  by at least  five orders  of magnitude ($\cO(10^5)$) for an example with moderate $h^{1,1}(X)$.

\subsection{Fixed Loci on the Calabi-Yau Hypersurface $X$}
\label{subsec:fixedX}

 After identifying the putative fixed point loci  on the ambient space $\cA$ as described in previous subsection,
 the next step is to verify whether each  point-wise fixed locus  lies on the Calabi-Yau hypersurface $X$ .  The definition of Stanley-Reisner ideal $\cI_{SR}$  can lead to a partitioning of $\cA$ into different patches \{$U_{i}$\}. In each patch \{$U_{i}$\}, the  $\cI_{SR}$  ideal can be trivially satisfied.  For a given fixed set, we compute the dimension of the ideal generated by the symmetry part of Calabi-Yau polynomial $P_{sym}$ and the fixed set generators $\mathcal{F}\equiv\{y_{1},...,y_{p}\}$ in each sector $U_i$ for divisor exchange involutions.  For reflection involution, we  use the original coordinate system $\{x_i\}$.
\bea
\label{eq:fixed}
\cI^{fixed}_{ip}=\langle U_{i},P_{sym},y_{1},...,y_{p}\rangle \, .
\eea
If  $\text{dim }\cI^{fixed}_{ip}<0$ for all $U_{i}$, then there is no  solution for $\cI^{fixed}_{ip}$ and so $\cF$ does not intersect $X$. 
For each subset of $\cF$ that is not discarded, we repeat this calculation for the ideal with one fixed set generator $\text{dim }\cI^{fixed}_{i1}$, and then with two generators  $\text{dim }\cI^{fixed}_{i2}$, etc. until { $\text{dim }\cI^{fixed}_{i\ell}=\text{dim }\cI^{fixed}_{ip}$} when adding more generators to the ideal no longer changes the dimension for any region $U_{i}$. Then, the intersection $\{y_{1}=\cdots =y_{\ell}=0\}$ of these generators gives the final point-wise fixed locus on $X$, with redundancies eliminated.  Furthermore, $\text{dim }\cI^{fixed}_{ip}\geq 0$ shows that the fix locus $\cF$ do intersect with our invariant Calabi-Yau manifold $P_{sym}$. 

Finally we check whether the invariant Calabi-Yau equation $P_{sym}$ vanishes at a given locus $\cF$. If $P_{sym}$ does not vanish, an $O3$-plane corresponds to a codimension-3 point-wise fixed  subvariety, an $O5$-plane corresponds to a  codimension-2 subvariety, and an $O7$-plane corresponds to a  codimension-1 subvariety.  If $P_{sym}$ does vanish at the fixed locus,  the for a fixed point $\mathcal{F}$ with complex co-dimension $n$,  $\mathcal{F}$ is an $Om$-plane with $m = 3 + 2(4 - n)$ \cite{Crino:2022zjk}.  
 
In cases where $P_{sym}$  vanishes at the fixed locus $\cF$,  it might seem puzzling how the intersection of four divisors can  describe an $O3$-plane on  $X$, while the intersection of two divisors   can described an $O7$-plane.  This situation primarily arises in reflection involutions. In divisor exchange involutions,  there may be  redundancy among the polynomial generators  $\{y_i\}$ of $\cF$ when consider the SR ideal $\cI_{SR}$ generated by the original coordinates $\{x_i\}$.  Such redundant generators $\{y_i\}$ should be excluded in the expression of O-plane on hypersurface, as demonstrated in the next section. However, in reflection cases, there is no additional constraint from the SR ideal, as it initially starts with the $\{x_i\}$ system in the first place.
For instance, consider an $O7$-plane. The   set $F_1 \equiv \{x_1 = x_2 = 0\}$  forces the invariant polynomial to vanish, $P_{sym}|_{F_1} =0$, creating   redundancy in describing the O-plane on $X$. On the other hand, neither $\{x_1=0\}$ nor $\{x_2=0\}$ alone can make $P_{sym}$ vanish automatically.  Setting $P_{sym}=0$ and either $x_1=0$ or $x_2 = 0$ does not force the other to vanish. 
Therefore we conclude that 
\bea
\label{eq:oplane}\{ P_{sym}=0,x_{1}=0, x_{2}=0\} 
\eea 
is not a complete intersection. 
Since omitting any of $x_i=0$ $(i=1, 2)$  makes  eq.(\ref{eq:oplane})  unsolvable while $P_{sym}=0$ is trivially satisfied on $F_1$, this $O7$-plane on $X$ is described by $F_1=  \{x_1 = x_2 = 0\}$ on the ambient space $\cA$.  Similarly, the $O3$-plane on $X$ is described by intersection of  four divisors on $\cA$.  This inconvenient description arises because  $\{x_i\}$ are the natural coordinates of  the ambient space rather than of the Calabi-Yau manifold. Finally, suppose  $F=\{x_1= \dots = x_n=0\}$ describes a  co-dimension $n$ fixed point,   the type of $Om$-plane can be summarized as \cite{Crino:2022zjk}:
\bea
\label{eq:type}
\left\{\begin{array}{rl}
m=3+2*(4-n),&{\rm if}\,  P_{sym}|_F =0\\
m=3+2*(3-n),&{\rm if}\,  P_{sym}|_F \neq 0\\
\end{array}\right .
\eea

Before we move on, let us give a final remark on the GLSM charge matrix again.  In fact, when consider the complex $\lambda$ system, linear row-operation would bring nothing different to fixed points set in our computation.  
Suppose there is a fixed locus
$\cF$, giving by  the solution 
$\boldsymbol{\lambda} = \{\lambda_1, \lambda_2, ... ,\lambda_r\} $, on a favorable Calabi-Yau with GLSM matrix 
$\bW_{r,r+4},\,\, r=h^{1,1}(X)$, we replace the m-th row of 
$\bW$ by  linear row-operations such as sum of n-th and m-th row, i.e.,
$\widehat{W_{m,r+4}}=W_{m,r+4}+W_{n,r+4}$, nothing will happen to the fixed locus $\cF$ except the  associated solution of 
\textbf{$\lambda$}  transformed accordingly as 
$\widehat{\lambda_n}=\lambda_n*\lambda_m^{-1}$ as shown below:
\bea
\label{eq:row}
\lambda_1^{W_{1,i}}\dots\lambda_n^{W_{n,i}}\dots\lambda_m^{W_{m,i}}\dots\lambda_r^{W_{r,i}}=\lambda_1^{W_{1,i}}\dots{\widehat\lambda_n^{W_{n,i}}}\dots{\lambda_m}^{\widehat{W_{m,i}}}\dots\lambda_r^{W_{r,i}}, \quad i= 1, \dots, r+4
\eea
Since $\cF$ is derived from  the solution of eq.(\ref{eq:system2}), which shares  the same left-hand side (lhs) as eq.(\ref{eq:row}),   it must also be solvable by replacing the  lhs of eq.(\ref{eq:system2})  with the right-hand side (rhs) of eq.(\ref{eq:row}). 
Therefore we can employ  linear row-operations to simplify the GLSM wighted matrix $\bW$ or $\tilde\bW$ in the complex $\lambda$ system.

\subsection{Smoothness and Free Action}
\label{subsec:smoothfree}

Smoothness is also an important feature for orientifold Calabi-Yau threefolds.  The general expression of Calabi-Yau hypersurface $P$ in eq.(\ref{eq:hypersurface}) is  smooth while  in defining of $P_{sym}$, some coefficients of $P$ are  changed, and then singularity may be introduced in $P_{sym}$. So we need to re-check the smoothness of invariant Calabi-Yau hypersurface defined by $P_{sym}=0$, which is important in determining whether an involution is a free action. 
We do this by checking if there is any solution to the condition $P_{sym}=dP_{sym}=0$ that is not ruled out by the Stanley-Reisner ideal. In practice, this is done by setting up the ideals
\begin{equation}
\label{eq:smooth}
\cI^{smooth}_{i}=\langle U_{i},\; P_{sym},\; \frac{\partial P_{sym}}{\partial x_{1}},\; ...,\; \frac{\partial P_{sym}}{\partial x_{k}}\rangle
\end{equation}
\noindent 
for each region $U_{i}$ allowed by the SR ideal, and computing the dimension. If $\text{dim }\cI^{smooth}_{i}<0$ for all $U_{i}$, then the invariant Calabi-Yau hypersurface is smooth. 
However, such computation is usually time-consuming and hard to be accomplished.  Instead, we computed  the value of $h^{2,1}_- $  by eq.(\ref{eq:h21split}), and the Calabi-Yau threefold is claimed to be smooth if this number is an  integer. Of course, this is only a necessary condition for smoothness.

A $\mathbb{Z}_{2}$ symmetry on hypersurface $X$ is said to be a free action, only when such symmetry deduce no O-planes and $X$ is smooth. In practice, there are two kinds of free actions. One is those found  in  \cite{Altman:2021pyc} that there is a fixed locus on the ambient space $\cA$ that does not intersect with the Calabi-Yau manifold $X$. However, this procedure does not guarantee the absence of a fixed locus on the ambient space $\cA$. In fact, we have identified another type of free action where  there is no fixed locus on the ambient space from the first place.
By examining  eq.(\ref{eq:system}) and eq.(\ref{eq:system2}), we define a naive fix locus by checking whether a set  of $\cF$   is solvable in the $\{\lambda_{j}\}$ system. There exists a situation where these equations can be realized trivially  without any basis being  fixed, indicating no need for any  set $\cF$ to be fixed on the ambient space $\cA$. Thus,   the involution is again a $\IZ_2$ free action if the invariant hypersurface $P_{sym}$ is smooth.  This type of free action  appears only in multiple reflection involutions and we will demonstrate  the details with  examples in the next section.

        \section{Concrete Construction}
        \label{sec:example}
        In this section, we demonstrate two explicit examples of finding and classifying the point-wise fixed sets of a Calabi-Yau orientifold threefolds, following the method described in the previous section. 
        \subsection{Example A}
        \label{subsec:example}
	
	We first choose an example with ({\rm Polyid: 545},  {\rm Tri\_id: 5})  in our database (\dburl). 
		This example with Hodge number  $h^{1,1}=7, h^{2,1}=37$ contains following  vertex in the dual-polytope $\Delta^*$ in the Kreuzer-Skarke list \cite{Kreuzer:2000xy}:
	\begin{equation}
		\label{eq:vertex}
		\begin{array}{|c|c|c|c|c|}
			\cline{1-5}
			v_{1} & v_{2} & v_{3} & v_{4} & v_{5} \\\cline{1-5}
			1 & 0 & 0 & 0 & -2       \\\cline{1-5}
			0 & 1 & 1 & 1 & -3       \\\cline{1-5}
			0 & 0 & 2 & 0 & -2     \\\cline{1-5}
			0 & 0 & 0 & 0 & -2      \\\cline{1-5}
		\end{array}
	\end{equation}
 This example defines an MPCP desingularized ambient toric variety with GLSM weight matrix $\mathbf{W}$ given by:
	\begin{equation}
		\label{example ori wm}
		\begin{array}{|c|c|c|c|c|c|c|c|c|c|c|c}
			\cline{1-11}
			x_{1} & x_{2} & x_{3} & x_{4} & x_{5} & x_{6} & x_{7} & x_{8} & x_{9} &x_{10} &x_{11} &\\\cline{1-11}
			1 & 0 & 1 & 0 & 2 & 0 & 0 & 0 & 0 & 2 & 0 &   \lambda_{1}  \\\cline{1-11}
			0 & 1 & 1 & 0 & 0 & 0 & 0 & 0 & -2 & 0 & 0 &  \lambda_{2}   \\\cline{1-11}
			0 & 0 & -1 & 1 & 0 & 0 & 0 & 0 & 2 & -2 & 0 & \lambda_{3}    \\\cline{1-11}
			0 & 0 & 1 & 0 & 1 & 1 & 0 & 0 & -1 & 1 & 0 &  \lambda_{4}   \\\cline{1-11}
			0 & 0 & 0 & 0 & 1 & 0 & 1 & 0 & 0 & 1 & 0 &  \lambda_{5}   \\\cline{1-11}
			0 & 0 & 0 & 0 & 1 & 0 & 0 & 1 & 1 & 0 & 0 &  \lambda_{6}   \\\cline{1-11}
			0 & 0 & -1 & 0 & 0 & 0 & 0 & 0 & 1 & -1 & 1 &  \lambda_{7}  \\\cline{1-11}
		\end{array}
	\end{equation}
	with $r=7$ independent torus $\IC^*$ actions:
	\bea
(x_1, x_2, x_3,  x_4,  x_5,  x_6, x_7, x_8, x_9, x_{10}, x_{11}) \sim ( \lambda_1\, x_1, \lambda_{2} \,x_2, 
\lambda_{1} \,\lambda_{2} \,\lambda_{3}^{-1} \,\lambda_{4} \,\lambda_{7}^{-1}  \,x_3, \lambda_3  \,x_4,  \nonumber\\   \lambda_{1}^{2} \,\lambda_{4} \,\lambda_{5} \,\lambda_{6} \, x_5, \lambda_4 \, x_6, \lambda_5 \, x_7, \lambda_6  \,x_8, \lambda_{2}^{-2} \,\lambda_{3}^{2} \,\lambda_{4}^{-1} \,\lambda_{6} \,\lambda_{7} \, x_9,  \lambda_{1}^{2} \,\lambda_{3}^{-2} \,\lambda_{4} \,\lambda_{5} \,\lambda_{7}^{-1}  \, x_{10}, \lambda_{7}  \,x_{11} )
	\eea
	\noindent 
and  Stanley-Reisner ideal:
	\bea
	\label{eq:sr}
	\cI_{SR}=\langle & x_1 x_2,\, x_1 x_3,\, x_1 x_4,\, x_1 x_9,\, x_1 x_{10},\, x_1 x_{11},\, x_2 x_3,\, x_2 x_4,\, x_2 x_7,\, x_2 x_8, \, x_2 x_{11},\\\notag &  x_3 x_4,\, x_3 x_6,\, x_3 x_8,\, x_3 x_{10},\, x_6 x_{11},\, x_7 x_{10},\,  x_8 x_9,\, x_8 x_{10},\, x_8 x_{11},\, x_{10} x_{11}, \\\notag&  \, x_4 x_5 x_6 x_7,\,  x_4 x_5 x_6 x_9,\,  x_4 x_5 x_7 x_9,\, x_5 x_6 x_7 x_8,\, x_5 x_6 x_9 x_{10}, \, x_5 x_7 x_9 x_{11} \rangle.
	\eea
The Calabi-Yau manifold $X$ is defined by the anti-canonical hypersurface in the ambient space $\cA$ with polynomial degree $||6,0,0,3,3,3,0||$. The Hodge numbers of the corresponding individual toric divisors $D_i \equiv \{x_i = 0\} $ are:
	\begin{equation}
		\label{eq:hodgex}
		\begin{array}{|c|c|c|c|c|c|c|c|c|c|c|c|l}
			\cline{1-12}
			h^\bullet(D_i)&D_{1} & D_{2} & D_{3} & D_{4} & D_{5} & D_{6} & D_{7} & D_{8} & D_{9} &D_{10} &D_{11} \\\cline{1-12}
			h^{0,0}(D_i)&1 & 1 & 1 & 1 & 1 & 1 & 1 & 1 & 1 & 1 & 1    \\\hline
			h^{0,1}(D_i)&0 & 0 & 0 & 0 & 0 & 1 & 1 & 1 & 1 & 1 & 1    \\\hline
			h^{0,2}(D_i)&0 & 0 & 0 & 0 & 4 & 0 & 0 & 0 & 0 & 0 & 0    \\\hline
			h^{1,1}(D_i)&7 & 7 & 7 & 16 & 44 & 11 & 11 & 2 & 11 & 2 & 2    \\\hline
		\end{array}
	\end{equation}
	\subsubsection{Divisor Exchange Involutions}
	Now we first consider the proper divisor exchange involutions. All the detail expressions in the calculation for   this example are presented in Appendix.\ref{ap:example}.

	From eq.(\ref{eq:hodgex}), there are three sets of divisors with the same topology, each containing three divisors, i.e., $D_{1,2,3} = dP_6$,  exact-Wilson divisors $D_{6,7,9} $ and $D_{8,10,11} $. Therefore we can define  involutions $\sigma_i,\, (i=1,2,3)$ to exchange these pairs of non-trival identical divisors (NIDs) in each  set and combine them together:
	\bea 
        &\sigma_1 \in \{\emptyset, x_1 \leftrightarrow x_2, x_2 \leftrightarrow x_3, x_1 \leftrightarrow x_3\}\notag\\
	&\sigma_2\in \{\emptyset, x_6 \leftrightarrow x_7, x_7 \leftrightarrow x_9, x_6 \leftrightarrow x_9\}\notag\\
	&\sigma_3\in \{\emptyset, x_8 \leftrightarrow x_{10}, x_{10} \leftrightarrow x_{11}, x_8 \leftrightarrow x_{11}\}
	\eea 
	and their non-trivial compositions:
	\bea
	 \sigma = \sigma_1 \circ \sigma_2  \circ \sigma_3
	 \eea
	
        Naively, we can consider 63 involutions  of these NIDs: $9$ single pairs of  divisors exchange involutions such as $\{x_1 \leftrightarrow x_3\}$, $27$ double pairs of  divisor exchange involutions such as $\{x_1 \leftrightarrow x_2, x_6 \leftrightarrow x_7\}$ and   27 triple pairs divisors involutions such as $\{x_1 \leftrightarrow x_2, x_6 \leftrightarrow x_7, x_8 \leftrightarrow x_{10}\}$.  
 However, not all of them are consistent with the Stanley-Reisner (SR) ideal and linear ideal. For instance, consider the single pair of divisor exchange  $\{x_1 \leftrightarrow x_3\}$. The SR ideal  changes to include $\langle \dots x_3 x_9, x_3 x_{11} \dots \rangle$, which is inconsistent with eq.(\ref{eq:sr}).  Furthermore, this involution fails to keep the defining polynomial homogeneous  without setting some coefficients to zero.  For example the monomial $x_3^2 x_5^2 x_7 x_9 x_{11} $ in the original defining polynomial with degree $||{6,0,0,3,3,3,0}||$ changes to monomial $x_1^2 x_5^2 x_7 x_9 x_{11} $ with degree  $||{6,-2,2,1,3,3,2}||$ after the involution. This violates the linear ideal $\cI_{lin}$ and alters the triple intersection number.  In fact, all  single pairs of divisor exchange involutions are ruled out by  the requirement of proper involution described in Section {\ref{subsec:involution}}. In the end,  only one proper non-trivial identical divisor (NID) exchange involution remains:
        \bea
        \sigma_e :  x_1\leftrightarrow x_2, x_7 \leftrightarrow x_9, x_8 \leftrightarrow x_{10}.
        \label{eq:invol}
        \eea
        From eq.(\ref{eq:hodgex}) we see that this is an exchange of one pair of dP$_{6}$ divisors and two  pairs of exact-Wilson divisors.

For a consistent orientifold, the volume form in eq.(\ref{eq:3form}) must have a definite parity under $\sigma_e$. 
With the explicit expression for $\mathcal{Q}$ in terms of the $x_{i}$  as shown in Appendix.\ref{ap:example}, we can prove that under the $\sigma_e$ the resulting form is precisely the negative of $\mathcal{Q}$. Consequently, the volume form has odd parity under $\sigma_e$, i.e., $\sigma_e^* \Omega_3 = - \, \Omega_3$. Therefore we would expect that  any fixed points under this involution  should correspond to $O3$ or $O7$-planes, or both.
       
        Using this information of weighted matrix and $\sigma_e$, we can determine the equation of the Calabi-Yau manifold.  We get the general expression of Calabi-Yau manifold  using eq.(\ref{eq:hypersurface}) and then restrict this expression to the invariant polynomial $P_{sym}$:
	\begin{align}
            \label{example psym}
		P_{sym}&= a_1 x_1^6x_6^3x_7^3x_8^3 + a_2 x_1^4x_2^2x_6^3x_7^2x_8^2x_9x_{10} + a_3 x_1^2x_2^4x_6^3x_7x_8x_9^2x_{10}^2 + a_4 x_2^6x_6^3x_9^3x_{10}^3 + a_5 x_1^4x_3^2x_6^2x_7^3x_8^2x_9x_{11}
     \nonumber    \\\notag
          &   + a_6 x_1^4x_4^2x_6^2x_7^2x_8^3x_{10}x_{11} + a_7 x_1^3x_2x_3x_4x_6^2x_7^2x_8^2x_9x_{10}x_{11} + a_8 x_1^2x_2^2x_3^2x_6^2x_7^2x_8x_9^2x_{10}x_{11}  \\\notag
		&+ a_9x_1^2x_2^2x_4^2x_6^2x_7x_8^2x_9x_{10}^2x_{11} + a_{10} x_1x_2^3x_3x_4x_6^2x_7x_8x_9^2x_{10}^2x_{11} + a_{11} x_2^4x_3^2x_6^2x_7x_9^3x_{10}^2x_{11} \notag\\
		& + a_{12} x_2^4x_4^2x_6^2x_8x_9^2x_{10}^3x_{11} + a_{13} x_1^2x_3^4x_6x_7^3x_8x_9^2x_{11}^2 + a_{14} x_1^2x_3^2x_4^2x_6x_7^2x_8^2x_9x_{10}x_{11}^2\\\notag
		& + a_{15} x_1x_2x_3^3x_4x_6x_7^2x_8x_9^2x_{10}x_{11}^2 + a_{16} x_2^2x_3^4x_6x_7^2x_9^3x_{10}x_{11}^2 + a_{17} x_1^2x_4^4x_6x_7x_8^3x_{10}^2x_{11}^2\\\notag
		& + a_{18} x_1x_2x_3x_4^3x_6x_7x_8^2x_9x_{10}^2x_{11}^2 + a_{19} x_2^2x_3^2x_4^2x_6x_7x_8x_9^2x_{10}^2x_{11}^2 +a_{20}x_2^2x_4^4x_6x_8^2x_9x_{10}^3x_{11}^2\\\notag
		& + a_{21} x_3^6x_7^3x_9^3x_{11}^3 + a_{22} x_3^4x_4^2x_7^2x_8x_9^2x_{10}x_{11}^3 + a_{23} x_3^2x_4^4x_7x_8^2x_9x_{10}^2x_{11}^3 + a_{24} x_4^6x_8^3x_{10}^3x_{11}^3 \\\notag
		&+ a_{25} x_1^4x_5x_6^2x_7^2x_8^2 + a_{26} x_1^2x_2^2x_5x_6^2x_7x_8x_9x_{10} + a_{27} x_2^4x_5x_6^2x_9^2x_{10}^2  + a_{28} x_1^2x_3^2x_5x_6x_7^2x_8x_9x_{11}\\\notag
		& + a_{29} x_1^2x_4^2x_5x_6x_7x_8^2x_{10}x_{11} + a_{30} x_1x_2x_3x_4x_5x_6x_7x_8x_9x_{10}x_{11}+ a_{31} x_2^2x_3^2x_5x_6x_7x_9^2x_{10}x_{11} \\\notag
		&  + a_{32} x_2^2x_4^2x_5x_6x_8x_9x_{10}^2x_{11} + a_{33} x_3^4x_5x_7^2x_9^2x_{11}^2 + a_{34} x_3^2x_4^2x_5x_7x_8x_9x_{10}x_{11}^2 + a_{35} x_4^4x_5x_8^2x_{10}^2x_{11}^2\\\notag
		& + a_{36} x_1^2x_5^2x_6x_7x_8 + a_{37} x_2^2x_5^2x_6x_9x_{10} + a_{38} x_3^2x_5^2x_7x_9x_{11} + a_{39} x_4^2x_5^2x_8x_{10}x_{11}
		+ a_{40} x_5^3\notag
	\end{align}
	\noindent where $a_{i}\in\mathbb{C}$ are arbitrary coefficients.
	
	The next step is to figure out the  fixed loci of the involution $\sigma_e$ in the ambient space $\cA$ and reduce them to O-plane structure on Calabi-Yau hypersurface $X$. As described in previous section, for computational convenience, we need to perform a version of the Segre embedding to ensure that $\sigma(y_i)/y_i \in \{-1,1\}$, thereby simplifying the overall computation. The projective coordinates $x_{3}$, $x_{4}$, $x_{5}$,  $x_{6}$ and $x_{11}$ are not affected by the involution $\sigma_e$ and are therefore included in our list of (anti-)invariant polynomial generators:
\bea
\mathcal{G}_{0}=\{x_{3},x_{4},x_{5},x_{6},x_{11}\}\, .
\eea	
	Let us define the permutations $\sigma_{1}:x_{1} \leftrightarrow x_{2}$, $\sigma_{2}:x_{7} \leftrightarrow x_{9}$ and $\sigma_{3}:x_{8} \leftrightarrow x_{10}$, such that $\sigma_e=\sigma_{1}\circ\sigma_{2}\circ\sigma_{3}$, then we have several sub-involution such as $\sigma_{1,2,3},\,\, \sigma_{1}\circ\sigma_{2}, \sigma_{2}\circ\sigma_{3}, \sigma_{1}\circ\sigma_{3}$ and $\sigma_e=\sigma_{1}\circ\sigma_{2}\circ\sigma_{3}$. 
	
	For $\sigma_1: x_{1} \leftrightarrow x_{2}$, because we only consider non-trivial identical divisors (NIDs), $x_1$ and $x_2$ have different weights and cannot be combined into a homogenous binomial. Thus, we are left  with the invariant monomial $\cG_+ = \{x_1 x_2\}$ and $\cG_- = \emptyset$. The same applies to $\sigma_{2,3}$, resulting in $\cG_+ = \{x_1 x_2, x_7 x_9, x_8 x_{10}\}$. 
	
	For $\sigma_1\circ\sigma_2: x_{1} \leftrightarrow x_{2}, x_{7} \leftrightarrow x_{9}$, we can consider binomial generators of the form $x_{1}^{m}x_{7}^{n} \pm x_{2}^{m}x_{9}^{n}$ for $m, n \in \IZ$ like eq.(\ref{eq:segre}). However, there is no solution for $m, n $ that satisfies eq.(\ref{eq:homo}) to maintain the  homogeneity of the polynomial.  This is also the case for $\sigma_1\circ\sigma_3$ and $\sigma_2\circ\sigma_3$.  As a result, there is no contribution to $\cG_\pm$ in these cases.
	
	For $\sigma_e: x_1\leftrightarrow x_2, x_7 \leftrightarrow x_9, x_8 \leftrightarrow x_{10}$, we should consider binomial generators of the form:
	\bea
	x_1^{m} x_7^{n} x_8^{p} \pm x_2^{m} x_9^{n} x_{10}^{p}
	\eea
for $m,n,p \in \IZ$.	The homogeneity of this binomial is determined by the following condition on the weights
\bea
m\, (\bW_{i,1}- \bW_{i,2}) + n\, (\bW_{i,7}- \bW_{i,9}) + p\, (\bW_{i,8}- \bW_{i,10})  = \mathbf{0}
\eea
The kernel is generated by the vector $(m,n,p)=(2,1,1)$, so that our binomial generators are given by $x_{1}^2 x_{7}x_8 \pm x_{2}^2x_{9}x_{10}$. This implies that $\cG_+ = \{x_{1}^2 x_{7}x_8 + x_{2}^2x_{9}x_{10}\}$ and $\cG_-=\{x_{1}^2 x_{7}x_8 - x_{2}^2x_{9}x_{10}\}$. 

Therefore,  all of the (anti-)invariant polynomial generations in $\mathcal{G}=\mathcal{G}_{0}\cup\mathcal{G}_{+}\cup\mathcal{G}_{-}$ are given by
\begin{equation}\label{eq:segre2}
\begin{gathered}
y_{1}=x_{8} x_{10},\;\;\; y_{2}=x_{7} x_9,\;\;\; y_{3}=x_{3},\;\;\; y_{4}=x_{4},\;\;\; y_{5}=x_{5}, \;\;\;y_{6}=x_{6},\\
y_7 = x_1 x_2,\;\;\;y_{8}=x_{1}^2 x_{7}x_8 + x_{2}^2x_{9}x_{10},\;\;\; y_{9}=x_{1}^2 x_{7}x_8 - x_{2}^2x_{9}x_{10},\;\;\; y_{10} = x_{11}\, .
\end{gathered}
\end{equation}

This coordinate transformation defines the Segre embedding with consistency condition $y_{8}^{2}=y_{9}^{2}+4 y_2 y_1 y_7^2$ and new weight matrix $\tilde\bW$:
\begin{equation}
		\label{eq:newmatrix}
		\begin{array}{|c|c|c|c|c|c|c|c|c|c|c}
			\cline{1-10}
			y_{1} & y_{2} & y_{3} & y_{4} & y_{5} & y_{6} & y_{7} & y_{8} & y_{9} &y_{10}  &\\\cline{1-10}
			2 & 0 & 1 & 0 & 2 & 0 & 1 & 2 & 2 & 0 & \tilde \lambda_{1}    \\\cline{1-10}
			0 & -2 & 1 & 0 & 0 & 0 & 1 & 0 & 0 & 0 & \tilde \lambda_{2}     \\\cline{1-10}
			-2 & 2 & -1 & 1 & 0 & 0 & 0 & 0 & 0 & 0 & \tilde \lambda_{ 3}     \\\cline{1-10}
			1 & -1 & 1 & 0 & 1 & 1 & 0 & 0 & 0 & 0 &\tilde \lambda_{4}      \\\cline{1-10}
			1 & 1 & 0 & 0 & 1 & 0 & 0 & 1 & 1 & 0 & \tilde \lambda_{5}     \\\cline{1-10}
			1 & 1 & 0 & 0 & 1 & 0 & 0 & 1 & 1 & 0 & \tilde\lambda_{ 6}      \\\cline{1-10}
			-1 & 1 & -1 & 0 & 0 & 0 & 0 & 0 & 0 & 1 & \tilde \lambda_{7}    \\\cline{1-10}
		\end{array}
	\end{equation}
where $\tilde\lambda_i = e^{i \pi \tilde u_i} \in \IC^*$ is the torus actions.
	
	\subsubsection*{Search fixed locus in real $\lambda$ system}
As described in Section \ref{subsubsec:newalgorithm}, we will first calculate the fixed locus in the real $\tilde\lambda$ system where $\tilde\lambda = \pm 1$ and then look for new fixed locus in the complex space $\tilde\lambda_i = e^{i \pi \tilde u_i} \in \IC^*$ with $0 \leq \tilde u_i < 2$. 
 Since solving $\tilde\lambda$  system in real space is sensitive to the parity of  the torus action,  we should apply the new GLSM matrix eq.(\ref{eq:newmatrix}) to calculate the fixed loci although there are some redundancy. The involution $\sigma_e$ can be rewritten simply as $y_{9}\mapsto -y_{9}$ in the new coordinate system. Therefore we have transform the divisor exchange involution to reflection.   Later, we will see  that $F_{1}=\{y_{9}=0\}$ is a point-wise fixed, codimension-1 subvariety and  it defines an $O7$-plane on the orientifold Calabi-Yau $X$.

Here we show the parity of the new coordinates under the exchange involution $\sigma_e$ in terms of  the  seven torus  actions $\tilde \lambda_i$:
\bea
\label{eq:parity}
(y_1, y_2, y_3,  y_4,  y_5,  y_6, y_7, y_8, y_9, y_{10}) \sim ( \tilde\lambda_{ 1}^{2}\tilde\lambda_{ 3}^{-2}\tilde\lambda_{ 4}\tilde\lambda_{5}\tilde\lambda_{ 6}\tilde\lambda_{7}^{-1}y_{1}, \tilde\lambda_{ 2}^{-2}\tilde\lambda_{ 3}^{2}\tilde\lambda_{4}^{-1}\tilde\lambda_{5}\tilde\lambda_{6}\tilde\lambda_{7}y_{2}, \nonumber\\ 
\tilde\lambda_{1}\tilde\lambda_{2}\tilde\lambda_{3}^{-1}\tilde\lambda_{4}\tilde\lambda_{7}^{-1}y_{3},   \tilde\lambda_{ 3}y_{4},\tilde\lambda_{ 1}^{2}\tilde\lambda_{4}\tilde\lambda_{5}\tilde\lambda_{6}y_{5}, \tilde\lambda_{ 4}y_{6}, \tilde\lambda_{ 1}\tilde\lambda_{ 2}y_{7}, \tilde\lambda_{1}^{2}\tilde\lambda_{ 5}\tilde\lambda_{6}y_{8} , \tilde\lambda_{ 1}^{2}\tilde\lambda_{ 5}\tilde\lambda_{6}y_{9} ,\tilde\lambda_{ 7}y_{10} ) \\
= (y_1, y_2, y_3,  y_4,  y_5,  y_6, y_7, y_8, - y_9, y_{10})  \nonumber
\eea
 It has been shown in Section \ref{subsec:newalgorithm} that whether there is a fix locus on $\cA$ depends on the solution of $\tilde\lambda$ system eq.(\ref{eq:system}). 
Therefore we need to find a solution for the choice of $\tilde\lambda_{i}$ such that coordinates not in the possible fixed set $\cF$ exhibit precise parity under the involution eq.(\ref{eq:parity}). Since point $F_{1}$ is chosen to be fixed, i.e., $F_1= \{y_{9} = 0\}$, the parity constraint $\tilde\lambda_{1}^{2}\tilde\lambda_{ 5}\tilde\lambda_{6} *y_{9} =-y_{9}$ are trivially satisfied without any restriction on  $\tilde\lambda_{1,5,6}$. 
Then we must use the seven $\mathbb{C}^{*}$ actions to neutralize the odd parity of $y_{9}$ while leaving the other 9 coordinates invariant. This constraint is defined by the toric equivalence class:
  \begin{align}
\label{eq:equiv}
 \tilde \lambda_{ 1}^{2}*\tilde\lambda_{ 3}^{-2}*\tilde\lambda_{ 4}*\tilde\lambda_{ 5}*\tilde\lambda_{ 6}*\tilde\lambda_{ 7}^{-1}*y_{1} &=  y_{1},& 
  \tilde\lambda_{ 2}^{-2}*\tilde\lambda_{ 3}^{2}*\tilde\lambda_{ 4}^{-1}*\tilde\lambda_{ 5}*\tilde\lambda_{ 6}*\tilde\lambda_{ 7}*y_{2}  &=  y_{2} \nonumber\\
 \tilde \lambda_{ 1}*\tilde\lambda_{ 2}*\tilde\lambda_{ 3}^{-1}*\tilde\lambda_{ 4}*\tilde\lambda_{ 7}^{-1}*y_{3} &= y_3,&
  \tilde\lambda_{ 3}*y_{4} &=  y_4 \nonumber \\
 \tilde \lambda_{ 1}^{2}*\tilde\lambda_{ 4}*\tilde\lambda_{ 5}*\tilde\lambda_{ 6}*y_{5} &= y_5,&
 \tilde \lambda_{ 4}*y_{6} &= y_{6} \nonumber\\
 \tilde \lambda_{ 1}*\tilde\lambda_{ 2}*y_{7} &= y_7,&
 \tilde \lambda_{ 1}^{2}*\tilde\lambda_{ 5}*\tilde\lambda_{ 6}*y_{8} &= y_{8} \nonumber \\
\tilde  \lambda_{ 7}*y_{10} &= y_{10}. 
\end{align}
where $\tilde\lambda_{i}  \in \{-1, 1\}$ for $i=1,\dots, 7$. It is obvious that eq.(\ref{eq:equiv}) does have solutions and $F_{1}=\{y_{9}=0\}$  indeed is a naive fix point locus in the ambient space $\cA$.

By taking advantage of the toric degrees of freedom,  there may be additional non-trivial fixed loci beyond $F_{1}$.  Thus, we need to  check whether any subset $\mathcal{F}$ of the generators can neutralize the odd parity of $y_{9}$, and become fixed  in the process. As mentioned earlier, if the simultaneous vanishing locus of a set of generators is not fixed, then 
neither is the vanishing of any subset which contained in this non-fixed point set in ambient space. We therefore begin our scan with the largest set of polynomial generators $\{y_i\}$ and work our way down.

Consider the subset $\{y_{4},y_{5},y_{8},y_{10}\}\equiv\mathcal{F}_{2}\subset\mathcal{G}$. In order for the locus $F_{2}=\{y_{4}=y_{5}=y_{8}=y_{10}=0\}$ to be fixed, we must use the $\mathbb{C}^{*}$ actions to neutralize the odd parity of $y_{9}$ while leaving everything else invariant (as $y_{9}$ is the only non-zero generator with negative parity). This constraint is defined by the toric equivalence class followed by eq.(\ref{eq:system}):
 \begin{align}
    \label{example fixed point equations ori}
y_1 : \tilde\lambda_{ 1}^{2}*\tilde\lambda_{ 3}^{-2}*\tilde\lambda_{ 4}*\tilde\lambda_{ 5}*\tilde\lambda_{ 6}*\tilde\lambda_{ 7}^{-1} &=  1,& 
 y_2 :\tilde \lambda_{ 2}^{-2}*\tilde\lambda_{ 3}^{2}*\tilde\lambda_{ 4}^{-1}*\tilde\lambda_{ 5}*\tilde\lambda_{ 6}*\tilde\lambda_{ 7}  &=1, \nonumber\\
 y_3 : \tilde\lambda_{ 1}*\tilde\lambda_{ 2}*\tilde\lambda_{ 3}^{-1}*\tilde\lambda_{ 4}*\tilde\lambda_{ 7}^{-1}  &= 1,&
 y_6 : \tilde\lambda_{ 4} &=1 \nonumber\\
 y_7 : \tilde\lambda_{1y 1}*\lambda_{ 2} &= 1&
 y_9: \tilde\lambda_{ 1}^{2}*\tilde\lambda_{ 5}*\lambda_{ 6} & = -1
    \end{align}   
  Obviously, $\{\tilde\lambda_{1}=1,\tilde\lambda_{2}=1,\tilde\lambda_{3}=-1, \tilde\lambda_{4}=1,\tilde\lambda_{5}=1, \tilde\lambda_{6}=-1,\tilde\lambda_{7}=-1\}$ is one of the solutions. Then we may test whether a subset of $\mathcal{F}_{2}$, which contained in $\mathcal{F}_2$, is still a fixed point. This is an important step which may change the type of the corresponding O-plane. 
	For example,  if we assume the fix locus is ${F}_{2}' =\{y_{4}=y_{5}=y_{8}=0\}$,  we have to add an additional constrain  $\tilde\lambda_7 * y_{10} = y_{10}$ from eq.(\ref{eq:equiv}), resulting in $ \tilde\lambda_7 = 1$, which will lead to a contradiction to the solution of $\tilde\lambda$ system  eq.(\ref{example fixed point equations ori}).   The same will happen to other subsets of $\mathcal{F}_{2}$ and we conclude the naive fix locus on the ambient space is indeed $F_2 = \{y_{4}=y_{5}=y_{8}=y_{10}=0\}$ itself.
  
By now those fix-loci are calculated on the ambient space $\cA$, and we must check  whether the fixed sets intersect the Calabi-Yau hypersurface transversally. Here we take $F_2$ as an example, fixed point $F_{2}=\{y_{4}=y_{5}=y_{8}={y_{10}}=0\}$ can be written in terms of the original projective coordinates $\{x_{4}=x_{5}=x_{11}=0\}\cap\{x_{1}^2 x_{7}x_8 + x_{2}^2x_{9}x_{10}\}$. If we make these substitutions in $P_{sym}$, it reduces to
\bea
P_{sym}&= a_1 x_1^6x_6^3x_7^3x_8^3 + a_2 x_1^4x_2^2x_6^3x_7^2x_8^2x_9x_{10} + a_3 x_1^2x_2^4x_6^3x_7x_8x_9^2x_{10}^2 + a_4 x_2^6x_6^3x_9^3x_{10}^3.
\eea
Consider the subset where $x_{11} = 0$, part of the SR ideal $\langle \dots, \, x_1x_{11}, \,x_2x_{11}, \, x_3x_{4}, \, \\ x_6x_{11},\, x_8x_{11},\, x_{10}x_{11} \dots \rangle$ forbids $x_1 = x_2 = x_3 = x_6=x_8 = x_{10} = 0$. So the only way for $P_{sym} = 0$ is to require  $\{x_7 = 0,x_9 = 0\}$, which is $\{y_8=0\}$. Hence, $P_{sym}=0$ implies { $y_{8}=0$} due to the SR constraints, and $y_{8}$ is redundant when restricting the fixed set to $X$. The reduced set is then $F_{2}=\{y_{4}=y_{5}=y_{10}=0\}$, which is an $O3$-plane, consistent with our parity of the holomorphic three-forms under the involutions $\sigma* \Omega_3 = - \Omega_3$.

In practice, we combine the transversality and SR ideal checks by performing Groebner basis calculations to check the dimension of the ideal $\cI^{fixed}_{ij}$ as in eq.(\ref{eq:fixed}) for $U_{i}$ a region allowed by the Stanley-Reisner ideal. 
\bea
\cI^{fixed}_{ij} =\langle U_{i},\; P_{sym},\; F_{j}\rangle
\eea
If the dimension $\text{dim }\cI^{fixed}_{ij} >0$, and removing any generator from $F_{j}$ changes this value, then we know that $F_{j}$ intersects $X$ transversally and is allowed by the SR ideal. In this case, given the $\mathcal{I}_{SR}$ ideal as eq.(\ref{eq:sr}), there are 24 sectors $\{U_i\}, \, i= 1, \dots, 24$  we need to check for a single putative fixed locus $F_j$. The detail of these sectors is collected in the Appendix.\ref{subsec:dataexample}

Finally, we can determine the type of the fixed loci by examining the number of intersecting codimension-1 subvarieties in each fixed set from eq.(\ref{eq:type}). Specifically,  $F_1$ and $F_2$ have complex codimensions  1 and 3 in $X$ respectively.
 This implies that $F_{1}$ is an $O7$-plane, while $F_{2}$ and other computed fixed loci  are $O3$-planes.  After scanning all  possible combination of $\cF \subset \cG$ in real space,  besides the one-generator fixed locus $\cF_1 = \{y_9\} $ and the three-generators fixed locus ${\cF}_2  = \{y_{4},  y_{5},  y_{10}\}$, we identify other fixed points: $\cF_3= \{y_{3}, y_{5}, y_{10}\}, \,\cF_4 = \{y_{4}, y_{6}, y_{8}\}$. These fixed point sets intersect the respective $\sigma$-invariant hypersurface so that we get a number of $O7$ and $O3$-planes as follows:
\begin{table}[ht!]
        \centering
		{\footnotesize
			\centering
			\begin{tabular}{c|c|c}
                \hline
                \centering Type of O-planes & Fixed Loci on $X$  & Homotopy Class  \\\hline\hline
            
             $\cF_1:$     O7   & $\{y_{9}\} \equiv \{ x_1^2 x_7 x_8 - x_2^2 x_9 x_{10} \}$ & $ 2 D_1 + D_7 + D_8$ \\\hline
             
          $\cF_2:$       O3 & $\{y_{4}, y_{5}, y_{10}\} \equiv \{x_4, x_5, x_{11}  \}$  &  $D_4, D_5, D_{11}$ \\\hline
         $\cF_3:$      O3 & $\{y_{3}, y_{5}, y_{10}\} \equiv \{ x_3, x_5, x_{11} \} $ & $D_3, D_5,D_{11}$  \\\hline
          $\cF_4:$      O3 & $\{y_{4}, y_{6}, y_{8}\} \equiv \{x_4, x_6, x_1^2 x_7 x_8 + x_2^2 x_9 x_{10} \}$ & $D_4,D_6, 2 D_1 + D_7 + D_8$ \\\hline
			\end{tabular}
			\caption{Putative Fixed Loci on $X$}
			\label{tab:O-plane}
		}
	\end{table}
 
 Therefore we get four fixed locus $\cF_i$ on the Calabi-Yau hypersurface $X$.  The next step is to check whether we miss some fixed locus in complex $\tilde\lambda$ system eq.(\ref{eq:system}) since there could be loci which are solvable in complex $\tilde\lambda \in \IC^*$ space but not in real space. So we need to re-search the fixed loci in the complex space,  and the previously computed results are very helpful in this process, as they allow us to exclude loci that containing these fixed loci sets.

\subsubsection*{Search fixed locus in complex $\lambda$ system}

The process for searching fixed loci in complex space is  similar to what we did before in real space. 
Here, we illustrate how the two types of four classes sets discussed  in Section \ref{subsubsec:reduce} assist in identifying fixed loci. The first type of sets comprises three classes of sets that  can be excluded  before solving the $\tilde\lambda$ system.  Any subsets of generators $\cG$ containing them are also excluded in the scanning in the first place as shown in Table.\ref{tab: type-1}. 
The second type, the non-fixed point sets, includes sets for which any subsets are excluded,  as shown in Table.\ref{tab:type-2}.

\begin{table}[ht!]
{\footnotesize
  \centering
	\begin{tabular}{|p{4.7cm}|p{4cm}|p{4.7cm}|}
                \hline
                 SR Ideal &  Fixed Loci &  Generators Linear Relations \\\hline
              $\{y_1 y_2 y_3 \},\, \{y_1 y_2 y_8 \},\, \{y_1 y_2 y_9 \},\,$ $ \{y_1 y_2 y_{10} \},\, \{y_1 y_3 y_4 \},\, \{y_1 y_3 y_5 \},\,$ $ \{y_1 y_3 y_6 \},\, \{y_1 y_3 y_7 \},\, \{y_1 y_3 y_8 \},\,$ $ \{y_1 y_3 y_9 \},\, \{y_1 y_3 y_{10} \},\, \{y_1 y_4 y_7 \},\, $ $\{y_1 y_4 y_8 \},\, \{y_1 y_4 y_9 \},\, \{y_1 y_4 y_{10} \},\,$ $ \{y_1 y_5 y_8 \},\, \{y_1 y_5 y_9 \},\, \{y_1 y_5 y_{10} \},\,$ $ \{y_1 y_6 y_8 \},\, \{y_1 y_6 y_9 \},\, \{y_1 y_6 y_{10} \},\,$ $ \{y_1 y_7 y_8 \},\, \{y_1 y_7 y_9 \},\, \{y_1 y_7 y_{10} \},\,$ $ \{y_1 y_8 y_9 \},\, \{y_1 y_8 y_{10} \},\, \{y_1 y_9 y_{10} \},\,$ $ \{y_2 y_3 y_4 \},\, \{y_2 y_3 y_6 \},\, \{y_2 y_3 y_7 \},\,$ $ \{y_2 y_4 y_7 \},\, \{y_2 y_6 y_{10} \},\, \{y_2 y_7 y_8 \},\,$ $ \{y_2 y_7 y_9 \},\, \{y_2 y_7 y_{10} \},\, \{y_3 y_4 y_5 \},\,$ $ \{y_3 y_4 y_6 \},\, \{y_3 y_4 y_7 \},\, \{y_3 y_4 y_8 \},\,$ $ \{y_3 y_4 y_9 \},\, \{y_3 y_4 y_{10} \},\, \{y_3 y_5 y_6 \},\,$ $ \{y_3 y_5 y_7 \},\, \{y_3 y_6 y_7 \},\, \{y_3 y_6 y_8 \},\,$ $ \{y_3 y_6 y_9 \},\, \{y_3 y_6 y_{10} \},\, \{y_3 y_7 y_8 \},\,$ $ \{y_3 y_7 y_9 \},\, \{y_3 y_7 y_{10} \},\, \{y_4 y_5 y_7 \},\,$ $ \{y_4 y_6 y_7 \},\, \{y_4 y_6 y_{10} \},\, \{y_4 y_7 y_8 \},\,$ $ \{y_4 y_7 y_9 \}$.
              &$\{y_9\}, \{y_{4} y_{5} y_{10}\}$, $ \{y_{3} y_{5} y_{10}\}$, $\, \{y_{4} y_{6} y_{8}\}$.
              &$\{y_8\neq0, y_1=0, y_9=0\}$,
              $\{y_8 \neq 0, y_2=0, y_9=0\}$,
              $\{y_8 \neq 0, y_7=0, y_9=0\}$,
              $\{y_9 \neq 0, y_1=0, y_8=0\}$, 
              $\{y_9 \neq 0,\, y_2=0, \,y_8=0\}$, 
              $\{y_9 \,\,\,\neq \,\,0, \,\,y_7\,\,= \,\,0, \,\,y_8\,\,=\,\,0\}$.\\\hline
	\end{tabular}
	\caption{Type one sets (and those containing them) could be excluded in the scan.}
	\label{tab: type-1}
		}
	\end{table}
	
\begin{table}
\centering
\begin{tabular}{|c|}
\hline
\text{Non-fixed loci }\\
\hline
$\{y_1, y_2, y_5, y_6, y_7\},\, \{y_1, y_2, y_4, y_5, y_6\},\, \{y_5, y_8, y_{10}\},\, \{y_5, y_6, y_8\},$\,\\
$\{y_4, y_8, y_{10}\},\, \{y_4, y_5, y_8\},\, \{y_3, y_8, y_{10}\},\, \{y_3, y_5, y_8\},\,\{y_2, y_5, y_{10}\},$\,\\
$ \{y_2, y_4, y_{10}\},\, \{y_2, y_3, y_{10}\},\, \{y_2, y_3, y_5\},\, \{y_7, y_{10}\},\, \{y_6, y_{10}\},\, \{y_4, y_7\},$\, \\
$\{y_3, y_7\},\, \{y_3, y_6\},\, \{y_3, y_4\},\, \{y_1, y_{10}\},\, \{y_1, y_3\}$\\
\hline
\end{tabular}
\caption{Type two sets  (and their subsets) could be excluded in the scan.}
\label{tab:type-2}
\end{table}

Given the ten polynomial generators  shown in eq.(\ref{eq:newmatrix}), the original number of  possible fixed locus we need to test is $2^{10}=1,024$. By applying the first type of sets to exclude the subsets of $\cG$, we only need to test 20 loci, all of which  turn out to be  non-fixed loci shown in Table.\ref{tab:type-2} and need to be excluded. Therefore the results of this example in complex space are the same with Table.\ref{tab:O-plane} in real space. 
	
Now we can summarize how efficiently our new algorithm  speeds up the calculation. Initially,  the complexity of  solving  the $\lambda$ system for each possible fix locus required scanning a maximum of $77,760$  lattice points in the complex number field, as indicated by eq.(\ref{eq:old}). This has been reduced to scanning   $2^{7} = 128$ points  in  real number field.   By using the methods described above to reduce the possible fixed loci,  we only need to check  20 sets  instead  $2^{10}=1,024$  to find new solutions in the complex $\tilde\lambda$ system. Consequently,   we have  effectively reduced the complexity of finding fixed loci in the complex $\lambda$ system by five order  of magnitude ($\cO(10^5)$).

The structure of O-plane system can contributes to the D3-tadpole through the calculation of   relevant topological quantites.
The Euler characteristic for $O7$-plane is $\chi(D(O7_{F_1})) = 27$,  while the number of $O3$-planes are determined by the triple intersection number:
\bea
\label{eq:o3}
&O3_{F_2}: D_4 D_5 D_{10} =  3, \quad O3_{F_3}: D_3 D_5 D_{11} =  3 \nonumber\\
&O3_{F_4}: D_4 D_6 (2 D_{1}+D_7+D_8) =  3
\eea
So there are in total  9 $O3$-planes.  Using eq.(\ref{eq:tadpole}) the contribution to the $D3$-brane tadpole is
\bea
 N_{D3} + \frac{N_{\text{flux}}}{2}+ N_{\rm gauge}= \frac{N_{O3}}{4}+\frac{\chi(D_{O7})}{4} = \frac{9 + 27}{4} = 9 \,.
\eea
Thus $Q_{D3}^{loc} = -9 $ and this is a \lq\lq naive orientifold Type~IIB string vacua".    
    
The involution considered in eq.(\ref{eq:invol}) will result in the Hodge number splitting on the orientifold Calabi-Yau.  We choose a basis in $H^{1,1}(X;\mathbb{Z})$  given by $J_{1} = D_{5}, J_{2} = D_{6}, J_{3} = D_{7}, J_{4} = D_{8}, J_{5} = D_{9}, J_{6} = D_{10},  J_{7} = D_{11}$. 
In this example the involution acts on the divisor classes as:
\bea
	\sigma^{*} : D_{1} \leftrightarrow D_{2}, D_{7} \leftrightarrow D_{9}, D_{8} \leftrightarrow D_{10}
\eea
thus all three of the exchanges in this example are of non-shrinkable rigid divisors.
This case is favorable, and we can thus expand the K\"ahler form in terms of these divisor classes
	$J = t_{1}J_{1} + t_{2}J_{2} + t_{3}J_{3} + t_{4}J_{4} + t_{5}J_{5} + t_{6}J_{6} + t_7 J_7$,
with $t_{1}, ... t_{7} \in \mathbb{Z}$. The constraint that the K\"ahler form must only have components in $H^{1,1}_{+}(X)$ implies
\begin{align}
	\label{eq:kformswapex4}
	J&=\sigma^{*}J	= t_{1}D_{5}+t_{2}D_{6}+t_{3}D_{9}+t_{4}D_{10}+t_{5}D_{7}+t_{6}D_{8}+t_{7}D_{11}.
\end{align}

As in previous examples, we rewrite $D_{1}, D_{2}, D_{3},$ and $D_{4}$ in terms of our chosen basis using the linear ideal. Performing the algebra, and plugging the relations  into eq.(\ref{eq:kformswapex4}), we get:
\bea
 t_{3} = t_{5}, \quad\quad t_4 = t_6
\eea
Hence $h^{1,1}_{+}(X/\sigma^{*}) = 5$ and $h^{1,1}_{-}(X/\sigma^{*}) = 2$. 
 By using Lefschetz fixed point theorem, we can further get the splitting of $h^{2,1}_{\pm}(X/\sigma^{*}) $ under the involutions:
    \bea
h^{2,1}_- (X/\sigma^{*}) = h^{1,1}_- (X/\sigma^{*})  + \frac{L(\sigma, X) -\chi(X)}{4} -1 = 2 + \frac{36+ 60}{4} -1=  25\, ,
\eea
and we collect the results of Hodge number  splitting as:
\bea
\label{eq:split}
h^{1,1}_+  = 5, \,\, h^{1,1}_-  = 2; \quad h^{2,1}_+   = 12, \,\, h^{2,1}_-   = 25\, .
\eea

Finally we check whether the locus $\{P_{sym}=0\}$ is smooth by %
computing the dimension $\text{dim }\cI^{smooth}_{i}$ as eq.(\ref{eq:smooth}) for each disjoint region $U_{i}, i = 1, \dots, 24$ allowed by the Stanley-Reisner ideal. We find that the maximum dimension is $-1$, so that $\{P_{sym}=0\}$ is indeed smooth. 
Since the manifold is smooth, there is no ambiguity in defining $h^{2,1}_- (X/\sigma^{*}) $ and eq.(\ref{eq:split}) gives the true Hodge number splitting.

\subsubsection{Multi-divisor Reflection Involutions }
\label{subsubsec:reflection}

There are many possible multi-divisor reflection involutions in the same example. We constrain  ourselves to maximal triple divisor reflections,  resulting in at most $11+ 55 +165 = 231 $ cases  to consider. Due to  time constraints,  we will consider $11$ single reflections and  randomly choose  $15$ double divisor reflections and  $15$ triple divisor reflections  to obtain the orientifold Calabi-Yau.

For reflection involutions, there are no constraints from SR ideal and linear ideal, and no change of the triple intersection form, except for a sign. The primary objective is to determine the fixed loci. 
    The algorithm for determining fixed loci is similar to divisor exchange involutions since after the Segre embedding,  we have already transform the divisor exchange to reflection in the new coordinates. Here, we present  a simple example of triple divisor reflection, and summarize the results for other reflections in Table.\ref{tab:reflectionex}.
Here we choose one triple reflection as: 
\bea
\sigma_r : \,\,x_3 \leftrightarrow -x_3,\ x_8 \leftrightarrow -x_8, \ x_9 \leftrightarrow -x_9.
\eea
The expression of invariant Calabi-Yau hypersurface $P_{sym}$ could  easily be obtained by requiring the sum of the power exponent of reflected divisors is an even number. 
\begin{align}
        \label{eq:sym2}
{P}_{sym}=&-56x_1^3x_2x_3x_4x_6^2x_7^2x_8^2x_9x_{10}x_{11} + 16x_1x_2^3x_3x_4x_6^2x_7x_8x_9^2x_{10}^2x_{11} 
- 50x_1x_2x_3^3x_4x_6x_7^2x_8x_9^2x_{10}x_{11}^2      \nonumber \\\notag
&+ 48x_1x_2x_3x_4^3x_6x_7x_8^2x_9x_{10}^2x_{11}^2 
+ 28x_1^4x_5x_6^2x_7^2x_8^2 + 22x_1^2x_2^2x_5x_6^2x_7x_8x_9x_{10} \\\notag
&+ 39x_2^4x_5x_6^2x_9^2x_{10}^2 + 22x_1^2x_3^2x_5x_6x_7^2x_8x_9x_{11}+ 78x_1^2x_4^2x_5x_6x_7x_8^2x_{10}x_{11} \\\notag
&- 74x_{2}^2x_3^2x_5x_6x_7x_9^2x_{10}x_{11}- 72x_2^2x_4^2x_5x_6x_8x_9x_{10}^2x_{11} - 44x_3^4x_5x_7^2x_9^2x_{11}^2 \\\notag
&+ 63x_3^2x_4^2x_5x_7x_8x_9x_{10}x_{11}^2 - 31x_4^4x_5x_8^2x_{10}^2x_{11}^2 - 5x_5^3,
\end{align}
with random complex structures.
Here we show the parity of the original coordinates under the reflection $\sigma_r$  in terms of the seven independent $\IC^*$ actions  based on matrix $\bW$ eq.(\ref{example ori wm}):
\bea
(x_1, x_2, x_3,  x_4,  x_5,  x_6, x_7, x_8, x_9, x_{10}, x_{11}) \sim ( \lambda_1\, x_1, \lambda_{2} \,x_2, 
\lambda_{1} \,\lambda_{2} \,\lambda_{3} \,\lambda_{4} \,\lambda_{7}  \,x_3, \lambda_3  \,x_4,  \nonumber\\   \lambda_{1}^{2} \,\lambda_{4} \,\lambda_{5} \,\lambda_{6} \, x_5, \lambda_4 \, x_6, \lambda_5 \, x_7, \lambda_6  \,x_8, \lambda_{2}^{-2} \,\lambda_{3}^{2} \,\lambda_{4} \,\lambda_{6} \,\lambda_{7} \, x_9,  \lambda_{1}^{2} \,\lambda_{3}^{-2} \,\lambda_{4} \,\lambda_{5} \,\lambda_{7}  \, x_{10}, \lambda_{7}  \,x_{11} ) \nonumber \\
= (x_1, x_2, -x_3,  x_4,  x_5,  x_6, x_7, -x_8, -x_9, x_{10}, x_{11}) 
\eea

Since we have shown the process in the exchange involution case above,  here we would omit some detail and list the fixed loci on $X$  that are solvable in the real number system when $\lambda\, \in\, \{1,-1\}$, i.e.,  four two-divisors fixed loci and one quartet-divisors fixed point:
\bea
\label{eq:locus1}
\mathcal{F}_{1}=\{x_1, x_5\}, \mathcal{F}_{2}=\{x_2, x_5\},\mathcal{F}_{3}=\{x_3, x_5\}, \mathcal{F}_{4}=\{x_4, x_{5}\}, \mathcal{F}_{5}=\{x_5, x_6, x_7, x_9\}.
\eea
Those fixed loci describe  $O7$-planes and $O3$-planes  since the  invariant polynomial vanish trivially on these locus $P_{sym}|_{\cF_i}=0, i = 1,\dots, 5$ as shown in eq.(\ref{eq:type}) and \cite{Crino:2022zjk}. 

To find complex solutions  $\lambda \in \IC^*$ for all possible fixed loci sets, we test from larger to smaller sets, demonstrating how known fixed points in the real number filed, 
 SR ideal  sets and  non-fixed loci sets can expedite the time-consuming process  of solving  eq.(\ref{eq:system2}). 
We first consider the largest possible fixed locus in the ambient space described by vanishing all the 11 divisors ${F'} = \{x_1= x_2=x_3=x_4=x_5=x_6=x_7=x_8=x_9=x_{10}=x_{11}=0\}$  and then put the three restrictions into consideration. 
For instance, the possible fixed locus $\{x_1 =  x_2=0\}$ is ruled out by considering the SR ideal eq.(\ref{eq:sr}). Since ${F'}$ is  part of $\{x_1= x_2=0\}$, it can be skipped in our scanning. Similarly, all the 10-divisors sets, 9-divisors sets and so on could be ruled out in this way by SR ideal until we meet 4-divisors sets. There is one  quartet-divisors fixed locus, denoted as  $F_6 =\{x_4=x_6=x_7=x_{9}=0\}$, which remain after applying the restriction of SR ideal. Thus we only need to test $F_6$ and find it indeed has a solution in the complex $\lambda$ system eq.(\ref{eq:system2}),   revealing a new putative fixed locus $\cF_6 = \{x_4, x_6, x_7, x_9\}$ in the ambient space $\cA$.

When comes to  3-divisors sets, beside those locus ruled out by fixed locus and SR ideal, we found  two non-fixed locus as $\cF_1^{non-fixed}=\{x_5, x_9, x_{11}\}$ and $\cF_2^{non-fixed}=\{x_3, x_7, x_{11}\}$. So their subsets $\{\{x_3\}, \{x_7\}, \{x_5\}, \{x_9\},\{x_{11}\}, \{x_5, x_9\}, \{x_9, x_{11}\}, \{x_5,x_{11}\},\{x_3,x_7\},\{x_7,x_{11}\},$ $\{x_3,x_{11}\}\}$ are all non-fixed points and we do not need to test them.

Repeat this process and we will get the fixed locus in complex space, besides those fixed locus we found in real $\lambda$ system eq.(\ref{eq:locus1}), there are additional new putative fixed locus in the ambient space which do have complex solution $\lambda \in \IC^*$: 
\bea
\label{eq:locus2}
& \mathcal{F}_{6}=\{x_4, x_6, x_7, x_9\},\,\, \mathcal{F}_{7}=\{x_3, x_8,x_9\}, \\
& \mathcal{F}_{8}=\{x_6, x_7,x_8\}, \,\,\mathcal{F}_{9}=\{x_6, x_9,x_{10}\}, \,\,\mathcal{F}_{10}=\{x_7,x_9 ,x_{11}\}, . \nonumber
\eea
However, all of these five putative fixed loci do not intersect with the invariant Calabi-Yau hypersurface defined by $P_{sym}=0$ since ${\rm dim}\,\cI_{\cF_i}^{fixed} < 0, (i=6,\dots 10)$ according to eq.(\ref{eq:fixed}).  Finally,  only the  five fixed loci on the Calabi-Yau hypersurface $X$  described in eq.(\ref{eq:locus1}) remain. As discussed in Section \ref{subsec:fixedX}, these fixed loci such as $\mathcal{F}_{i}, (i= 1, \dots, 5)$  can not be further reduced  as we did in previous subsection for  quartet-divisors fixed points in the divisor exchange involution. 
Moreover, $P_{sym}$ vanishes trivially on these fixed loci, $P_{sym}|_{F_{i}} = 0$ for $i=1,\dots,5$, 
indicating that the $O3,O7$-planes on the Calabi-Yau $X$ are described by the intersection of four and two divisors in the ambient space respectively. Their contribution to D3-tadpole  $Q_{D3}^{loc}$ cancellation is calculated accordingly.
\begin{table}[ht!]
            \centering
		{\footnotesize
			\centering
			\begin{tabular}{c|c|c}
                \hline
                \centering Type of O-planes & Fixed Loci on $X$  & Contribution to $Q_{D3}^{loc}$ \\\hline\hline
            
             $\cF_1:$     O7   & $\{x_1, x_5\} $ & $ 3$ \\\hline
             
          $\cF_2:$       O7 & $\{x_2, x_5\}$  &  $3$ \\\hline
         $\cF_3:$      O7 & $\{x_3, x_5\} $ & $3$  \\\hline
          $\cF_4:$      O7 & $\{x_4, x_{5}\}  $ & $6$ \\\hline
          $\cF_5:$      O3 & $\{x_5, x_6, x_7, x_9\} $ & $1$ \\\hline
			\end{tabular}
			\label{tab:ref2}
			\caption{ Fixed loci on $X$}
		}
	\end{table}
	
\noindent	
Using eq.(\ref{eq:tadpole}) the contribution to the $D3$-brane tadpole locally is 
 \bea
 N_{D3} + \frac{N_{\text{flux}}}{2}+ N_{\rm gauge}= \frac{N_{O3}}{4}+\frac{\chi(D_{O7})}{4} = \frac{9 + 15}{4} = 6 \,.
\eea
Thus $Q_{D3}^{loc} = 4$ which indicate that it is also an \lq\lq naive orientifold Type~IIB string vacua".   
The Hodge number splitting is followed:
    \bea
&h^{2,1}_-  = \frac{h^{1,2}- h^{1,1}}{2}-1+\frac{\chi(O_\sigma)}{4} = 14+4 =  18\, , \nonumber\\
&h^{1,1}_+ = 7, \,\,h^{1,1}_-=0,\,\, h^{2,1}_+ = 19.
\eea
Since $h^{2,1}_\pm$ are integer, the invariant  hypersurface $P_{sym}$ is supposed to be smooth.

We have collected all the results for these 11 single reflections, 15 randomly selected double reflections and 15 randomly selected triple reflections in Table.\ref{tab:reflectionex} in Appendix.\ref{sec:ref}. From this data,  we identified  three free actions which we will discuss in detail shortly.

 \subsubsection{New Type of Free Action}
 \label{subsub:free}

 There are two type of free actions. One is that it do have a fixed locus on the ambient space $\cA$, but not intersect with Calabi-Yau hypersurface $X$ as described in  \cite{Altman:2021pyc}. The other one is a new type of free action that it does not exist fixed locus on $\cA$ in the beginning. 
 Here we give an example of the new type of free action. Consider the reflection as:
\bea
\sigma_r : \,\, x_1 \leftrightarrow -x_1, \,x_2 \leftrightarrow -x_2
\label{eq:ref2}
\eea
Here we show the parity of the coordinates under the reflection $\sigma_r$ eq.(\ref{eq:ref2}) in terms of the seven independent $\IC^*$ actions  from weighted matrix  $\bW$ eq.(\ref{example ori wm}):
\bea
(x_1, x_2, x_3,  x_4,  x_5,  x_6, x_7, x_8, x_9, x_{10}, x_{11}) \sim ( \lambda_1 x_1, \lambda_{2} \,x_2, 
\lambda_{1} \,\lambda_{2} \,\lambda_{3} \,\lambda_{4} \,\lambda_{7}  \,x_3, \lambda_3  \,x_4,  \nonumber\\   \lambda_{1}^{2} \,\lambda_{4} \,\lambda_{5} \,\lambda_{6} \, x_5, \lambda_4 \, x_6, \lambda_5 \, x_7, \lambda_6  \,x_8, \lambda_{2}^{-2} \,\lambda_{3}^{2} \,\lambda_{4} \,\lambda_{6} \,\lambda_{7} \, x_9,  \lambda_{1}^{2} \,\lambda_{3}^{-2} \,\lambda_{4} \,\lambda_{5} \,\lambda_{7}  \, x_{10}, \lambda_{7}  \,x_{11} ) \nonumber \\
= (- x_1, - x_2, x_3,  x_4,  x_5,  x_6, x_7, x_8, x_9, x_{10}, x_{11}) 
\eea
However,  those transformation are trivially satisfied by setting $\boldsymbol{\lambda}=\{-1, -1, 1, 1, 1, 1, 1\} $ without forcing any locus to vanish. This indicates there is no fixed loci sets on the corresponding ambient space $\cA$ and we conclude that such reflections $\{x_1\ \leftrightarrow\ -x_1,\ x_2\ \leftrightarrow\ -x_2\}$ is a $\IZ_2$ free action on Calabi-Yau hypersurface $X$.  The Hodge number splitting under  the double divisor reflections is:
  \bea
&h^{2,1}_-  = \frac{h^{1,2}- h^{1,1}}{2}-1+\frac{\chi(O_\sigma)}{4} = 14 + 0 =  14\, , \nonumber\\
&h^{1,1}_+ = 7, \,\,h^{1,1}_-=0,\,\, h^{2,1}_+ = 23.
\eea
By calculating the dimension of ideal eq.(\ref{eq:smooth}) in all patches, we confirmed that $P_{sym}$ is  smooth and thus $\sigma_r$ is indeed a free action.

Following the similar discussion, we can identify another two free actions with the same type for those reflections we considered. They are:
\bea
&\sigma_r' : x_1 \leftrightarrow -x_1, \,x_3 \leftrightarrow -x_3, \\\nonumber 
&\sigma_r'':  x_5 \leftrightarrow -x_5, \,x_8 \leftrightarrow -x_8, \, x_9 \leftrightarrow -x_9
\eea
which are sumarized in Table.\ref{tab:reflectionex} in Appendix.\ref{sec:ref}.

\subsection{Example B}

In previous example, it may give people one illusion that there is no new fixed locus on the Calabi-Yau manifold when solving the complex $\lambda$ system. This is not always the case. It is well known that certain systems may lack solutions in the real number field but  possess solutions in the  complex number field.  Therefore, we may miss some fixed loci if we only restrict our analysis in the real  system.

Here we give a simple example with $h^{1,1}=3, h^{2,1}=37$ (${\rm Polyid:} 79, {\rm Tri\_id:} 0$) to show the differences. The GLSM charge matrix of this example is:
	\begin{equation}
		\label{eq:h113example}
		\begin{array}{|c|c|c|c|c|c|c|l}
			\cline{1-7}
			x_{1} & x_{2} & x_{3} & x_{4} & x_{5} & x_{6} & x_{7} &\\\cline{1-7}
                  1   &   0   &   1   &   1   &   0   &   0   &   0   & \lambda_{1} \\\cline{1-7}   
                  0   &   1   &   0   &   1   &   1   &   2   &   0   & \lambda_{2} \\\cline{1-7}   
                  0   &   0   &   0   &   0   &   0   &   1   &   1   & \lambda_{3} \\\cline{1-7}   
		\end{array}
	\end{equation}
	\noindent 
 and Stanley-Reisner ideal:
	\bea
	\label{eq:sr, h11=3 two algorithm different}
	\cI_{SR}=\langle x_1 x_3 x_4,\, x_1 x_3 x_7,\, x_2 x_4 x_5,\, x_2 x_5 x_6,\, x_6 x_7 \rangle.
	\eea
The Calabi-Yau manifold $X$ is defined by the anti-canonical hypersurface in the ambient space $\cA$ with polynomial degree $||3,5,2||$. The Hodge numbers of the corresponding individual toric divisors $D_i \equiv \{x_i = 0\} $ are:
	\begin{equation}
		\label{eq:divisorh113example}
		\begin{array}{|c|c|c|c|c|c|c|c|l}
			\cline{1-8}
			h^\bullet(D_i)&D_{1} & D_{2} & D_{3} & D_{4} & D_{5} & D_{6} & D_{7}  \\\cline{1-8}
			h^{0,0}(D_i)  &  1   &   1   &   1   &   1   &   1   &   1   &   1  \\\hline
			h^{0,1}(D_i)  &  0   &   0   &   0   &   0   &   0   &   1   &   1  \\\hline
			h^{0,2}(D_i)  &  1   &   1   &   1   &   4   &   1   &   3   &   0  \\\hline
			h^{1,1}(D_i)  &  21  &  21   &  21   &  46   &   21  &   40  &   2  \\\hline
		\end{array}
	\end{equation}
	There is no proper NIDs exchange involution and we only need to consider reflections. For simplicity, we consider the reflection as:
	\bea
	\sigma_r:  x_6\, \leftrightarrow - x_6
	\eea
	Such reflection gives the parity of coordinates followed by the weighted matrix  eq.(\ref{eq:h113example}) as:
\bea
\label{eq:parity, h11=3 two algorithm different}
(x_1, x_2, x_3,  x_4,  x_5,  x_6, x_7) &\sim &( \lambda_{1}\, x_1, \lambda_{2} \,x_2, 
\lambda_{1} \,x_3, \lambda_{1}  \,\lambda_{2} \,x_4,  \lambda_{2} \, x_5, \lambda_{2}^2 \,\lambda_{3} \, x_6, \lambda_{3} \, x_7 ) \nonumber \\
&= &(x_1, x_2, x_3,  x_4,  x_5,  -x_6, x_7) 
\eea
By analyzing the fixed locus in the real $\lambda$ system as described in previous examples, we can identify the presence of  two
 $O7$-planes: 
 \bea
 F_1:  \{x_6=0\} \quad &{\rm from \,\,\, solution}& \quad  \boldsymbol{\lambda}=  \{1,1,1 \}\\\nonumber
 F_2:  \{x_7=0\} \quad &{\rm from \,\,\, solution}& \quad  \boldsymbol{\lambda}=  \{1,1,-1 \}\
 \eea
 If we directly take these results as  the final determination of the  fixed locus on  $X$,  we reproduce the findings of  \cite{Crino:2022zjk}. 
 Consequently,  this orientifold Calabi-Yau threefold  can support a naive orientifold Type IIB string vacuum,
\begin{table}[ht!]
            \centering
		{\footnotesize
			\centering
			\begin{tabular}{c|c|c}
                \hline
                \centering Type of O-planes & Fixed Loci on $X$  & Contribution to $Q_{D3}^{loc}$ \\\hline\hline
            
           $ \cF_1:$   O7   & $\{x_6\} $ & $ 4$ \\\hline
             
         $\cF_2:$      O7   & $\{x_7\}$  &  $48$ \\\hline
			\end{tabular}
			\label{table: ref3}
			\caption{ Fixed loci on $X$ in real number system}
		}
	\end{table}
	
	\noindent
Because the local D3-tadpole cancellation condition can  be satisfied: 
 \bea
 N_{D3} + \frac{N_{\text{flux}}}{2}+ N_{\rm gauge}= \frac{N_{O3}}{4}+\frac{\chi(D_{O7})}{4} = \frac{52}{4} = 13 \,.
\eea
The Hodge number splitting follows:
    \bea
h^{2,1}_- (X/\sigma^{*}) = \frac{h^{1,2}- h^{1,1}}{2}-1+\frac{\chi(O_\sigma)}{4} = 35+13 =  48\, .
\eea

However,  new fixed loci do exist on the Calabi-Yau threefolds $X$ when we extend the solution space to the complex field $\IC^*$. To identify these loci, consider the potential fixed locus  $\cF_3 = \{x_1, x_2, x_3, x_5\}$ by setting these coordinates to zero simultaneously while keep the other  coordinates invariant under parity. Then  eq.(\ref{eq:system2}) reads:
\bea
\label{eq:parity2}
(x_4,  x_6, x_7) \sim (  \lambda_{1}  \,\lambda_{2} \,x_4, \lambda_{2}^2 \,\lambda_{3} \, x_6, \lambda_{3} \, x_7 ) = ( x_4, -x_6, x_7) 
\eea
Clearly, there are no solutions of $\lambda_i = \pm 1$ in eq.(\ref{eq:parity2}). However, eq.(\ref{eq:parity2}) do have solutions  $ \boldsymbol{\lambda}=  \{-i,i,-1 \}$ once we expand the solution space to the complex field $ \IC^*$, as required by definition of toric variety. Further analysis confirms that this locus  intersect with $P_{sym}$. Specifically,  $P_{sym}$ vanishes trivially on $F_3$ such that $P_{sym}|_{F_3} =0$,  indicating $\cF_3$ is an $O3$-plane followed from eq.(\ref{eq:type}):
\begin{table}[ht!]
            \centering
		{\footnotesize
			\centering
			\begin{tabular}{c|c|c}
                \hline
                \centering Type of O-planes & Fixed Loci on $X$  & Contribution to $Q_{D3}^{loc}$ \\\hline\hline

          $\cF_1:$     O7   & $\{x_6\} $ & $ 4$ \\\hline
             
            $\cF_2:$    O7   & $\{x_7\}$  &  $48$ \\\hline
          $\cF_3:$       O3   & $\{x_1,x_2,x_3,x_5\} $ & $ \frac{1}{2}$ \\\hline
               
			\end{tabular}
			\label{table: ref4}
			\caption{Fixed loci on $X$ in complex number $\IC^*$ system}
		}
	\end{table}

\noindent
Therefore the D3-tadpole cancelation condition can not be satisfied:
 \bea
 N_{D3} + \frac{N_{\text{flux}}}{2}+ N_{\rm gauge}= \frac{N_{O3}}{4}+\frac{\chi(D_{O7})}{4} = \frac{52.5}{4} = 13.125 \,,
\eea
and the Hodge number split as
    \bea
h^{2,1}_- (X/\sigma^{*}) = \frac{h^{1,2}- h^{1,1}}{2}-1+\frac{\chi(O_\sigma)}{4} = 35+13.125 =  48.125\, ,
\eea
So we do not have a string vacua under such reflection.


 \section{Scanning Results}
 \label{sec:result}
 The calculation tools we applied for preparing the initial data  including PALP \cite{Kreuzer:2002uu} for desingularize the polytopes,  SAGE \cite{sage} and CYTools \cite{Demirtas:2022hqf} for triangulations and SINGULAR \cite{DGPS} for singularity check. 
 
 First, a systematic scan was performed on Kreuzer-Skarke database \cite{Kreuzer:2000xy} up to $h^{1,1}(X)=7$. We analyzed all those polytopes and found $71,190$ favorable polytopes, from which  $6,637,231$ MPCP triangulations are founded.  The number of triangulations we considered is one order magnitude larger than those considered in our previous work \cite{Altman:2021pyc}.  Second, a limited scan was performed up to $ h^{1,1}(X)=12$.  We  choose the favorable polytope for each $\{h^{1,1}, h^{2,1}\}$ considered in \cite{Crino:2022zjk}, followed with $7,934$ polytopes and $59,483$ MPCP triangulations. These results are listed in Table.\ref{tab:tri7} and Table.\ref{tab:tri8}.
 
         According to the definition of orientifold projection eq.(\ref{eq:orientifold}), each of the proper involutions will lead to an orientifold Calabi-Yau threefold.  As a result, we will classify various properties of orientifold Calabi-Yau threefolds in the $\IZ_2$ orbifold limit according to different kinds of  divisor exchange involutions and multi-divisor reflections.
 
\begin{table}[h!]
		{\footnotesize
			\centering
			\begin{tabular}{|r|r||c|c|c|c|c|c||c|}
				\hline
				\multicolumn{2}{|c||}{\parbox[c][2em][c]{5cm}{\centering$\mathbf{h^{1,1}(X)}$}}  & \textbf{2} & \textbf{3} & \textbf{4} & \textbf{5} & \textbf{6} & \textbf{7}  & \bf{Total} \\\hline
				\multicolumn{2}{|r||}{\parbox[c][3em][c]{5cm}{\centering\textbf{\# of Favorable Polytopes}}}     & 36    & 243   & 1185  & 4897  & 16608 & 48221 & 71190\\    \hline
				\multicolumn{2}{|r||}{\parbox[c][3em][c]{5cm}{\centering\textbf{\# of Favorable Triangulations}}}    & 48    & 525   & 5330  & 56714 & 584281 & 5990333 & 6637231\\\hline
			\end{tabular}
			\caption{The favorable polytopes,  triangulations for $h^{1,1} \leq 7$. }
			\label{tab:tri7}
		} 
	\end{table}

 	\begin{table}[h!]
		{\footnotesize
			\centering
			\begin{tabular}{|r|r||c|c|c|c|c||c|}
				\hline
				\multicolumn{2}{|c||}{\parbox[c][2em][c]{5cm}{\centering$\mathbf{h^{1,1}(X)}$}} & \textbf{8} & \textbf{9} & \textbf{10} & \textbf{11} & \textbf{12} & \bf{Total} \\\hline
				\multicolumn{2}{|r||}{\parbox[c][3em][c]{5cm}{\centering\textbf{\# of Favorable Polytopes}}}  &  1486        &  1581        &  1554        &  1621        &  1692         & 7934  \\    \hline
				\multicolumn{2}{|r||}{\parbox[c][3em][c]{5cm}{\centering\textbf{\# of Favorable Triangulations}}}  &  6847        &  9833        &  12185       &  15638       &  14980        & 59483     \\\hline
			\end{tabular}
			\caption{Limited  triangulations for partial favorable polytopes $8 \leq h^{1,1} \leq 12$. }
			\label{tab:tri8}
		}
	\end{table}
	
        \subsection{Divisor Exchange Involutions}
        \label{subsec:NID}
        
\subsubsection{Classification of Proper Involution}
        
For divisor exchange involutions, we have identified the proper involution of the so called non-trivial identical divisors (NIDs). These proper involutions must satisfy the symmetry requirements of the Stanley-Reisner ideal $\cI_{SR}$ and the linear ideal $\cI_{lin}$. This entails that the involution should be an automorphism of the ambient space $\cA$, preserving the homogeneity of the polynomials. Additionally,  the triple intersection form should remain invariant under the proper involutions.  In Table.\ref{tab:NID7}, we provide a count  of  polytopes and triangulations that contain  proper NIDs exchange involutions. Specifically,  $8,664$ out of the $71,190$ favorable polytopes contain the proper NIDs exchange involutions for $h^{1,1}(X)\leq 7$,  whereas only $158$ out of $7,934$ favorable polytopes contain the proper NIDs exchange involutions for $8 \leq h^{1,1}(X)\leq 12$. The low percentage ($2\%-3\%$) of polytopes containing such proper NIDs exchange involutions for higher $h^{1,1}(X)$ is attributed to  the fact that not  all triangulations for a given polytope where exhaustively considered for higher $h^{1,1}(X)$. A similar trend is observed  when counting the number of triangulations containing proper NIDs exchange involutions:   $147,767$ out of $6,637,231$ for $h^{1,1}(X)\leq 7$   and $435$ out of $59,483$ for  $8 \leq h^{1,1}(X) \leq 12$. This indicates that  polytopes or triangulations with  proper NIDs exchange involutions are rare when $h^{1,1}(X)$ increases, which is a promising signal for machine learning to provide  predictions. The total number of proper NIDs exchange involutions is $156,244$  for $h^{1,1}(X)\leq 7$   and  $465$ for $8 \leq h^{1,1}(X)\leq 12$ respectively.  These numbers align closely with the number of triangulations containing proper NIDs exchange involutions, suggesting  that for large $h^{1,1}(X)$, each   triangulation may  contain only one  proper involution. Since each of the proper NIDs exchange involution defines an orientifold Calabi-Yau, we conclude with a total of $156,709$ orientifolds.

Each  proper NIDs exchange involutions may exchange several pairs of topologically distinguished divisors.  we enumerate the number of different pairs of NIDs for each of the involution, as shown in Table.\ref{tab:pairs7} and Table.\ref{tab:pairs8}. Only divisors with specific topological characteristics are counted, hence the total number of divisors does not match the number of involutions. Each row starts with the divisor type, indicating that the involution includes that particular type of divisor. For example, if an involution exchanges both Wilson divisors and K3 divisors, the row labeled \lq\lq K3 and Wilson" will be incremented by one, corresponding to the respective $h^{1,1}(X)$. It is noteworthy that it is very rare for an involution to simultaneously exchange  del Pezzo, K3, and exact-Wilson divisors  for small $h^{1,1}(X)$.

	\begin{table}[h!]
		{\footnotesize
			\centering
			\begin{tabular}{|r|r||P{0.7cm}|P{1cm}|P{1cm}|P{1cm}|P{1cm}|P{1cm}||c|}
				\hline
				\multicolumn{2}{|c||}{\parbox[c][2em][c]{5cm}{\centering$\mathbf{h^{1,1}(X)}$}} &  \textbf{2} & \textbf{3} & \textbf{4} & \textbf{5} & \textbf{6} & \textbf{7}  & \bf{Total} \\\hline
				\multicolumn{2}{|r||}{\parbox[c][3em][c]{5cm}{\centering\textbf{\# of Polytopes \\ contains  proper involutions}}}   &  2                   &  28                  &  171                 &  712                 &  2172                &  5579                 & 8664   \\\hline

				\multicolumn{2}{|r||}{\parbox[c][3em][c]{5cm}{\centering\textbf{\# of Triangulations contains proper involutions}}}          &  2                   &  36                  &  410                 &  3372                &  21566               &  122381               & 147767 \\\hline
			
				\multicolumn{2}{|r||}{\parbox[c][3em][c]{5cm}{\centering\textbf{{ \# of  proper involutions}}}}         & 12          &  61          &  548         &  4085        &  23805       &  127733       & 156244   \\\hline\hline
				\multicolumn{2}{|c||}{\parbox[c][2em][c]{5cm}{\centering$\mathbf{h^{1,1}(X)}$}} &  \textbf{8} & \textbf{9} & \textbf{10} & \textbf{11} & \textbf{12} && \bf{Total} \\\hline
				
				
    		\multicolumn{2}{|r||}{\parbox[c][3em][c]{5cm}{\centering\textbf{\# of Polytopes \\ contains proper Involutions}}}    &  38          &  40          &  34          &  22          &  24        &   & 158 \\\hline 

				\multicolumn{2}{|r||}{\parbox[c][3em][c]{5cm}{\centering\textbf{\# of Triangulations contains proper Involutions}}}     &  80          &  112         &  79          &  85          &  79           & &435  \\\hline 
				\multicolumn{2}{|r||}{\parbox[c][3em][c]{5cm}{\centering\textbf{{ \# of proper Involutions}}}}    &  86          &  122         &  82          &  86          &  89         &  & 465     \\\hline

			\end{tabular}
			\caption{Number of the proper NID  exchange involutions in favorable polytopes and triangulations.}
			\label{tab:NID7}
		}
	\end{table}

	\begin{table}[h!]
	
		{\footnotesize
			\centering
			\hspace*{-1.2cm}
			\begin{tabular}{|r||c|c|c|c|c|c||c|}
				\hline
				\multicolumn{8}{|c|}{\parbox[c][3em][c]{15cm}{\centering\textbf{Number of pairs of Non-trivial Identical Divisors (NIDs) under involutions}}} \\
				\hline
				\parbox[c][2em][c]{6cm}{\centering$\mathbf{h^{1,1}(X)}$} & \textbf{2} & \textbf{3} & \textbf{4} & \textbf{5} & \textbf{6} & \textbf{7} & {\bf Total} \\
				\hline
				\parbox[c][2em][c]{6cm}{\centering\textbf{\# of  Involutions}}  &  12          &  61          &  548         &  4085        &  23805       &  127733       & 156244   \\\hline
				\hline
				\parbox[c][2em][c]{6cm}{\centering\textbf{del Pezzo surface }$\mathbf{dP_{n}}$, $\mathbf{n\leq 8}$}& 0           &  12          &  238         &  2192        &  13550       &  82174       &  98166\\\hline
				\parbox[c][2em][c]{6cm}{\centering\textbf{Rigid surface} $\mathbf{dP_{n}}$, $\mathbf{n > 8}$}& 6           &  24          &  360         &  3381        &  20498       &  110385      &  134654 \\\hline
				\parbox[c][2em][c]{6cm}{\centering\textbf{(exact-)Wilson}\textbf{ surface}}  & 0           &  7  (0)          &  44   (7)       &  176   (80)       &  744  { (411)}        &  3965   (1944)     &  4936 (2442)\\\hline
				\parbox[c][2em][c]{6cm}{\centering\textbf{K3 surface}} & 0           &  33          &  170         &  481         &  1821        &  6528        &  9033 \\\hline
				\parbox[c][2em][c]{6cm}{\centering\textbf{SD1 surface}} & 0           &  5           &  41          &  407         &  2185        &  9977        &  12615 \\\hline
				\parbox[c][2em][c]{6cm}{\centering\textbf{SD2 surface}}& 6           &  6           &  33          &  109         &  459         &  1343        &  1956 \\\hline
					\parbox[c][2em][c]{6cm}{\centering\textbf{del Pezzo and K3 }}& 0           &  7           &  84          &  391         &  1741        &  6472        &  8695  \\\hline
              \parbox[c][2em][c]{6cm}{\centering\textbf{K3 and (exact-)Wilson}} & 0           &  4  (0)         &  4      (4)     &  16    (5)      &  70    (4)      &  98   (35)       &  192 (48)\\\hline

				\parbox[c][2em][c]{6cm}{\centering\textbf{del Pezzo and (exact-)Wilson }}& 0           &  3           &  38  (5)        &  176  (80)        &  744   (411)    &  3965 (1944)        &  4926  (2440)\\\hline
				\parbox[c][3em][c]{6cm}{\centering\textbf{del Pezzo, K3 and (exact-)Wilson }}& 0           &  0           &  2  { (2)}         &  21  (5)        &  74     (4)     &  133    (35)      &  230 (46) \\\hline
              				\end{tabular}
			\caption{Number of pairs of NIDs exchanged under proper involutions with $ h^{1,1}\leq 7$}
			\label{tab:pairs7}
		}
	\end{table}%

\begin{table}[h!]
        
		{\footnotesize
			\centering
			\hspace*{-.5cm}
			\begin{tabular}{|r||c|c|c|c|c||c|}
				\hline
				\multicolumn{7}{|c|}{\parbox[c][3em][c]{15cm}{\centering\textbf{Number of pairs of Non-trivial Identical Divisors (NIDs) under involutions}}} \\
		\hline
				\parbox[c][2em][c]{6cm}{\centering$\mathbf{h^{1,1}(X)}$} & \textbf{8} & \textbf{9} & \textbf{10} & \textbf{11} & \textbf{12}  & {\bf Total} \\
				\hline
				\parbox[c][2em][c]{6cm}{\centering\textbf{\# of  Involutions}}  &  86          &  122         &  82          &  86          &  89           & 465      \\\hline\hline
				\parbox[c][2em][c]{6cm}{\centering\textbf{del Pezzo surface }$\mathbf{dP_{n}}$, $\mathbf{n\leq 8}$}& 62          &  72          &  54          &  70          &  87          &  345\\\hline
				\parbox[c][2em][c]{6cm}{\centering\textbf{(exact-)Wilson}\textbf{ surface}}   & 24  { (20)}         &  29  { (19)}        &  15  { (11)}       &  13   { (9)}        &  50   { (28)}       &  131 { (87)}\\\hline
				\parbox[c][2em][c]{6cm}{\centering\textbf{K3 surface}}& 16          &  4           &  12          &  4           &  7           &  43 \\\hline
				\parbox[c][2em][c]{6cm}{\centering\textbf{SD1 surface}} & 1           &  0           &  3           &  4           &  8           &  16\\\hline
				\parbox[c][2em][c]{6cm}{\centering\textbf{SD2 surface}} & 0           &  0           &  0           &  1           &  2           &  3 \\\hline
					\parbox[c][3em][c]{6cm}{\centering\textbf{del Pezzo and K3 } } & 16          &  4           &  12          &  4           &  7           &  43\\\hline
                \parbox[c][2em][c]{6cm}{\centering\textbf{K3 and (exact-)Wilson}}& 3   { (2)}        &  1 (0)          &  1   (0)        &  0  (0)         &  7      (1)     &  12 (3)\\\hline
			
				\parbox[c][3em][c]{6cm}{\centering\textbf{del Pezzo and (exact-)Wilson}} & 24   { (20)}        &  29  { (19)}        &  15  { (11)}         &  13   { (9)}        &  50     { (28)}     &  131 { (87)}  \\\hline
				\parbox[c][3em][c]{6cm}{\centering\textbf{del Pezzo, K3 and (exact-)Wilson}}& 3   { (2)}        &  1  (0)         &  1    (0)       &  0    (0)       &  7     (1)      &  12 (3)\\\hline
    			\end{tabular}
			\caption{Number of pairs of NIDs exchanged under proper involutions with  $8\leq h^{1,1}\leq 12$}
			\label{tab:pairs8}
		}
	\end{table}%
	\subsubsection{Classification of O-planes and String Landscape}

	In addition to determining the proper NIDs exchange involution of the toric variety, we have computed all  fixed point sets of the involution to identify the emergence of O-planes. These planes can potentially resolve  anomalies if they have appropriate configurations. Through this process, we have successfully identified all  O-planes on the Calabi-Yau hypersurface $X$. 
The fixed sets allowed by a  proper NIDs exchange involution include individual $O3$, $O5$, or $O7$-planes, or combinations of $O3$ and $O7$-planes. In every case, the parity of the volume form under $\sigma$ is in agreement with the orientifold planes found by our algorithm, i.e.,  $\sigma^* \Omega = - \Omega$ for $O3$, $O7$, and $O3/O7$ cases, and  $\sigma^* \Omega = \Omega$ for $O5$ configurations.  As shown in Table.\ref{tab:oplane1}, for $h^{1,1}(X)\leq 7$, there are $123,237$ out of $156,244$ involutions that contain both $O3$ and $O7$-planes. For $8 \leq h^{1,1}(X)\leq 12$, $285$ out of $465$ involutions contain both $O3$ and $O7$-planes. If there is no O-plane and the invariant Calabi-Yau hypersurface $P_{sym}$ is smooth, we identify such involution as a free action. In our scan, there are only triangulations with $h^{1,1}=6$ that contain such  free actions for proper  NIDs  exchange involution. It shows that under the proper involutions one end up with majority the $O3/O7$-planes system and most of them ($105,758$ out of $156,709$ orientifold CY) will further admit a naive Type IIB string vacua.  Here we  only consider the string vacua at the case of $O3/O7$ plane using the assumption that eight $D7$-branes are placing on the top of $O7$-plane just like \cite{Altman:2021pyc}, and  D3-tadpole condition is satisfied if the total charge of $Q_{D3}^{loc}$ is an integer. After consider the D3-tadpole cancellation, the number of naive orientifold Type IIB string vacua is summarized in Table.\ref{tab:stringvacua7}.
	 
\begin{figure}[ht!]
\centering
\includegraphics[width=7.5cm]{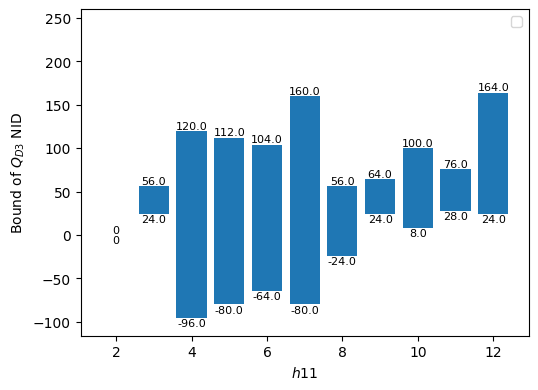}\quad
\includegraphics[width=7.5cm]{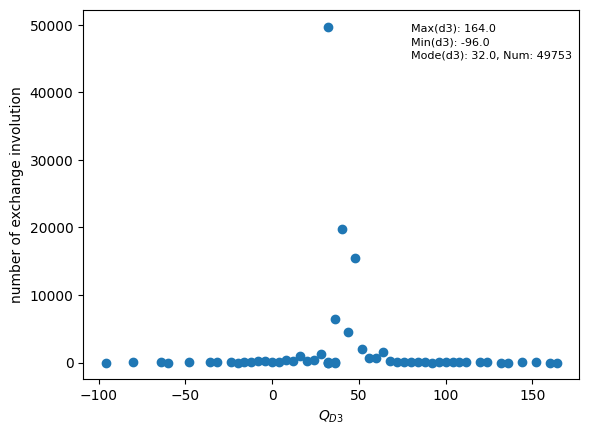}
\caption{Distribution of $Q_{D3}^{loc}$ under  exchange involutions for naive orientifold Type IIB string vacua.}
\label{figure}
\end{figure}

	 For these $105,758$ naive orientifold Type IIB string vacua with  $O3/O7$-system, the distribution of $Q_{D3}^{loc}$  is shown in Fig.~\ref{figure}.  The data indicates   that most of the involutions result in  orientifold Calabi-Yau threefolds with $Q_{D3}^{loc} $ around $32$ in our scan. The smallest and largest $Q_{D3}^{loc}$ values are $-96$ and $164$ respectively.

	\begin{table}[h!]
		{\footnotesize
			\centering
			\begin{tabular}{|P{3.5cm}||P{1cm}|P{1cm}|P{1cm}|P{1.2cm}|P{1.3cm}|P{1.4cm}||P{1.5cm}|}
				\hline
				
				\multicolumn{8}{|c|}{\parbox[c][2em][c]{15cm}{\centering\textbf{Classification of O-plane fixed point locus}}} \\
				\hline
				
				\parbox[c][2em][c]{3.5cm}{\centering$\mathbf{h^{1,1}(X)}$}& \textbf{2} & \textbf{3} & \textbf{4} & \textbf{5} & \textbf{6} & \textbf{7} & {\bf Total} \\\hline	
				\parbox[c][2em][c]{3.5cm}{\centering\textbf{\# of  Involutions}}       &  12          &  61          &  548         &  4085        &  23805       &  127733       & 156244   \\\hline\hline
				
				\parbox[c][2em][c]{3.5cm}{\centering\textbf{ Only O3}}                    &  0           &  0           &  31          &  74          &  359         &  727          & 1191    \\\hline
				
				\parbox[c][2em][c]{3.5cm}{\centering\textbf{Only O5}}                    &  12          &  22          &  173         &  1006        &  3283        &  10921        & 15417   \\\hline
				
				\parbox[c][2em][c]{3.5cm}{\centering\textbf{Only O7}}                  &  0           &  30          &  122         &  432         &  2121        &  13679        & 16384   \\\hline
				
				\parbox[c][2em][c]{3.5cm}{\centering\textbf{O3 and O7}}            &  0           &  9           &  222         &  2573        &  18027       &  102406       &  123237   \\\hline
				
				\parbox[c][2em][c]{3.5cm}{\centering\textbf{Free Action}}  &  0                &  0           &  0           &  0           &  15          &  0            & 15  \\\hline\hline
				\parbox[c][2em][c]{3.5cm}{\centering$\mathbf{h^{1,1}(X)}$}& \textbf{8} & \textbf{9} & \textbf{10} & \textbf{11} & \textbf{12} && {\bf Total} \\\hline
				\parbox[c][2em][c]{3.5cm}{\centering\textbf{\# of  Involutions}}    &  86          &  122         &  82          &  86          &  89      &     & 465      \\\hline\hline
				
				\parbox[c][2em][c]{3.5cm}{\centering\textbf{Only O3}}   &  4           &  0           &  1           &  0           &  0        &    & 5    \\\hline
				\parbox[c][2em][c]{3.5cm}{\centering\textbf{Only O5}}    &  17          &  32          &  10          &  5           &  12       &    & 76  \\\hline
				
				\parbox[c][2em][c]{3.5cm}{\centering\textbf{Only O7}}   &  13          &  33          &  16          &  20          &  17       &    & 99    \\\hline
				
				\parbox[c][2em][c]{3.5cm}{\centering\textbf{O3 and O7}}   &  52          &  57          &  55          &  61          &  60     &      & 285    \\\hline
				\parbox[c][2em][c]{3.5cm}{\centering\textbf{Free Action}}  &  0           &  0           &  0           &  0           &  0        &    & 0  \\\hline
			\end{tabular}
			\caption{ O-planes and free actions under proper divisor exchange involutions.  
			}
			\label{tab:oplane1}
		}
	\end{table}

	\begin{table}[ht!]
		{\footnotesize
			\centering
			\begin{tabular}{|P{4.5cm}||P{1cm}|P{1cm}|P{1cm}|P{1cm}|P{1.5cm}|P{1cm}||P{1.5cm}|}
				\hline
				\multicolumn{8}{|c|}{\parbox[c][2em][c]{15cm}{\centering\textbf{Naive Orientifold Type IIB String Vacua with $O3/O7$-system}}} \\
				\hline
				\parbox[c][2em][c]{4.5cm}{\centering$\mathbf{h^{1,1}(X)}$}  & \textbf{2} & \textbf{3} & \textbf{4} & \textbf{5} & \textbf{6} & \textbf{7} & {\bf Total} \\\hline
                \hline
				\parbox[c][2em][c]{4.5cm}{\centering\textbf{\# of  Involutions}}    &  12          &  61          &  548         &  4085        &  23805       &  127733       & 156244  \\\hline
				\hline
				\parbox[c][2em][c]{3.5cm}{\centering\textbf{Contains O3 \& O7}}     &  0           &  7           &  195         &  2259        &  14396       &  73325        & 90182  \\
				\hline
				\parbox[c][2em][c]{3.5cm}{\centering\textbf{Contains Only O3}}      &  0           &  0           &  31          &  74          &  359         &  725          & 1189   \\
				\hline
				\parbox[c][2em][c]{3.5cm}{\centering\textbf{Contains Only O7}}    &  0           &  30          &  117         &  381         &  1867        &  11742        & 14137  \\
				\hline
				\parbox[c][2em][c]{3.5cm}{\centering\textbf{Total String Vacua}}  &  0           &  37          &  343         &  2714        &   16637      &  85792        & 105538   \\
				\hline
				 \hline
				 \parbox[c][2em][c]{4.5cm}{\centering$\mathbf{h^{1,1}(X)}$}  & \textbf{8} & \textbf{9} & \textbf{10} & \textbf{11} & \textbf{12}  & & {\bf Total} \\\hline
				\parbox[c][2em][c]{4.5cm}{\centering\textbf{\# of  Involutions}}  &  86          &  122         &  82          &  86          &  89        &   & 465    \\\hline
				\hline
				\parbox[c][2em][c]{3.5cm}{\centering\textbf{Contains O3 \& O7}}    &  29          &  34          &  30          &  14          &  21        &   & 128    \\
				\hline
				\parbox[c][2em][c]{3.5cm}{\centering\textbf{Contains Only O3}}   &  4           &  0           &  1           &  0           &  0       &     & 5     \\
				\hline
				\parbox[c][2em][c]{3.5cm}{\centering\textbf{Contains Only O7}}   &  12          &  33          &  16          &  15          &  11      &     & 87     \\
				\hline
				\parbox[c][2em][c]{3.5cm}{\centering\textbf{Total String Vacua}}   &  45          &  67          &  47          &  29          &  32     &      & 220    \\
				\hline
			\end{tabular}
			\caption{Naive orientifold Type IIB string vacua under proper  divisor exchange involutions}
			\label{tab:stringvacua7}
		}
	\end{table}%

	\subsubsection{Hodge Number Splitting}

	Finally,  we discuss the decomposition of the K\"ahler moduli space into odd and even parity equivariant cohomology $H^{1,1}(X/\sigma^{*})=H^{1,1}_{+}(X/\sigma^{*})\oplus H^{1,1}_{-}(X/\sigma^{*})$. The constraint that the K\"ahler form must be invariant $\sigma^{*}J=J$ ensures that we can always determine the dimension of the even parity space.  By deduction, we can then ascertain the dimension of the odd party space $h^{1,1}_{-}(X/\sigma^{*})$, which, as   discussed, must be non-trivial in NIDs exchange involutions. The results of this K\"ahler moduli space splitting are presented  in Table \ref{tab:split1}. 
	The largest $h^{1,1}_-$ observed is $3$ for $h^{1,1}(X)\leq 7$ and $6$ for $h^{1,1}(X)\leq 12$.
 	 By utilizing the Lefschetz fixed point theorem as indicated in eq.(\ref{eq:h21split}), we can further determine the splitting of $h^{2,1}_{\pm}(X/\sigma^*)$  in the orbifold limit.  The value of $h^{2,1}_{\pm}(X/\sigma^*)$ may be altered  by a possible conifold resolution, while $h^{1,1}_-(X/\sigma^*)$ remains robust.  In the case of  divisor reflection, $h^{1,1}_-(X) \equiv 0$ and therefore we do not present the hodge number splitting for this scenario.
	
	\begin{table}[h!]
		{\footnotesize
			\centering
			\begin{tabular}{|r|r||P{1cm}|P{1cm}|P{1cm}|P{1cm}|P{1cm}|P{1cm}||P{1.5cm}|}
				\hline
				\multicolumn{9}{|c|}{\parbox[c][2em][c]{14cm}{\centering\textbf{Hodge numbers splitting}}} \\
				\hline
				\multicolumn{2}{|c||}{\parbox[c][2em][c]{3.5cm}{\centering$\mathbf{h^{1,1}(X)}$}}& \textbf{2} & \textbf{3} & \textbf{4} & \textbf{5} & \textbf{6} & \textbf{7} & {\bf Total} \\\hline

				\hline
				\multicolumn{2}{|r||}{\parbox[c][2em][c]{3.5cm}{\centering\textbf{\# of  Involution}}}     &  12          &  61          &  548         &  4085        &  23805       &  127733       & 156244   \\
				\hline\hline
				\multirow{2}[12]{*}{\parbox[c][2em][c]{2.5cm}{\centering\textbf{\# of} $\mathbf{h^{1,1}_{-}}$}} & \parbox[c][2em][c]{0.5cm}{\centering\textbf{1}}  			    & 12           & 61           & 495          & 3683         & 21019        & 109175       & 134445\\
				\cline{2-9} & \parbox[c][2em][c]{0.5cm}{\centering\textbf{2}}     & --       & --          & 53           & 402          & 2588         & 17106        & 20149 \\
				\cline{2-9} & \parbox[c][2em][c]{0.5cm}{\centering\textbf{3}}     & --       & --       & --        & --        & 198          & 1452         & 1650\\
				\hline\hline
				\multicolumn{2}{|c||}{\parbox[c][2em][c]{3.5cm}{\centering$\mathbf{h^{1,1}(X)}$}}& \textbf{8} & \textbf{9} & \textbf{10} & \textbf{11} & \textbf{12} && {\bf Total} \\\hline

				\hline
				\multicolumn{2}{|r||}{\parbox[c][2em][c]{3.5cm}{\centering\textbf{\# of  Involution}}}   &  86          &  122         &  82          &  86          &  89         &  & 465  \\
				\hline\hline
				\multirow{6}[12]{*}{\parbox[c][3em][c]{2.5cm}{\centering\textbf{\# of} $\mathbf{h^{1,1}_{-}}$}} 
                            & \parbox[c][2em][c]{0.5cm}{\centering\textbf{1}}  & 48           & 76           & 53           & 49           & 32         &  & 258\\
				\cline{2-9} & \parbox[c][2em][c]{0.5cm}{\centering\textbf{2}}  & 22           & 26           & 12           & 18           & 12       &    & 90\\
				\cline{2-9} & \parbox[c][2em][c]{0.5cm}{\centering\textbf{3}}  & 14           & 7            & 10           & 6            & 37      &     & 74\\
				\cline{2-9} & \parbox[c][2em][c]{0.5cm}{\centering\textbf{4}} & 2            & 13           & 4            & 13           & 4        &    & 36\\
				\cline{2-9} & \parbox[c][2em][c]{0.5cm}{\centering\textbf{5}} & --            & --            & 3            & --            & 2         &   & 5\\
				\cline{2-9} & \parbox[c][2em][c]{0.5cm}{\centering\textbf{6}}  & --            & --            & --            & --            & 2       &     & 2\\
				\hline

			\end{tabular}
			\caption{  $h^{1,1}(X/\sigma^{*})$ splitting under the  proper divisor exchange involutions}
			\label{tab:split1}
		}
	\end{table}%

	
        \subsection{Reflections}
        \label{results reflection}
        
        There are two new observations regarding reflection involutions. Firstly, we must verify whether $P_{sym}$ vanishes trivially on the fixed locus $\cF_i$ and subsequently apply eq.(\ref{eq:type}) to determine the type of O-plane.  Secondly,  we identify a new type of free action, as described in Section \ref{subsub:free}, which occurs in  reflections with more than one divisor.
                
        For each of the triangulation of polytope with $h^{1,1}(X)\leq 6$, we computed  all of the single divisor reflections, double divisor reflections and triple divisor reflections.  For  cases with $h^{1,1}=7$, we computed all  single reflections and  randomly selected 15 double reflections and 15 triple reflections, as shown in Table.\ref{tab:reflection1}. 
For $8 \leq h^{1,1}(X)\leq 12$ cases, we compute all single reflections and  randomly selected 15 triple reflections, classifying the type of O-plane as presented in Table.\ref{tab:reflection2}.
When considering the D3-tadpole cancelation condition,  the naive Orientifold Type IIB string vacua with $O3/O7$ system are  compiled in Table.\ref{tab:reflectionvacua1} and Table.\ref{tab:reflectionvacua2}.
It is interesting to note that for single reflections, we obtain the same fixed locus  as \cite{Crino:2022zjk} when restricting  to real number solutions. 

There are numerous  free actions for multi-reflections. Many of them are of the new type, characterized by the absence of a fixed locus in the ambient space $\cA$ initially. We denote the previous type of free action as "type one," indicating those that do have a fixed locus on $\cA$ but do not intersect with the Calabi-Yau manifold $X$. 

\begin{figure}[ht!]
\centering
\includegraphics[width=7.5cm]{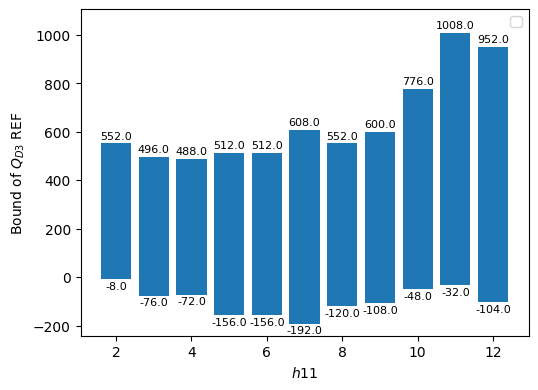}\quad
\includegraphics[width=7.5cm]{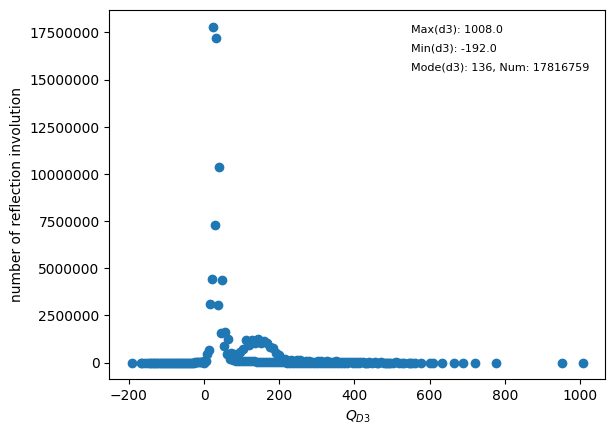}
\caption{Distribution of $Q_{D3}^{loc}$ under  reflections for naive orientifold Type IIB string vacua.}
\label{figure2}
\end{figure}

  For the $97,639,633$ naive orientifold Type IIB string vacua with an $O3/O7$-system, the distribution of $Q_{D3}^{loc}$  is shown in Fig.~\ref{figure2}.  Most  involutions result in  orientifold Calabi-Yau threefolds with $Q_{D3}^{loc} $ around $136$ in our scan. The smallest and largest $Q_{D3}^{loc}$ are $-192$ and $1008$ respectively.   This result aligns with our discussion in  \cite{Gao:2022fdi}, where we used the Lefschetz fixed point theorem to demonstrate  that the D3-tadpole can be bounded above by  $-Q_3 \leq 252$  for the Kreuzer-Skarke database.

        \begin{table}[h!]
		{\footnotesize
			\centering
			\begin{tabular}{|P{3.5cm}||P{1cm}|P{1cm}|P{1cm}|P{1.2cm}|P{1.3cm}|P{1.4cm}||P{1.5cm}|}
				\hline
				
				\multicolumn{8}{|c|}{\parbox[c][2em][c]{15cm}{\centering\textbf{Classification of O-plane fixed point locus}}} \\
				\hline
				
				\parbox[c][2em][c]{3.5cm}{\centering$\mathbf{h^{1,1}(X)}$}& \textbf{2} & \textbf{3} & \textbf{4} & \textbf{5} & \textbf{6} & \textbf{7} & {\bf Total} \\\hline
				
				\multicolumn{8}{|c|}{\parbox[c][2em][c]{15cm}{\centering\textbf{Single reflection}}} \\\hline
				\parbox[c][2em][c]{3.5cm}{\centering\textbf{\# of  Reflection}}   &  184         &  3019        &  39251       &  488352      &  5659818     &  64203360     & 70393984    \\\hline
				
				\parbox[c][2em][c]{3.5cm}{\centering\textbf{O3}}     &  2           &  55          &  776         &  10437       &  100804      &  760675       & 872749   \\\hline
				\parbox[c][2em][c]{3.5cm}{\centering\textbf{O7}}     &  64          &  871         &  6350        &  49243       &  356844      &  2658260      & 3071632   \\\hline
				
				\parbox[c][2em][c]{3.5cm}{\centering\textbf{O3 and O7}}   &  118         &  2093        &  32125       &  428672      &  5202170     &  60784425     & 66449603      \\\hline
				
                \parbox[c][2em][c]{3.5cm}{\centering\textbf{Free Action}}      &  0           &  0           &  0           &  0           &  0           &  0            & 0 \\\hline
				\multicolumn{8}{|c|}{\parbox[c][2em][c]{15cm}{\centering\textbf{Double divisor reflection}}} \\\hline
				
				\parbox[c][2em][c]{3.5cm}{\centering\textbf{\# of  Reflection}}    &  270         &  6066        &  94555       &  1347475     &  17653586    &  70850383     & 89952335     \\\hline
				
				
				\parbox[c][2em][c]{3.5cm}{\centering\textbf{O5}}   &  219         &  4503        &  69013       &  945594      &  11830587    &  44782225     & 57632141           \\\hline
				
				
                \parbox[c][2.5em][c]{3.5cm}{\centering\textbf{Free Action\\ (type one)}}  &  22          &  515         &  7481        &  108385      &  1428245     &  5527373  (403)     & 7072021  (403)    \\\hline
				\multicolumn{8}{|c|}{\parbox[c][2em][c]{15cm}{\centering\textbf{Triple divisors reflection}}} \\\hline
				
				\parbox[c][2em][c]{3.5cm}{\centering\textbf{\# of  Reflections}}  &  418         &  12328       &  250141      &  4332469     &  65640277    &  88047584     & 158283217    \\\hline
				
				\parbox[c][2em][c]{3.5cm}{\centering\textbf{O3}}      &  96          &  2413        &  36515       &  449416      &  4335075     &  3296422      & 8119937    \\\hline
				
				\parbox[c][2em][c]{3.5cm}{\centering\textbf{O7}}     &  45          &  1360        &  17180       &  209827      &  2172172     &  2113777      & 4514361    \\\hline
				
				\parbox[c][2em][c]{3.5cm}{\centering\textbf{O3 and O7}}     &  261         &  8090        &  188076      &  3523818     &  56670874    &  78995670     & 139386789     \\\hline
    
			
                \parbox[c][2.5em][c]{3.5cm}{\centering\textbf{Free Action \\ (type one)}}   &  16          &  465         &  8370        &  149408  (4)    &  2462156   (395)  &  3641715 (1008)      & 6262130 (1407)   \\\hline
				
			\end{tabular}
			\caption{Classification of  O-planes  and free actions under reflections for $h^{1,1}\leq 7$}
			\label{tab:reflection1}
		}
	\end{table}
	
	   \begin{table}[h!]
		{\footnotesize
			\centering
			\begin{tabular}{|P{3.5cm}||P{1.4cm}|P{1.4cm}|P{1.4cm}|P{1.4cm}|P{1.4cm}||P{2cm}|}
				\hline
				\multicolumn{7}{|c|}{\parbox[c][2em][c]{15cm}{\centering\textbf{Classification of O-plane fixed point locus}}} \\
				\hline
				
				\parbox[c][2em][c]{3.5cm}{\centering$\mathbf{h^{1,1}(X)}$}& \textbf{8} & \textbf{9} & \textbf{10} & \textbf{11} & \textbf{12} & {\bf Total} \\\hline
				\multicolumn{7}{|c|}{\parbox[c][2em][c]{15cm}{\centering\textbf{Single reflection}}} \\\hline
				
				\parbox[c][2em][c]{3.5cm}{\centering\textbf{\# of  Reflection}}  &  78930       &  123420      &  153858      &  210130      &  230431       & 796769    \\\hline
				
				\parbox[c][2em][c]{3.5cm}{\centering\textbf{O3}}   &  107         &  58          &  21          &  76          &  37           & 299     \\\hline
				
				\parbox[c][2em][c]{3.5cm}{\centering\textbf{O7}}   &  7663        &  9732        &  9112        &  8650        &  9074         & 44231    \\\hline
				
				\parbox[c][2em][c]{3.5cm}{\centering\textbf{O3 and O7}}  &  71160       &  113630      &  144725      &  201404      &  221320       & 752239     \\\hline
				
				\multicolumn{7}{|c|}{\parbox[c][2em][c]{15cm}{\centering\textbf{Triple divisor reflections}}} \\\hline
				
				\parbox[c][2em][c]{3.5cm}{\centering\textbf{\# of  Reflection}}   &  98067       &  141432      &  155496      &  196174      &  211884       & 803053   \\\hline
				
				\parbox[c][2em][c]{3.5cm}{\centering\textbf{O3}}  &  714         &  657         &  249         &  338         &  403          & 2361  \\\hline
				
				\parbox[c][2em][c]{3.5cm}{\centering\textbf{O7}}    &  4792        &  5905        &  4903        &  4179        &  4525         & 24304    \\\hline
				
				\parbox[c][2em][c]{3.5cm}{\centering\textbf{O3 and O7}}    &  87526       &  127736      &  141764      &  180780      &  196656       & 734462     \\\hline
    
                \parbox[c][2.5em][c]{3.5cm}{\centering\textbf{Free Action \\ (type one) }}    &  5035 (480)        &  7134   (1117)     &  8580    (1499)    &  10877    (2820)    &  10300     (2811)    & 41926\,\,\,  (8727) \\\hline
    			\end{tabular}
			 \caption{Classification of  O-planes  and free actions under reflections for  $8 \leq h^{1,1}\leq 12$}
			\label{tab:reflection2}
		}
	\end{table}

        \begin{table}[h!]
		{\footnotesize
			\centering
			\begin{tabular}{|P{3.5cm}||P{1cm}|P{1cm}|P{1cm}|P{1.2cm}|P{1.3cm}|P{1.4cm}||P{1.5cm}|}
				\hline
				
				\multicolumn{8}{|c|}{\parbox[c][2em][c]{15cm}{\centering\textbf{Naive Orientifold Type IIB String Vacua with $O3/O7$-system}}} \\
				\hline
				\parbox[c][2em][c]{3.5cm}{\centering$\mathbf{h^{1,1}(X)}$}& \textbf{2} & \textbf{3} & \textbf{4} & \textbf{5} & \textbf{6} & \textbf{7} & {\bf Total} \\\hline    
				\multicolumn{8}{|c|}{\parbox[c][2em][c]{15cm}{\centering\textbf{Single reflection}}} \\\hline
				 \parbox[c][2em][c]{3.5cm}{\centering\textbf{\# of  Reflection}}  &  137         &  2052        &  24605       &  275905      &  2905604     &  30797733     & 34006036  \\\hline
				\parbox[c][2em][c]{3.5cm}{\centering\textbf{O3}}     &  0           &  3           &  49          &  771         &  7403        &  54261        & 62487   \\\hline
				
				\parbox[c][2em][c]{3.5cm}{\centering\textbf{O7}}  &  52          &  688         &  4893        &  35970       &  257887      &  1887150      & 2186640   \\\hline
				
				\parbox[c][2em][c]{3.5cm}{\centering\textbf{O3 and O7}}   &  85          &  1361        &  19663       &  239164      &  2640314     &  28856322     & 31756909  \\\hline
				\multicolumn{8}{|c|}{\parbox[c][2em][c]{15cm}{\centering\textbf{Triple divisor reflection}}} \\\hline
				
				\parbox[c][2em][c]{3.5cm}{\centering\textbf{\# of  Reflection}}   &  271         &  6815        &  125209      &  1942434     &  26823480    &  34289949     & 63188158    \\\hline
				
				\parbox[c][2em][c]{3.5cm}{\centering\textbf{O3}} &  35          &  762         &  9142        &  93922       &  764601      &  611384       & 1479846    \\\hline
				
				\parbox[c][2em][c]{3.5cm}{\centering\textbf{O7}}   &  42          &  1088        &  12592       &  143910      &  1406394     &  1288578      & 2852604      \\\hline
				
				\parbox[c][2em][c]{3.5cm}{\centering\textbf{O3 and O7}}   &  178         &  4500        &  95105       &  1555194     &  22190329    &  28748272     & 52593578     \\\hline
				
                \parbox[c][2.5em][c]{3.5cm}{\centering\textbf{Free Action \\ (type one)}}   &  16          &  465         &  8370        &  149408  (4)    &  2462156   (395)  &  3641715 (1008)      & 6262130 (1407)   \\\hline

			\end{tabular}
			\caption{Classification of  naive orientifold Type IIB string vacua under reflections for $ h^{1,1}\leq 7$}
			\label{tab:reflectionvacua1}
		}
	\end{table}

    \begin{table}[h!]
		{\footnotesize
			\centering
			\begin{tabular}{|P{3.5cm}||P{1.4cm}|P{1.4cm}|P{1.4cm}|P{1.4cm}|P{1.4cm}||P{2cm}|}
				\hline
				\multicolumn{7}{|c|}{\parbox[c][2em][c]{15cm}{\centering\textbf{Naive Orientifold Type IIB String Vacua with $O3/O7$-system}}} \\
				\hline
				
				\parbox[c][2em][c]{3.5cm}{\centering$\mathbf{h^{1,1}(X)}$}& \textbf{8} & \textbf{9} & \textbf{10} & \textbf{11} & \textbf{12} & {\bf Total} \\\hline
                \multicolumn{7}{|c|}{\parbox[c][2em][c]{15cm}{\centering\textbf{Single reflection}}} \\\hline
				
				\parbox[c][2em][c]{3.5cm}{\centering\textbf{\# of  Reflection}}   &  25611       &  36638       &  43540       &  56870       &  67709        & 230368  \\\hline
				
				\parbox[c][2em][c]{3.5cm}{\centering\textbf{O3}}    &  0           &  0           &  6           &  0           &  8            & 14    \\\hline
				
				\parbox[c][2em][c]{3.5cm}{\centering\textbf{O7}}    &  3930        &  4669        &  5167        &  4665        &  5463         & 23894   \\\hline
				
				\parbox[c][2em][c]{3.5cm}{\centering\textbf{O3 and O7}}    &  21681       &  31969       &  38367       &  52205       &  62238        & 206460 \\\hline
				
				\multicolumn{7}{|c|}{\parbox[c][2em][c]{15cm}{\centering\textbf{Triple divisor reflection}}} \\\hline
				
				\parbox[c][2em][c]{3.5cm}{\centering\textbf{\# of  Reflection}}   &  29395       &  38854       &  40651       &  49782       &  56389        & 215071    \\\hline
				
				\parbox[c][2em][c]{3.5cm}{\centering\textbf{O3}}  &  455         &  449         &  100         &  71          &  304          & 1379    \\\hline
				
				\parbox[c][2em][c]{3.5cm}{\centering\textbf{O7}}   &  2153        &  2478        &  2439        &  1836        &  2058         & 10964  \\\hline
				
				\parbox[c][2em][c]{3.5cm}{\centering\textbf{O3 and O7}}  &  21752       &  28793       &  29532       &  36998       &  43727        & 160802   \\\hline
    
                  \parbox[c][2.5em][c]{3.5cm}{\centering\textbf{Free Action \\ (type one) }}    &  5035 (480)        &  7134   (1117)     &  8580    (1499)    &  10877    (2820)    &  10300     (2811)    & 41926\,\,\,  (8727) \\\hline
					\end{tabular}
			\caption{Classification of  naive orientifold Type IIB string vacua under the  reflections for $8 \leq h^{1,1}\leq 12$}
			\label{tab:reflectionvacua2}
		}
	\end{table}

        \section{Conclusions and Outlook}
        \label{Conclusions and Outlook}
	\label{sec:conc}
	
	In this paper, we significantly broaden our scope on orientifold Calabi-Yau threefolds in three key areas. Firstly, we expand the scale of our constructions by three orders of magnitude, increasing from  $\cO(10^6)$ to $\cO(10^9)$ orientifold Calabi-Yau threefolds compared to previous work  \cite{Altman:2021pyc}.
	We extend our  construction of orientifold Calabi-Yau threefolds up to $h^{1,1}(X)=12$ in the Calabi-Yau database constructed from the Kreuzer-Skarke list~\cite{Kreuzer:2000xy}.  For  $h^{1,1}(X) \leq 7$,  we expand our analysis to hypersurfaces in all possible maximal projective crepant partial (MPCP) desingularizations. For $8 \leq h^{1,1}(X) \leq 12$, we randomly choose some favorable polytopes for each  Hodge number as in ~\cite{Crino:2022zjk}.  The number of toric triangulations we  analyze increases by one order of magnitude,  from $653,062$ \cite{Altman:2021pyc} to $6,696,714$.
 We expand our classification to cover both divisor exchange involutions and multi-divisor reflection involutions, which include $156,709$ proper NIDs exchange involutions and $320,229,358$ different types of reflections.
Since each of  these involutions  results in an orientifold Calabi-Yau manifolds,  combined those two types of involutions,  we constructed a total of $320,386,067$  orientifold Calabi-Yau threefolds in our new database (\dburl), which is three orders of   magnitude ($\cO(10^3)$) larger than the database in  \cite{Altman:2021pyc},  where only exchange involution for small $h^{1,1}(X)$ was considered. We will continue to update our website regularly with  new results not included in our current database.

Secondly, we have significantly improved the efficiency of our new algorithm for calculating putative fixed loci under divisor exchanges and reflection involutions.  This enhancement enables us to identify the positions of various types of O-planes, which are crucial for D-brane constructions. The upgraded algorithm  reduces the calculation complexity required to determine fixed loci by more than five orders of magnitude $\cO(10^5)$ in certain simple examples, even for  small values such as$h^{1,1}(X)= 7$. 
We categorize freely acting involutions on Calabi-Yau threefolds into two types: first, those without a fixed locus in the ambient space under the involutions, and second, those with a fixed locus in the ambient space that does not intersect with the Calabi-Yau threefold. In orientifold Calabi-Yau threefolds featuring the $O3/O7$-system, we proceed to classify the so-called  \lq\lq naive orientifold Type IIB string vacua" by considering the D3-tadpole cancellation condition when placing eight $D7$-branes on top of $O7$-plane. 
 
 Thirdly, we clarify some ambiguities in the construction of orientifold Calabi-Yau threefolds.   One such ambiguity is 
 in determining the type of O-plane using  eq.(\ref{eq:type}).	We should be more careful to accurately determine their contribution to the D3-tadpole condition, as described in Section \ref{subsec:fixedX}. 
 Additionally,  when solving the $\lambda$ system eq.(\ref{eq:system}-\ref{eq:system2}) to determine the putative fixed loci, we must ensure that we do not overlook any  solutions in the complex $\lambda$ system for $\lambda = e^{i \pi u_i} \in \IC^*$.
 
The systematic calculations presented  in this paper  pave the way for  extending these analyses beyond our current computational limit into the region $h^{1,1}(X) \geq 12$ of the Kreuzer-Skarke database. 
Given  the extensive size of the orientifold Calabi-Yau database, it is  natural to anticipate that in addition to the formal progress, supervised machine learning techniques will be necessary to understand the landscape.   In \cite{Gao:2021xbs},  we  utilized machine learning technique to classify the polytopes which can result in the orientifold Calabi-Yau with divisor exchange involutions.  That study demonstrated that  high accuracy could be achieved by training on just 10\% of the data for higher $h^{1,1}(X)$. Since we have only calculated part of the polytopes for $8 \leq h^{1,1}(X)\leq 12$,  it would be great for us to predict the desired polytopes using  machine learning techniques on the limited data  available and validate these predictions.

In addition to the Kreuzer-Skarke and CICY databases, it has been discovered that relaxing the condition for non-negative entries in the configuration or weighted matrix leads to a new class of Calabi-Yau manifolds, known as \lq\lq generalized Complete Intersection Calabi-Yau" (gCICYs) \cite{Anderson:2015iia, Anderson:2015yzz}, along with their toric variations \cite{Berglund:2016yqo, Berglund:2016nvh}. We have also employed  machine learning techniques to generate more gCICY \cite{Cui:2022cxe}.   investigating involutive or more general quotient symmetries  of these new manifolds could yield further intriguing insights.

In the context of swampland conjecture,  fundamental doubts about the the existence of de-Sitter space have triggered intense discussion. The KKLT scenario \cite{Kachru:2003aw} has withstood attacks related to the stability  of the anti-D3 uplift. Especially, in \cite{Gao:2020xqh} we asserted that  a singular-bulk problem arises because one is forced to glue a large throat into a fairly small Calabi-Yau \cite{Carta:2019rhx} to get a de-Sitter space. On the other hand, people have explored a landscape of de-Sitter vacua in the large complex structure region of mirror Calabi-Yau manifolds with all  order $\alpha'$ corrections \cite{McAllister:2024lnt}.  Various other corrections, such as warping correction, loop and $\alpha'$ correction in the ordinary Type IIB orientifold compactifications was also studied.  For example, in the Large Volume Scenario (LVS), considering the warping correction imposes  a parameter constraint that provides a lower bound of the number of tadpole that the compactified geoemtry   must satisfy  \cite{Junghans:2022exo, Gao:2022fdi,  Gao:2022uop, Junghans:2022kxg}. Combined with the tadpole conjecture which gives an upper bound on the number of tadpole \cite{Bena:2020xrh}, this creates a window specifying the types of oreintifold Calabi-Yau compactifications that are feasible. It would be very interesting to investigate the constraints on orientifold Calabi-Yau threefolds imposed by these and other corrections.

\clearpage

	\section*{Acknowledgments}
        \label{Acknowledgments}
        
        We would like to thank  Wei Cui, Andreas Schachner, Pramod Shukla, Juntao Wang,  Yinan Wang,  Hao Zou for helpful discussions and correspondence. This work was supported in part by the NSFC under grant numbers 12375065.

        \appendix
      

        \section{Fixed locus for reflections of Example A}
        \label{sec:ref}
        Table.\ref{tab:reflectionex} shows the fixed locus we get from the single, double and triple reflections for Example A  in Section~\ref{subsubsec:reflection}. The type of O-plane is determined by eq.(\ref{eq:type}) depending on whether $P_{sym}$ vanish trivially on the fixed sets. These results also collected in Appendix.\ref{subsubsec:formatref} in the format on the website (\dburl). Notice  there are three free actions in these reflections we considered.

          \begin{table}[h!]
		{\footnotesize
			\centering
			\begin{tabular}{|P{2cm}|P{14cm}|}
				\hline
				
				\parbox[c][1.5em][c]{2cm}{\centering \textbf{ Reflection}}&  \textbf{Fixed Locus of O-plane on $X$ }\\\hline
				
				\parbox[c][1.5em][c]{1.5cm}{\centering $x_1$  } & $ [x_{6}, x_{7}, x_{9} ]  [x_{1} ]  [x_{2} ]  [x_{3} ]  [x_{4} ]   $    \\\hline
		     \parbox[c][1.5em][c]{1.5cm}{\centering $x_2$} & $ [x_{6}, x_{7}, x_{9} ]  [x_{2} ]  [x_{1} ]  [x_{3} ]  [x_{4}]$ \\\hline
		         \parbox[c][1.5em][c]{1.5cm}{\centering $x_3$} &   $[x_{6}, x_{7}, x_{9} ]  [x_{3} ]  [x_{1} ]  [x_{2} ]  [x_{4} ]$\\\hline
		 \parbox[c][1.5em][c]{1.5cm}{\centering $x_4$} &  $ [x_{6}, x_{7}, x_{9} ]  [x_{4} ]  [x_{1} ]  [x_{2} ]  [x_{3} ]$\\\hline
		  \parbox[c][1.5em][c]{1.5cm}{\centering $x_5$}  &   $[x_{1}, x_{6}, x_{7}, x_{8} ]  [x_{2}, x_{6}, x_{9}, x_{10} ]  [x_{3}, x_{7}, x_{9}, x_{11} ]  [x_{4}, x_{6}, x_{7}, x_{8} ]  [x_{4}, x_{6}, x_{9}, x_{10} ]  [x_{4}, x_{7}, x_{9}, x_{11} ]  [x_{5} ] $ \\\hline
		      \parbox[c][1.5em][c]{1.5cm}{\centering $x_6$}  & $  [x_{3}, x_{7}, x_{9} ]  [x_{1}, x_{5}, x_{7}, x_{8} ]  [x_{2}, x_{5}, x_{9}, x_{10} ]  [x_{4}, x_{5}, x_{7}, x_{8} ]  [x_{4}, x_{5}, x_{9}, x_{10} ]  [x_{6} ]  [x_{5}, x_{11} ]$\\\hline
                 \parbox[c][1.5em][c]{1.5cm}{\centering $x_7$}  & $  [x_{2}, x_{6}, x_{9} ]  [x_{1}, x_{5}, x_{6}, x_{8} ]  [x_{3}, x_{5}, x_{9}, x_{11} ]  [x_{4}, x_{5}, x_{6}, x_{8} ]  [x_{4}, x_{5}, x_{9}, x_{11} ]  [x_{7} ]  [x_{5}, x_{10} ]$\\\hline
               \parbox[c][1.5em][c]{1.5cm}{\centering $x_8$} &  $ [x_{2}, x_{6}, x_{10} ]  [x_{3}, x_{7}, x_{11} ]  [x_{4}, x_{6}, x_{10} ]  [x_{4}, x_{7}, x_{11} ]  [x_{1}, x_{5}, x_{6}, x_{7} ]  [x_{8} ]  [x_{5}, x_{9} ]$\\\hline
                 \parbox[c][1.5em][c]{1.5cm}{\centering $x_9$} &   $[x_{1}, x_{6}, x_{7} ]  [x_{2}, x_{5}, x_{6}, x_{10} ]  [x_{3}, x_{5}, x_{7}, x_{11} ]  [x_{4}, x_{5}, x_{6}, x_{10} ]  [x_{4}, x_{5}, x_{7}, x_{11} ]  [x_{9} ]  [x_{5}, x_{8} ]$\\\hline
                \parbox[c][1.5em][c]{1.5cm}{\centering $x_{10}$}  &  $ [x_{1}, x_{6}, x_{8} ]  [x_{3}, x_{9}, x_{11} ]  [x_{4}, x_{6}, x_{8} ]  [x_{4}, x_{9}, x_{11} ]  [x_{2}, x_{5}, x_{6}, x_{9} ]  [x_{10} ]  [x_{5}, x_{7} ]$\\\hline
                  \parbox[c][1.5em][c]{1.5cm}{\centering $x_{11}$}  &  $ [x_{1}, x_{7}, x_{8} ]  [x_{2}, x_{9}, x_{10} ]  [x_{4}, x_{7}, x_{8} ]  [x_{4}, x_{9}, x_{10} ]  [x_{3}, x_{5}, x_{7}, x_{9} ]  [x_{11} ]  [x_{5}, x_{6} ]$\\\hline\hline
	    \parbox[c][1.5em][c]{1.5cm}{\centering	    $ x_{7} x_{9} $} &  $ [x_{7}, x_{9} ]  [x_{1}, x_{6} ]  [x_{2}, x_{6} ]  [x_{4}, x_{6} ]  [x_{3}, x_{5}, x_{11} ]  [x_{4}, x_{5}, x_{11} ]  [x_{5}, x_{7}, x_{8} ]  [x_{5}, x_{9}, x_{10} ]$\\\hline
              \parbox[c][1.5em][c]{1.5cm}{\centering    $  x_{1} x_{2} $} & Free Action \\\hline
          \parbox[c][1.5em][c]{1.5cm}{\centering         $ x_{2} x_{6}$}  &  $ [x_{2}, x_{6} ]  [x_{1}, x_{6} ]  [x_{4}, x_{6} ]  [x_{7}, x_{9} ]  [x_{3}, x_{5}, x_{11} ]  [x_{4}, x_{5}, x_{11} ]  [x_{5}, x_{7}, x_{8} ]  [x_{5}, x_{9}, x_{10} ]$\\\hline
             \parbox[c][1.5em][c]{1.5cm}{\centering   $    x_{3} x_{10} $} & $  [x_{2}, x_{10} ]  [x_{4}, x_{10} ]  [x_{6}, x_{8} ]  [x_{9}, x_{11} ]  [x_{1}, x_{5}, x_{7} ]  [x_{3}, x_{5}, x_{7} ]  [x_{4}, x_{5}, x_{7} ]  [x_{5}, x_{6}, x_{9} ]$\\\hline
              \parbox[c][1.5em][c]{1.5cm}{\centering   $   x_{3} x_{6}$}  & $  [x_{1}, x_{6} ]  [x_{2}, x_{6} ]  [x_{4}, x_{6} ]  [x_{7}, x_{9} ]  [x_{3}, x_{5}, x_{11} ]  [x_{4}, x_{5}, x_{11} ]  [x_{5}, x_{7}, x_{8} ]  [x_{5}, x_{9}, x_{10} ]$\\\hline
              \parbox[c][1.5em][c]{1.5cm}{\centering   $   x_{1} x_{3} $} & Free Action \\\hline
               \parbox[c][1.5em][c]{1.5cm}{\centering  $   x_{2} x_{5} $} &  $ [x_{2}, x_{5} ]  [x_{1}, x_{5} ]  [x_{3}, x_{5} ]  [x_{4}, x_{5} ]  [x_{6}, x_{7}, x_{8} ]  [x_{6}, x_{9}, x_{10} ]  [x_{7}, x_{9}, x_{11} ]$\\\hline
              \parbox[c][1.5em][c]{1.5cm}{\centering   $   x_{1} x_{7} $} &  $ [x_{1}, x_{7} ]  [x_{3}, x_{7} ]  [x_{4}, x_{7} ]  [x_{6}, x_{9} ]  [x_{2}, x_{5}, x_{10} ]  [x_{4}, x_{5}, x_{10} ]  [x_{5}, x_{6}, x_{8} ]  [x_{5}, x_{9}, x_{11} ]$\\\hline
              \parbox[c][1.5em][c]{1.5cm}{\centering   $   x_{6} x_{7}  $}&  $ [x_{6}, x_{7} ]  [x_{2}, x_{9} ]  [x_{3}, x_{9} ]  [x_{4}, x_{9} ]  [x_{5}, x_{10} ]  [x_{1}, x_{5}, x_{8} ]  [x_{4}, x_{5}, x_{8} ]  [x_{5}, x_{7}, x_{11} ]$\\\hline
              \parbox[c][1.5em][c]{1.5cm}{\centering   $   x_{5} x_{10} $} &  $ [x_{5}, x_{10} ]  [x_{2}, x_{6}, x_{9} ]  [x_{4}, x_{6}, x_{9} ]  [x_{7} ]$\\\hline
             \parbox[c][1.5em][c]{1.5cm}{\centering    $   x_{1} x_{5} $} &  $ [x_{1}, x_{5} ]  [x_{2}, x_{5} ]  [x_{3}, x_{5} ]  [x_{4}, x_{5} ]  [x_{6}, x_{7}, x_{8} ]  [x_{6}, x_{9}, x_{10} ]  [x_{7}, x_{9}, x_{11} ]$\\\hline
             \parbox[c][1.5em][c]{1.5cm}{\centering    $   x_{5} x_{6}$}  &  $ [x_{5}, x_{6} ]  [x_{1}, x_{7}, x_{8} ]  [x_{2}, x_{9}, x_{10} ]  [x_{4}, x_{7}, x_{8} ]  [x_{4}, x_{9}, x_{10} ]  [x_{11} ]$\\\hline
                         \parbox[c][1.5em][c]{1.5cm}{\centering     $   x_{3} x_{5} $} &  $ [x_{3}, x_{5} ]  [x_{1}, x_{5} ]  [x_{2}, x_{5} ]  [x_{4}, x_{5} ]  [x_{6}, x_{7}, x_{8} ]  [x_{6}, x_{9}, x_{10} ]  [x_{7}, x_{9}, x_{11} ]$\\\hline
            \parbox[c][1.5em][c]{1.5cm}{\centering     $   x_{1} x_{8}$}  & $  [x_{1}, x_{8} ]  [x_{4}, x_{8} ]  [x_{6}, x_{10} ]  [x_{7}, x_{11} ]  [x_{2}, x_{5}, x_{9} ]  [x_{3}, x_{5}, x_{9} ]  [x_{4}, x_{5}, x_{9} ]  [x_{5}, x_{6}, x_{7} ]$\\\hline
             \parbox[c][1.5em][c]{1.5cm}{\centering     $  x_{2} x_{7} $} &  $ [x_{1}, x_{7} ]  [x_{3}, x_{7} ]  [x_{4}, x_{7} ]  [x_{6}, x_{9} ]  [x_{2}, x_{5}, x_{10} ]  [x_{4}, x_{5}, x_{10} ]  [x_{5}, x_{6}, x_{8} ]  [x_{5}, x_{9}, x_{11} ]$\\\hline\hline
	   \parbox[c][1.5em][c]{1.5cm}{\centering   $	      x_{4} x_{5} x_{6} $} & $  [x_{1}, x_{5}, x_{6} ]  [x_{2}, x_{5}, x_{6} ]  [x_{3}, x_{11} ]  [x_{4}, x_{11} ]  [x_{7}, x_{8} ]  [x_{9}, x_{10} ]$\\\hline
              \parbox[c][1.5em][c]{1.5cm}{\centering   $     x_{3} x_{8} x_{9}$}  &  $ [x_{5}, x_{6}, x_{7}, x_{9} ]  [x_{1}, x_{5} ]  [x_{2}, x_{5} ]  [x_{3}, x_{5} ]  [x_{4}, x_{5} ]$\\\hline
              \parbox[c][1.5em][c]{1.5cm}{\centering   $     x_{4} x_{7} x_{11} $} & $  [x_{4}, x_{7}, x_{11} ]  [x_{2}, x_{6}, x_{10} ]  [x_{3}, x_{7}, x_{11} ]  [x_{4}, x_{6}, x_{10} ]  [x_{1}, x_{5}, x_{6}, x_{7} ]  [x_{8} ]  [x_{5}, x_{9} ]$\\\hline
             \parbox[c][1.5em][c]{1.5cm}{\centering   $      x_{1} x_{6} x_{7} $} & $  [x_{1}, x_{6}, x_{7} ]  [x_{2}, x_{5}, x_{6}, x_{10} ]  [x_{3}, x_{5}, x_{7}, x_{11} ]  [x_{4}, x_{5}, x_{6}, x_{10} ]  [x_{4}, x_{5}, x_{7}, x_{11} ]  [x_{9} ]  [x_{5}, x_{8} ]$\\\hline
            \parbox[c][1.5em][c]{1.5cm}{\centering   $       x_{5} x_{7} x_{11} $} & $  [x_{1}, x_{5}, x_{8} ]  [x_{4}, x_{5}, x_{8} ]  [x_{2}, x_{9} ]  [x_{3}, x_{9} ]  [x_{4}, x_{9} ]  [x_{6}, x_{7} ]$\\\hline
           \parbox[c][1.5em][c]{1.5cm}{\centering   $        x_{2} x_{7} x_{11} $} & $  [x_{2}, x_{6}, x_{10} ]  [x_{3}, x_{7}, x_{11} ]  [x_{4}, x_{6}, x_{10} ]  [x_{4}, x_{7}, x_{11} ]  [x_{1}, x_{5}, x_{6}, x_{7} ]  [x_{8} ]  [x_{5}, x_{9} ]$\\\hline
             \parbox[c][1.5em][c]{1.5cm}{\centering   $      x_{5} x_{6} x_{10} $} &  $ [x_{1}, x_{5}, x_{8} ]  [x_{4}, x_{5}, x_{8} ]  [x_{2}, x_{9} ]  [x_{3}, x_{9} ]  [x_{4}, x_{9} ]  [x_{6}, x_{7} ]$\\\hline
            \parbox[c][1.5em][c]{1.5cm}{\centering   $       x_{2} x_{5} x_{8} $} &  $ [x_{1}, x_{5}, x_{8} ]  [x_{4}, x_{5}, x_{8} ]  [x_{2}, x_{9} ]  [x_{3}, x_{9} ]  [x_{4}, x_{9} ]  [x_{6}, x_{7} ]$\\\hline
                  \parbox[c][1.5em][c]{1.5cm}{\centering   $         x_{1} x_{9} x_{11}  $} & $  [x_{1}, x_{6}, x_{8} ]  [x_{3}, x_{9}, x_{11} ]  [x_{4}, x_{6}, x_{8} ]  [x_{4}, x_{9}, x_{11} ]  [x_{2}, x_{5}, x_{6}, x_{9} ]  [x_{10} ]  [x_{5}, x_{7} ]$\\\hline
          \parbox[c][1.5em][c]{1.5cm}{\centering   $         x_{5} x_{8} x_{9} $} &  Free Action \\\hline
           \parbox[c][1.5em][c]{1.5cm}{\centering   $        x_{1} x_{5} x_{9} $} &  $ [x_{2}, x_{5}, x_{9} ]  [x_{3}, x_{5}, x_{9} ]  [x_{1}, x_{8} ]  [x_{4}, x_{8} ]  [x_{6}, x_{10} ]  [x_{7}, x_{11} ]$\\\hline
          \parbox[c][1.5em][c]{1.5cm}{\centering   $         x_{3} x_{6} x_{8} $} & $  [x_{1}, x_{6}, x_{8} ]  [x_{3}, x_{9}, x_{11} ]  [x_{4}, x_{6}, x_{8} ]  [x_{4}, x_{9}, x_{11} ]  [x_{2}, x_{5}, x_{6}, x_{9} ]  [x_{10} ]  [x_{5}, x_{7} ]$\\\hline
          \parbox[c][1.5em][c]{1.5cm}{\centering   $         x_{1} x_{5} x_{7} $} &  $ [x_{1}, x_{5}, x_{7} ]  [x_{3}, x_{5}, x_{7} ]  [x_{2}, x_{10} ]  [x_{4}, x_{10} ]  [x_{6}, x_{8} ]  [x_{9}, x_{11} ]$\\\hline
            \parbox[c][1.5em][c]{1.5cm}{\centering   $       x_{2} x_{4} x_{6}$ } & $  [x_{3}, x_{7}, x_{9} ]  [x_{1}, x_{5}, x_{7}, x_{8} ]  [x_{2}, x_{5}, x_{9}, x_{10} ]  [x_{4}, x_{5}, x_{7}, x_{8} ]  [x_{4}, x_{5}, x_{9}, x_{10} ]  [x_{6} ]  [x_{5}, x_{11} ]$\\\hline
            \parbox[c][1.5em][c]{1.5cm}{\centering   $       x_{4} x_{6} x_{7} $} & $  [x_{1}, x_{6}, x_{7} ]  [x_{2}, x_{5}, x_{6}, x_{10} ]  [x_{3}, x_{5}, x_{7}, x_{11} ]  [x_{4}, x_{5}, x_{6}, x_{10} ]  [x_{4}, x_{5}, x_{7}, x_{11} ]  [x_{9} ]  [x_{5}, x_{8} ]$\\\hline

									\end{tabular}
			\caption{Fixed locus under the reflections for the Example A in Section~\ref{subsubsec:reflection}.}
			\label{tab:reflectionex}
		}
	\end{table}

\clearpage
        
          \section{Pseudocode description of new algorithm}
	Here we present the pseudo-code for the new algorithm to find the putative fixed locus (Algorithm.\ref{algo 1}) and the optimized logical to reduce the possible fixed locus  needed to test (Algorithm.\ref{algo 2}).
 
        {\footnotesize
        \begin{algorithm}
        \label{algo 1}
        $m = W.rows$\\
        $n = \mathcal{F}.length$\\
        let $M$ be a matrix of size $m \times n$\\
        \For{$i \leftarrow 0$ \KwTo $n$}{
        	set column $i$ of $M$ to be the toric weight vector of $\mathcal{F}[i]$\\
        }
        	
        let $b$ be an array of zeros of size $n$\\
        \For{$i \leftarrow 0$ \KwTo $n$}{
        	\If{$\mathcal{F}[i] \in \mathcal{G}_{-}$}{
        		$b[i] = 1$\\
        	}
        }
        \tcc{Improvement}
        let $u_{iT} = 2^{m}$ \\
        let $u_i = 0$\\
        \While{$u_i < u_{iT}$}{
            let $odd = false$\\
            let $u\_real_i$ be the $n$ bit binary type of $u_i$\\
            \tcc{for $u_i=3$ and $n=5$, $u\_real_i=00011$, the i-th bit tells the value of $\mathcal{F}_i$}
            \For{$x \leftarrow 0$ \KwTo $n$}{
                $col=M[x]$\\
                let $value_x = \sum_{i \in \text{range}(m)} (u\_real_i[i] \cdot \text{col}[i]) + b[x]$\\
                \If{$value_x\%2 = 1$}{
                    $odd=True$\\
                    break
                    }
                }
            \If{not $odd$}{
                \KwRet true
                }
            $u_i+=1$
        }
        \KwRet false        
        \caption{$\textrm{WEIGHT\_FIXED\_REALSPACE}(\mathcal{F}, W, \mathcal{G}_{0}, \mathcal{G}_{+}, \mathcal{G}_{-})$}
        \end{algorithm}
        }
        

        {\footnotesize
        \begin{algorithm}
        \label{algo 2}
        $\mathcal{G} = \mathcal{G}_{0} \cup \mathcal{G}_{+} \cup \mathcal{G}_{-}$\\
        let $\mathcal{L}_f$ be list of fixed points computed before in real space\\
        let $\mathcal{L}_{fs}$ be a sum of SR ideal and $\mathcal{L}_f$\\
        let $\mathcal{L}_{uf}$ be an empty list\\
        let $\mathcal{S}$ be the power set of $\mathcal{G}$\\
        \tcc{$\mathcal{S}$ are sorted so the number of divisor of element start large which means small in geometry}
        \For{$\mathcal{F} \in \mathcal{S}$}{        	
        	skip = false\\
        	\For{$\mathcal{T} \in \mathcal{L}_{uf}$}{
        		\If{$\mathcal{F} \subset \mathcal{T}$}{
                        \tcc{Non-fixed locus $\mathcal{T}$ is a part of $\mathcal{F}$ in geometry, so $\mathcal{F}$ can not fixed }
        			skip = true\\
        			\Break
        		}
        	}
        	\If{skip}{
        		\Continue
        	}
                \For{$\mathcal{T} \in \mathcal{L}_{fs}$}{
        		\If{$\mathcal{T} \subset \mathcal{F}$}{
                        \tcc{$\mathcal{F}$ is a part of $\mathcal{T}$ in geometry, so $\mathcal{F}$ is ruled out by existing fixed point or singularity introduced by SR ideal}
        			skip = true\\
        			\Break
        		}
        	}
                \If{skip}{
        		\Continue
        	}
        	\If{not $\textrm{CONSISTENT}(\mathcal{F}, \mathcal{G}_{0}, \mathcal{G}_{+}, \mathcal{G}_{-})$}{
                    continue\\
                }
        	\If{not $\textrm{WEIGHT\_FIXED}(\mathcal{F}, W, \mathcal{G}_{0}, \mathcal{G}_{+}, \mathcal{G}_{-})$}{
                    \tcc{This WEIGHT\_FIXED algorithm is just the old one in complex space}
        		append $\mathcal{F}$ to $\mathcal{L}_{uf}$\\
                    \Continue
        	}
                append $\mathcal{F}$ to $\mathcal{L}_{f}$\\
                append $\mathcal{F}$ to $\mathcal{L}_{fs}$\\
        }
        \KwRet $\mathcal{L}_f$
        \caption{$\textrm{FIXED\_LOCI\_WEIGHTS\_NEW}(W, SR\ ideal,\mathcal{G}_{0}, \mathcal{G}_{+}, \mathcal{G}_{-})$}
        \end{algorithm}
        }

        	%
        	
        \section{Database Format}

        \subsection{Terminology}
        \label{subsec:dataexample}
        
        The entry for each involution will contain the following terminologes\footnote{  In this appendix we will describe the format of the data in the website (\dburl). For the coordinate system $\{x_i\}$ in the website, the index of   ``x" goes from 0, i.e., ``x0".  The coordinate system $\{z_i\}$ are labeled from ``z1". } :

\subsubsection{Divisor exchange involutions}
        \begin{itemize}
              {\footnotesize  
                \item \textbf{Polyid}: The index numbers labeling the polytope in the database.
          \item \textbf{Tri\textunderscore{}id}: The index of triangulation in a given polytope.
        \item \textbf{BP\textunderscore{}not\textunderscore{}in\textunderscore{}F}: Boundary points not interior to facets of a polytope, corresponding to the vertex in the dual-polytope $\Delta^{*}$ of the toric ambient space.
        \item \textbf{F\textunderscore{}Intsec}: Quartet intersection number on the ambient space $\cA$.
         \item \textbf{triple\textunderscore{}inters}: Triple intersection number on the Calabi-Yau threefolds $X$.
         \item \textbf{INVOL}: The involution considered in this example.
        \item \textbf{KK\textunderscore{}data}: The geometric data of this polytope mentioned in Kreuzer-Skarke database. 
        \item \textbf{Linear\textunderscore{}I}: Linear ideal of the polytope. 
        \item \textbf{OPLANES}: The locus and type of O-plane, followed by their contribution to D3-tadpole locally labeled as \lq\lq value".
         \begin{verbatim}    
            { type of O-plane: [[Fixed Locus_1], number of O-plane_1,...]
              ...
              "tadpole_cancel": True or False,
              "value": contribution to D3-tadpole
              "[h12p,h12n,smooth]" :  [h21+, h21-, whether P_sym is smooth]
             }
                \end{verbatim}    

        \item \textbf{P\textunderscore{}symm1}: The polynomail expression of invariant hypersurface under the involution.
        \item \textbf{Polys}: Generators of $\cG$ for exchange involutions.
        \item \textbf{Q\textunderscore{}parity}: The parity of the holomorhpic three form under involution.
        \item \textbf{Q\textunderscore{}str}: The expression of Q.
        \item \textbf{SR\textunderscore{}list}: The  Stanley-Reisner ideal.
        \item \textbf{Sectors}: Different patches of the polytope associated to the SR-ideal.
        \item \textbf{Wmatrix2}: GLSM weighted matrix  $\bW$ together with the degree of hypersurface.
        \item \textbf{divisor\textunderscore{}indenp}: Independent divisors chosen to be the basis of divisor class.
        \item \textbf{divisors\textunderscore{}Hodge}: The Hodge numbers of each divisors.
        \item \textbf{Hodge\textunderscore{}split\textunderscore{}p\textunderscore{}n}: The value of $[h^{1,1}_+, h^{1,1}_-]$.       
        \item \textbf{rwmat\textunderscore{}for\textunderscore{}y\textunderscore{}notreduced}: The new weight matrix $\tilde\bW$ after Segre embedding.
        }
        \end{itemize}
   
   \subsubsection{Reflections}
   The terminologes for reflections are  almost the same as  divisor exchange involution, except the single, double and triple reflections are labeled as  {\bf invol1},  {\bf invol2} and  {\bf invol3}.
    The new type of free action described in Section \ref{subsec:smoothfree} for multi-reflections are labeled  as {\rm \lq\lq total fixed"}. The format is as follows:
  \begin{verbatim}    
[Reflection]: type of O-plane, contribution to D3-tadpole, Fixed Locus
\end{verbatim}

        \begin{itemize}
              {\footnotesize  
     \item  \textbf{invol1}: Type of O-plane, its contribution to D3-tadpole  and the fixed locus  for single reflection.
      \item  \textbf{invol2}: Type of O-plane, its contribution to D3-tadpole  and the fixed locus  for double reflection.
       \item  \textbf{invol3}: Type of O-plane, its contribution to D3-tadpole  and the fixed locus  for triple reflection.
            }
     \end{itemize}    
        
        \subsection{Entry for Example A}
        \label{ap:example}

     \subsubsection{Divisor exchange involution}   
        \footnotesize{
        \begin{itemize}
          \item \textbf{polyid}: 545
  \item \textbf{tri\textunderscore{}id}: 5
        \item \textbf{BP\textunderscore{}not\textunderscore{}in\textunderscore{}F}: 
\begin{verbatim}
[
[-2, -3, -2, -2], [0, 1, 0, 0], [0, 1, 0, 2], [0, 1, 2, 0],
[1, 0, 0, 0], [-1, -1, -1, -1], [-1, -1, -1, 0], [-1, -1, 0, -1],
[0, 1, 0, 1], [0, 1, 1, 0], [0, 1, 1, 1]
]
\end{verbatim} 
        \item \textbf{F\textunderscore{}Intsec}: 
\begin{verbatim}
[
[[[2, 0, 0, 0, 0, 0, 0], [0, -2, 0, 0, 0, 0, 0], [0, 0, -2, 0, 0, 0, 0], 
[0, 0, 0, -2, 0, 0, 0], [0, 0, 0, 0, -2, 0, 0], [0, 0, 0, 0, 0, -2, 0], 
[0, 0, 0, 0, 0, 0, -2]],..., [[-2, 0, -2, 0, -2, 0, 0], 
[0, 0, 0, 0, 0, 0, 0], [-2, 0, -2, 0, -2, 0, 0], [0, 0, 0, 0, 0, 0, 0],
[-2, 0, -2, 0, -2, 0, 0], [0, 0, 0, 0, 0, 0, 0], [0, 0, 0, 0, 0, 0, -2]]]
]
\end{verbatim} 
  \item \textbf{triple\textunderscore{}inters}: 
\begin{verbatim}
{(1, 5, 7): 3, (1, 6, 7): 3, (1, 5, 6): 3, (1, 5, 8): 3, (1, 6, 8): 3,
...
, (0, 0, 5): 54, (0, 0, 2): 27, (0, 0, 3): 27, (0, 0, 4): 27, (0, 0, 0): 162}
\end{verbatim}     
        \item \textbf{INVOL}: 
\begin{verbatim}
[[7, 9], [6, 8], [0, 1]]
\end{verbatim}  
        \item \textbf{KK\textunderscore{}data}:
        \begin{verbatim}
4 5  M:12 5 N:40 5 H:37,7 [60]
        \end{verbatim}
        \item \textbf{Linear\textunderscore{}I}: 
        \begin{verbatim}
["1/2*z5 - 1/2*z6 - 1/2*z7 - 1/2*z8",
"1/2*z5 - 1/2*z6 - 1/2*z9 - 1/2*z10",
"1/2*z5 - 1/2*z7 - 1/2*z9 - 1/2*z11",
"1/2*z5 - 1/2*z8 - 1/2*z10 - 1/2*z11",
"z5",
"z6",
"z7",
"z8",
"z9",
"z10",
"z11"]
        \end{verbatim}
        \item \textbf{OPLANES}: 
        \begin{verbatim}
{
  "O3": [
    [
      "[x3, x4, x10]",
      3.0
    ],
    [
      "[x2, x4, x10]",
      3.0
    ],
    [
      "[x3, x5, x0^2*x6*x7 + x1^2*x8*x9]",
      3.0
    ]
  ],
  "O7": [
    [
      "[x0^2*x6*x7 - x1^2*x8*x9]",
      1
    ]
  ],
  "tadpole_cancel": true,
  "value": 36.0,
  "[h12p, h12n, smooth]": "[12.0, 25.0, True]"
}
        \end{verbatim}
        \item \textbf{P\textunderscore{}symm1}: 
        \begin{verbatim}
x0^6*x5^3*x6^3*x7^3 + x0^4*x1^2*x5^3*x6^2*x7^2*x8*x9 +
x0^2*x1^4*x5^3*x6*x7*x8^2*x9^2 + x1^6*x5^3*x8^3*x9^3 + 
... 
+ x2^2*x4^2*x6*x8*x10 + x3^2*x4^2*x7*x9*x10 + x4^3
        \end{verbatim}
        \item \textbf{Polys}: 
        \begin{verbatim}
[x7*x9, x6*x8, x0*x1, x0^2*x6*x7 + x1^2*x8*x9, x0^2*x6*x7 - x1^2*x8*x9]
        \end{verbatim}
        \item \textbf{Q\textunderscore{}parity}: -1
        \item \textbf{Q\textunderscore{}str}: \\
        \begin{verbatim}
+(1 z1z2z3z4z5z6z7 dz8dz9dz10dz11) +(1 z1z2z3z4z5z6z8 dz7dz9dz10dz11) 
-(1 z1z2z3z4z5z6z9 dz7dz8dz10dz11) -(1 z1z2z3z4z5z6z10 dz7dz8dz9dz11)
...
+(4 z4z6z7z8z9z10z11 dz1dz2dz3dz5) -(8 z5z6z7z8z9z10z11 dz1dz2dz3dz4)
        \end{verbatim}      
        \item \textbf{SR\textunderscore{}list}:         
        \begin{verbatim}
[
[z1, z2], [z1, z3], [z1, z4], [z1, z9], [z1, z10], [z1, z11], [z2, z3],
[z2, z4], [z2, z7], [z2, z8], [z2, z11], [z3, z4], [z3, z8], [z3, z10], 
[z4, z5, z6, z7], [z4, z5, z6, z9], [z5, z6, z7, z8], [z5, z6, z9, z10], 
[z5, z6, z11], [z7, z9], [z7, z10], [z8, z9], [z8, z10], [z8, z11], [z10, z11]
]
        \end{verbatim}
        \item \textbf{Sectors}: 
        \begin{verbatim}
[
[x0 - 1, x1 - 1, x2 - 1, x3 - 1, x7 - 1, x9 - 1, x10 - 1], 
[x0 - 1, x1 - 1, x2 - 1, x4 - 1, x5 - 1, x7 - 1, x9 - 1], 
[x0 - 1, x1 - 1, x2 - 1, x4 - 1, x6 - 1, x7 - 1, x10 - 1],
...
[x1 - 1, x2 - 1, x3 - 1, x5 - 1, x8 - 1, x9 - 1, x10 - 1],
[x1 - 1, x2 - 1, x3 - 1, x6 - 1, x8 - 1, x9 - 1, x10 - 1],
[x1 - 1, x2 - 1, x3 - 1, x7 - 1, x8 - 1, x9 - 1, x10 - 1]
]
        \end{verbatim}
        \item \textbf{Wmatrix2}: 
        \begin{verbatim}
[
[-6  1 -1  0  0  2  0  0  0  2  2  0]
[ 0  0  1  1  0  0  0  0  0 -2  0  0]
[ 0  0  1  0  1  0  0  0  0  0 -2  0]
[-3  0 -1  0  0  1  1  0  0  1  1  0]
[-3  0  0  0  0  1  0  1  0  0  1  0]
[-3  0  0  0  0  1  0  0  1  1  0  0]
[ 0  0  1  0  0  0  0  0  0 -1 -1  1]
]
        \end{verbatim}
        \item \textbf{divisor\textunderscore{}indenp}: 
\begin{verbatim}
['z5', 'z6', 'z7', 'z8', 'z9', 'z10', 'z11']
\end{verbatim}    
        \item \textbf{divisors\textunderscore{}Hodge}: 
\begin{verbatim}
[
'[1, 0, 0, 7]', '[1, 0, 0, 7]', '[1, 0, 0, 7]', '[1, 0, 0, 16]', 
'[1, 0, 4, 44]', '[1, 1, 0, 11]', '[1, 1, 0, 11]', '[1, 1, 0, 2]', 
'[1, 1, 0, 11]', '[1, 1, 0, 2]', '[1, 1, 0, 2]'
]
\end{verbatim}    
        \item \textbf{Hodge\textunderscore{}split\textunderscore{}p\textunderscore{}n}:
\begin{verbatim}
[5, 2]
\end{verbatim}    
      
        \item \textbf{rwmat\textunderscore{}for\textunderscore{}y\textunderscore{}notreduced}: 
        \begin{verbatim}
[
[ 2  0  1  0  2  0  1  2  2  0]
[ 0 -2  1  0  0  0  1  0  0  0]
[-2  2 -1  1  0  0  0  0  0  0]
[ 1 -1  1  0  1  1  0  0  0  0]
[ 1  1  0  0  1  0  0  1  1  0]
[ 1  1  0  0  1  0  0  1  1  0]
[-1  1 -1  0  0  0  0  0  0  1]
]
        \end{verbatim}       
        \end{itemize}
        }
        
        \subsubsection{Reflection}	
        \label{subsubsec:formatref}
                \begin{itemize}
                 \item \textbf{invol1:}
      \begin{verbatim}    
    [0]: O3: 3 [x5, x6, x8], O7: 9 [x0], O7: 9 [x1], O7: 9 [x2], O7: 18 [x3]
    [1]: O3: 3 [x5, x6, x8], O7: 9 [x1], O7: [x0], O7: 9 [x2], O7: 18 [x3]
    ...
    [10]: O3: 3 [x0, x6, x7], O3: 3 [x1, x8, x9], O3: 3 [x3, x6, x7], 
          O3: 3 [x3, x8, x9], O3: 1 [x2, x4, x6, x8], O7: 7 [x4, x5]
              \end{verbatim}
    \item  \textbf{invol2:}
              \begin{verbatim}
    [6, 8]: O5: [x6, x8], O5: [x0, x5], O5: [x1, x5], O5: [x3, x5], O5: [x2, x4, x10], 
            O5: [x3, x4, x10], O5: [x4, x6, x7], O5: [x4, x8, x9]
    [0, 1]: [[x0], [x1], [x2], [x3], [x4], [x5], [x6], [x7], [x8], [x9], [x10]] total fixed
    ...
    [1, 6]: O5: [x0, x6], O5: [x2, x6], O5: [x3, x6], O5: [x5, x8], O5: [x1, x4, x9], 
            O5: [x3, x4, x9], O5: [x4, x5, x7], O5: [x4, x8, x10]
          \end{verbatim}
              \item  \textbf{invol3:}
              \begin{verbatim}
    [3, 4, 5]: O3: 3 [x0, x4, x5], O3: 3 [x1, x4, x5], O7: 3 [x2, x10], 
               O7: 3 [x3, x10], O7: 4 [x6, x7], O7: 4 [x8, x9]
    [2, 7, 8]: O3: 1 [x4, x5, x6, x8], O7: 3 [x0, x4], O7: 3 [x1, x4], 
               O7: 3 [x2, x4], O7: 6 [x3, x4]
    ...
    [0, 4, 6]: O3: 3 [x0, x4, x6], O3: 3 [x2, x4, x6], O7: 3 [x1, x9], 
               O7: 3 [x3, x9], O7: 4 [x5, x7], O7: 4 [x8, x10]
          \end{verbatim}
       \end{itemize}


	\nocite{*}
	\bibliography{orintifoldCY}
	\bibliographystyle{utphys}

\end{document}